\documentclass[aps,prd,nofootinbib,twocolumn,superscriptaddress,letterpaper,preprintnumbers, longbibliography]{revtex4-1}

\usepackage{amssymb}
\usepackage{amsmath}
\usepackage{epsfig}
\usepackage[colorlinks=true
,urlcolor=blue
,anchorcolor=blue
,citecolor=blue
,filecolor=blue
,linkcolor=red
,menucolor=blue
,hyperfootnotes=false,
,linktocpage=true
,pdfproducer=medialab
,pdfa=true
]{hyperref}
\usepackage{array}
\usepackage{booktabs}
\usepackage{comment}
\usepackage{cleveref}
\usepackage{multirow}

\newcolumntype{L}[1]{>{\raggedright\let\newline\\\arraybackslash\hspace{0pt}}m{#1}}
\newcolumntype{C}[1]{>{\centering\let\newline\\\arraybackslash\hspace{0pt}}m{#1}}
\newcolumntype{R}[1]{>{\raggedleft\let\newline\\\arraybackslash\hspace{0pt}}m{#1}}

\usepackage{afterpage}
\usepackage{longtable}
\usepackage{capt-of}

\usepackage{siunitx}
\DeclareSIUnit\electronvolt{e\kern-.05em V}
\DeclareSIUnit\eV{e\kern-.05em V}
\DeclareSIUnit\parsec{pc}
\usepackage{graphicx}
\usepackage{gensymb}
\usepackage{subfigure}
\usepackage{blindtext}

\usepackage{listings}
\usepackage[dvipsnames]{xcolor}
\usepackage[T1]{fontenc}

\definecolor{codegreen}{rgb}{0,0.6,0}
\definecolor{codegray}{rgb}{0.5,0.5,0.5}
\definecolor{codepurple}{rgb}{0.58,0,0.82}
\definecolor{backcolour}{rgb}{0.95,0.95,0.92}

\lstdefinestyle{mystyle}{
    language = Python,  
    commentstyle=\color{codegreen},
    keywordstyle=\color{magenta},
    keywordstyle=[1]\color[rgb]{0,0,0.75},
    keywordstyle=[2]\color[rgb]{0.5,0.0,0.0},
    keywordstyle=[3]\color[rgb]{0.127,0.427,0.514},
    keywordstyle=[4]\color[rgb]{0.4,0.4,0.4},
    commentstyle=\color[rgb]{0.133,0.545,0.133},
    numberstyle=\tiny\color{codegray},
    stringstyle=\color{codepurple},
    basicstyle=\ttfamily,
    breakatwhitespace=false,         
    breaklines=true,                 
    captionpos=b,                    
    keepspaces=false,                 
    numbers=none,                    
    numbersep=5pt,                  
    showspaces=false,                
    showstringspaces=false,
    showtabs=false,                  
    tabsize=2,
    morekeywords={True, False, len},
    columns=flexible
}

\newcommand\beq{\begin{alignat}{1}}
\newcommand\eeq{\end{alignat}}
\newcommand{\dhis}{\texttt{DarkHistory} }

\newcommand*\bbar[1]{%
  \vbox{%
    \hrule height 0.5pt
    \kern-0.4ex
    \hbox{%
      \kern-0.2em
      \ifmmode#1\else\ensuremath{#1}\fi
      \kern-0.1em
    }
  }
}

\defcitealias{paperI}{Paper~I}

\begin{document}

\preprint{MIT-CTP/5524} 

\title{
	Exotic energy injection in the early universe II: \\
	CMB spectral distortions and constraints on light dark matter
}

\author{Hongwan Liu}
\email{hongwanl@princeton.edu}
\affiliation{Center for Cosmology and Particle Physics, Department of Physics, New York University, New York, NY 10003, U.S.A.}
\affiliation{Department of Physics, Princeton University, Princeton, New Jersey, 08544, U.S.A.}

\author{Wenzer Qin}
\email{wenzerq@mit.edu}
\affiliation{Center for Theoretical Physics, Massachusetts Institute of Technology, Cambridge, MA 02139, U.S.A.}

\author{Gregory W. Ridgway}
\email{gridgway@mit.edu}
\affiliation{Center for Theoretical Physics, Massachusetts Institute of Technology, Cambridge, MA 02139, U.S.A.}

\author{Tracy R. Slatyer}
\email{tslatyer@mit.edu}
\affiliation{Center for Theoretical Physics, Massachusetts Institute of Technology, Cambridge, MA 02139, U.S.A.}

\begin{abstract} 
	We calculate the post-recombination contribution to the Cosmic Microwave Background (CMB) spectral distortion due to general exotic energy injections, including dark matter (DM) decaying or annihilating to Standard Model particles.
	Upon subtracting the background distortion that would be present even without such energy injections, we find residual distortions that are still potentially large enough to be detectable by future experiments such as PIXIE. 
	The distortions also have a high-energy spectral feature that is a unique signature of the injection of high-energy particles.
	We present a calculation of the global ionization history in the presence of decaying dark matter with sub-keV masses, and also show that previous calculations of the global ionization history in the presence of energy injection are not significantly modified by these additional spectral distortions. 
	Our improved treatment of low-energy electrons allows us to extend calculations of the CMB anisotropy constraints for decaying DM down to arbitrarily low masses. 
	We also recast these bounds as constraints on the coupling of axion-like particles to photons.
\end{abstract}

\maketitle

\section{Introduction}

Exotic sources of energy injection, such as dark matter (DM) annihilating or decaying into Standard Model (SM) particles, can inject a significant amount of energy into the universe.
This additional energy can manifest as modifications to the global ionization history $x_\text{HII} \equiv n_\text{HII} / n_\text{H}$, where $n_\text{HII}$ and $n_\text{H}$ are the number densities of ionized hydrogen and all hydrogen nuclei respectively, and the intergalactic medium (IGM) temperature history $T_m (z)$; thus, observations of these quantities can be used to constrain the properties of DM~\cite{0906.1197, 0907.0719, 0907.3985, 1308.2578, 1408.1109, 1506.03811, 1603.06795, 1604.02457, 1610.06933, 1610.10051,1710.00700, 1803.03629, 1803.09739, 1803.09398, 1803.11169, 1803.09390, 1808.04367, 1911.05086, 2001.10018, 2002.05165, 2003.13698,2008.01084,2009.00016,2206.13520}.
Moreover, exotic energy injection can modify the background of low-energy photons.

Changes to the background spectrum of photons can have a number of interesting implications---for example, exotic energy injection could explain observed excesses, such as that of the cosmic optical background~\cite{Lauer:2022fgc, Bernal:2022wsu}.
In addition, excess Lyman-$\alpha$ radiation~\cite{Hirata:2005mz} can affect the global and inhomogeneous redshifted 21cm signal via the Wouthuysen-Field effect. 
Star formation is also affected by the background radiation; for example, the formation of molecular hydrogen, which is required to form the first stars, can be disrupted by photodetachment of intermediate states~\cite{Hirata:2006bt}, or photodissociation by photons in the Lyman-Werner band~\cite{1967ApJ...149L..29S,1992A&A...253..525A,1996ApJ...467..522H}.

The distortion of the Cosmic Microwave Background due to exotic energy injection is of particular interest.
COBE/FIRAS measured the CMB energy spectrum to be a perfect blackbody within about 
one part in $10^4$ over the frequency range of about 60 to 630 GHz~\cite{COBE_FIRAS_Mather, COBE_FIRAS_Fixsen}.
The balloon-borne ARCADE 2 experiment also measured the CMB spectrum between 3 to 90 GHz and reported a rise in the blackbody temperature at this low frequency tail of the spectrum~\cite{2011ApJ...734....5F,2011ApJ...734....6S}.
A proposed next-generation experiment that can improve upon these measurements is the Primordial Inflation Explorer (PIXIE)~\cite{2011JCAP...07..025K}; there are also a number of other proposed efforts which could improve sensitivity to the CMB energy spectrum by orders of magnitude~\cite{Chluba:2019nxa,Kogut:2019vqh,Maffei:2021xur,Chang:2022tzj}.

Distortions to the blackbody spectrum can arise when matter and radiation are driven out of thermal equilibrium after $z \lesssim 2 \times 10^6$, i.e. once thermalization is no longer efficient~\cite{Chluba:2019kpb}.
Spectral distortions from energy release in the early universe are often characterized by $\mu$ and $y$-type distortions. 
$\mu$-type distortions arise when Compton scattering is still fast enough to maintain the photons and electrons in kinetic equilibrium but photon number-changing processes are inefficient, causing the blackbody to develop a chemical potential~\cite{Illarionov_Sunyaev}.
$y$-type distortions form later, once Compton scattering is also inefficient and unable to hold the photon bath in full kinetic equilibrium~\cite{Zeldovich_Sunyaev}, and have magnitude controlled by the ``$y$-parameter'', which roughly describes the energy density exchanged by scattering off of electrons as a fraction of the total CMB energy density~\cite{Chluba:2018cww}.
The $y$-parameter can be non-zero even at very early redshifts, when the radiation and matter temperature are tightly coupled.
Even within the standard $\Lambda$CDM cosmological model, we expect such distortions to the CMB blackbody spectrum to arise due to reionization and structure formation; the overall size of the distortion from reionization could reach a $y$-parameter of up to $y \sim 10^{-6}$~\cite{Chluba:2016bvg, Chluba:2019kpb}. 

In addition to Compton scattering, distortions can be sourced by atomic transitions; in particular, such distortions arising from recombination have received much attention.
Earlier studies, such as Ref.~\cite{1993ASPC...51..548R} calculated these distortions using up to 10 excited states of hydrogen.
More recent studies, such as Refs.~\cite{Jens2006, Chluba:2006bc}, refined this calculation and included up to 100 energy levels of hydrogen.

Processes beyond $\Lambda$CDM such as DM annihilation or decay could lead to the injection of non-thermal, electromagnetically interacting particles into the early universe, potentially producing a large number of secondary photons as the injected particles cool and lose their energy. The effects of such energy injections, and DM interactions with the SM bath more broadly, on spectral distortions has been examined from a number of perspectives.
For example, limits on CMB distortions have been used to constrain DM scattering on SM particles~\cite{Ali-Haimoud:2015pwa}, energy injection from dark photons~\cite{1911.05086}, and axions and axion-like particles (ALPs)~\cite{Bolliet:2020ofj}.
Multiple studies have also used Green's functions to more generally study spectral distortions from heating~\cite{Chluba:2013vsa}, photon injection~\cite{Chluba:2015hma}, and electron injection~\cite{Acharya:2018iwh}. However, the focus of this latter set of studies has been the epoch prior to recombination, when the universe can be well-approximated as fully ionized, and photons scatter rapidly with electrons.

In addition to inducing spectral distortions, exotic energy injection can also affect CMB anisotropies by modifying the global ionization history~\cite{Adams:1998nr,Chen:2003gz,Padmanabhan:2005es}.
This has been studied in the context of both decaying~\cite{Zhang:2007zzh,1610.06933,1610.10051,Acharya:2019uba,Cang:2020exa} and annihilating DM~\cite{Galli:2009zc,0906.1197,0907.3985,Hisano:2011dc,Hutsi:2011vx,Galli:2011rz,2012PhRvD..85d3522F,Slatyer:2012yq,Galli:2013dna,Madhavacheril:2013cna,1506.03811,1506.03812}.
In many of these studies, the constraints were either applied only to a set of benchmark DM models; in studies where the constraints were calculated more generically, the limits only extended to DM masses corresponding to particle injection energies above a few keV. For example, in the work of Ref.~\cite{1506.03811}, the limits were not extended to sub-keV DM on the basis of approximations made in the analysis that were likely to break down for sufficiently low DM masses (approaching the hydrogen and helium ionization thresholds).

In a companion paper, hereafter referred to as~\citetalias{paperI}, we describe recent improvements made to the \dhis code package~\cite{DH} to better track the behavior of low-energy electrons and photons, either directly injected or arising as part of a secondary particle production cascade, in the universe at $z \lesssim 3000$. 
We treat the energy deposition as homogeneous, a common assumption also made in previous works~\cite{DH,2008.01084}.
In this work, we apply the tools that we developed in~\citetalias{paperI} to obtain several immediate and significant results. 
In particular, we calculate the CMB spectral distortions arising from exotic energy injections during the post-recombination epoch for the first time, and extend CMB anisotropy constraints based on the ionization history to lower DM masses. 
We also demonstrate how additional low-energy photons produced by exotic energy injections back-react on the process of recombination itself, finding a small effect on the global ionization history.  

In Section~\ref{sec:distortions}, we show the spectral distortions resulting from energy injection between $3000 > 1+z > 4$ and compare their amplitude to the sensitivity of future experiments, as well as contributions to the distortion from other redshift ranges.
In Section~\ref{sec:ionization}, we show that although these spectral distortions could in principle modify the ionization history, changes to the ionization history compared to the \dhis \texttt{v1.0} calculation are not significant.
In Section~\ref{sec:anisotropy}, we extend the CMB anisotropy constraints on DM decaying to photons, and also translate this into a constraint on the ALP-photon coupling.
We conclude in Section~\ref{sec:conclusion}. 
Throughout this work, we will use natural units, where $c = \hbar = k_B = 1$.

\begin{figure*}
	\includegraphics[width=0.8\textwidth]{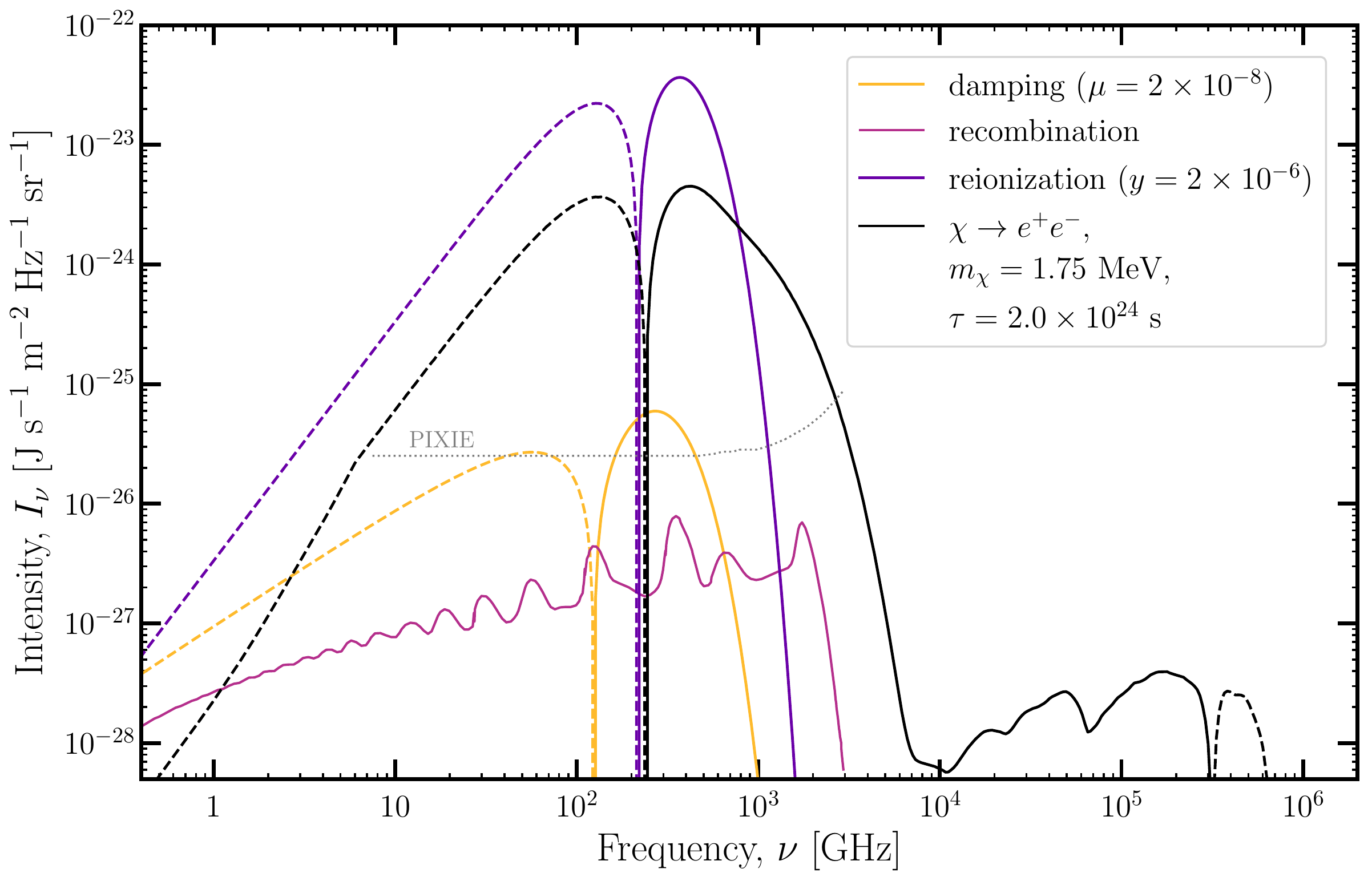}
	\caption{
		Predicted spectral distortions from $\Lambda$CDM, including the average $\mu$-distortion expected from Silk damping (yellow), the signal from recombination (magenta), and the $y$-distortion expected from reionization and structure formation (dark purple).
		Dashed lines indicate where the spectral distortions have negative values.
		The gray dotted line is the expected sensitivity of PIXIE~\cite{2011JCAP...07..025K, 2016SPIE.9904E..0WK}.
		Shown in black is the largest spectral distortion from DM energy injection in our study (the $\Lambda$CDM contributions, including contributions from a model of reionization, have been subtracted out from this line), restricting to scenarios that are not otherwise excluded.
	}
	\label{fig:LCDM}
\end{figure*}
%

\section{CMB Distortions}
\label{sec:distortions}

Hereafter, we define the CMB spectral distortion as the difference between the total photon intensity $I_\omega$ and the intensity of a blackbody at the CMB temperature $I_{\omega, \mathrm{CMB}}$:
\begin{equation}
	\Delta I_\omega = I_\omega - I_{\omega, \mathrm{CMB}} .
\end{equation}
Within $\Lambda$CDM cosmology, spectral distortions can be sourced by a number of processes~\cite{Chluba:2016bvg,Chluba:2019nxa}.
Prior to redshifts of $z \sim \text{few} \times 10^6$, any injected energy thermalizes very quickly and only changes the temperature of the CMB blackbody, leading to no discernible distortion.
Below this redshift, processes that do not conserve photon number become inefficient, while photons and electrons remain in thermal contact; energy injection therefore causes the photon bath to develop a chemical potential, $\mu$.
This leads to a $\mu$-type distortion, given by
\begin{equation}
	\Delta I_{\omega, \mu} = \frac{\mu \omega^3}{2\pi^2} \frac{e^{x}}{(e^{x}-1)^2} \left[ \frac{x}{\beta} - 1 \right],
\end{equation}
where $x= \omega / T_\text{CMB}$ and $\beta \simeq 2.1923$~\cite{Chluba:2013vsa}.
As an example, dissipation of primordial density perturbations through Silk damping is expected to contribute a $\mu$-type distortion with parameter $\mu \simeq 2 \times 10^{-8}$~\cite{Chluba:2016bvg}, whereas data from COBE/FIRAS constrains $\mu$ to be less than several times $10^{-5}$~\cite{COBE_FIRAS_Fixsen,Bolliet:2020ofj,Bianchini:2022dqh}.

Around $z \sim 5 \times 10^4$, Comptonization also becomes inefficient and energy injection typically results in $y$-type distortions, which are defined as
\begin{equation}
	\Delta I_{\omega, y} = \frac{y \omega^3}{2\pi^2} \frac{x e^{x}}{(e^{x}-1)^2} \left[ x \coth \left( \frac{x}{2} \right) - 4 \right],
\end{equation}
where the $y$ parameter characterizes the amplitude of the distortion.
For example, heating of the IGM during reionization and structure formation is expected to contribute a $y$-type distortion with amplitude of about $y \simeq 2 \times 10^{-6}$~\cite{Refregier:2000xz,Zhang:2004fh,Dolag:2015dta,Hill:2015tqa,DeZotti:2015awh,Chluba:2016bvg}.
COBE/FIRAS limits such distortions to $y \lesssim 10^{-5}$~\cite{COBE_FIRAS_Fixsen,Bolliet:2020ofj}.

There can also exist distortions that are neither $\mu$-type nor $y$-type~\cite{Chluba:2011hw,Khatri:2012tw,Chluba:2013vsa,Acharya:2018iwh}.
For example, during recombination, as electrons are captured and cascade to the ground state, they emit out-of-equilibrium radiation that appear today as broadened and redshifted atomic lines (helium recombination also contributes a spectral distortion, but this is expected to be an order of magnitude smaller than the signal from hydrogen recombination, hence we only consider hydrogen~\cite{Wong:2005yr,Rubino-Martin:2007tua,Chluba:2019nxa}).
Fig.~\ref{fig:LCDM} shows these predicted signals, as well as the expected sensitivity of PIXIE, which can measure spectral distortions down to $\mu \sim 10^{-8}$ or $y \sim 2\times10^{-9}$~\cite{2011JCAP...07..025K, 2016SPIE.9904E..0WK}.

Exotic energy injections can similarly distort the CMB frequency spectrum.
For example, DM can decay or annihilate to photons, or to SM particles that promptly decay producing photons; these photons could contribute directly to the distortion, but they can also produce secondary electrons and photons through processes such as pair production, Compton scattering, photoionization, or photoexcitation, with subsequent de-excitation and recombination.
Electrons (either produced directly by the energy injection mechanism, or as secondaries) can produce more secondaries through collisional ionization and excitation, and inverse Compton scattering (ICS) off CMB photons.
Especially for higher-energy photons, the timescale for cooling processes can be comparable to or longer than a Hubble time, so it is important to take into account the expansion of the universe during the development of the secondary particle cascade.
Calculating these rich particle cascades with the necessary amount of detail to accurately track the resulting distortion is nontrivial; the numerical methods we have developed for this purpose are described in Ref.~\cite{DH} and \citetalias{paperI}.

For the rest of this paper, we will focus on DM annihilation and decay as two main examples of exotic energy injection; we stress, however, that \texttt{DarkHistory} can handle general energy injection processes.
To include the effects of energy injection by DM in \texttt{DarkHistory}, one need only specify if DM decays or annihilates, the spectrum of decay/annihilation products, the DM mass $m_\chi$, and the lifetime $\tau$ or cross-section $\langle \sigma v \rangle$. 
When calculating the evolution of the temperature, ionization level, and spectral distortions, there are a number of options within \dhis that can be adjusted, depending on which effects one would like to include.
We describe a few of the most relevant options for this work below; see Ref.~\cite{DH} and \citetalias{paperI} for more details.

\begin{itemize}
	\item \texttt{backreaction}: If set to \texttt{False}, then the fraction of energy deposited into each energy deposition channel (e.g.\ heating, hydrogen ionization etc.) is computed assuming the baseline value of $x_e \equiv n_e / n_\text{H}$, the number density ratio of free electrons to all hydrogen nuclei, obtained without exotic energy injection. 
	Otherwise, if \texttt{True}, the effect on how energy is deposited due to modifications of the temperature and ionization histories away from the baseline solutions is taken into account. 
	This is crucial for an accurate determination of $T_m(z)$ for $z \lesssim 200$~\cite{DH}.
	
	\item \texttt{struct\_boost}: When structure formation begins, annihilation is enhanced due to the fact that the average of the squared density exceeds the square of the average density.
	With this option, the annihilation rate of DM is enhanced by a boost factor due to structure formation, based on the model first developed in Ref.~\cite{1604.02457}, and implemented in \dhis in Ref.~\cite{DH}. 
	
	\item \texttt{reion\_switch}: If \texttt{True}, then a model for reionization is included in the evolution.
	
	\begin{itemize}
		\item \texttt{reion\_method}: Specifies which reionization model to use.
		By default, we use \texttt{`Puchwein'}, which is the model given in Ref.~\cite{Puchwein:2018arm}.
	\end{itemize}
	
	\item \texttt{distort}: If \texttt{True}, then \dhis will track the background spectrum of photons and its evolution.
	To accurately calculate the hydrogen atom transitions that contribute to spectral distortions, we treat hydrogen as a Multi-Level Atom (MLA).
	
	\begin{itemize}
		\item \texttt{nmax}: Sets the maximum hydrogen principal quantum number to track in the MLA.
		
		\item \texttt{iterations}: The MLA recombination and photoionization rates are the most computationally expensive quantities to compute; hence, we have developed an iterative method for their calculation, described in \citetalias{paperI}. 
		This option sets the number of iterations over which to improve the calculation.
		
		\item \texttt{reprocess\_distortion}: The atomic transition rates for the MLA depend on the state of the background radiation and affect subsequent emissions and absorptions by hydrogen atoms.
		If this option is set to \texttt{False}, then the rates are calculated assuming the CMB is a perfect blackbody; i.e. spectral distortions are not ``reprocessed'' through the MLA.
		If \texttt{True}, then we also account for the effect of spectral distortions on these rates.
		See \citetalias{paperI} for more details.
	\end{itemize}
\end{itemize}
For clarity, we will be explicit about how these flags are set in the results described below.

\subsection{Results}
\label{sec:dist_results}

%
\begin{figure*}
	\includegraphics[width=\textwidth]{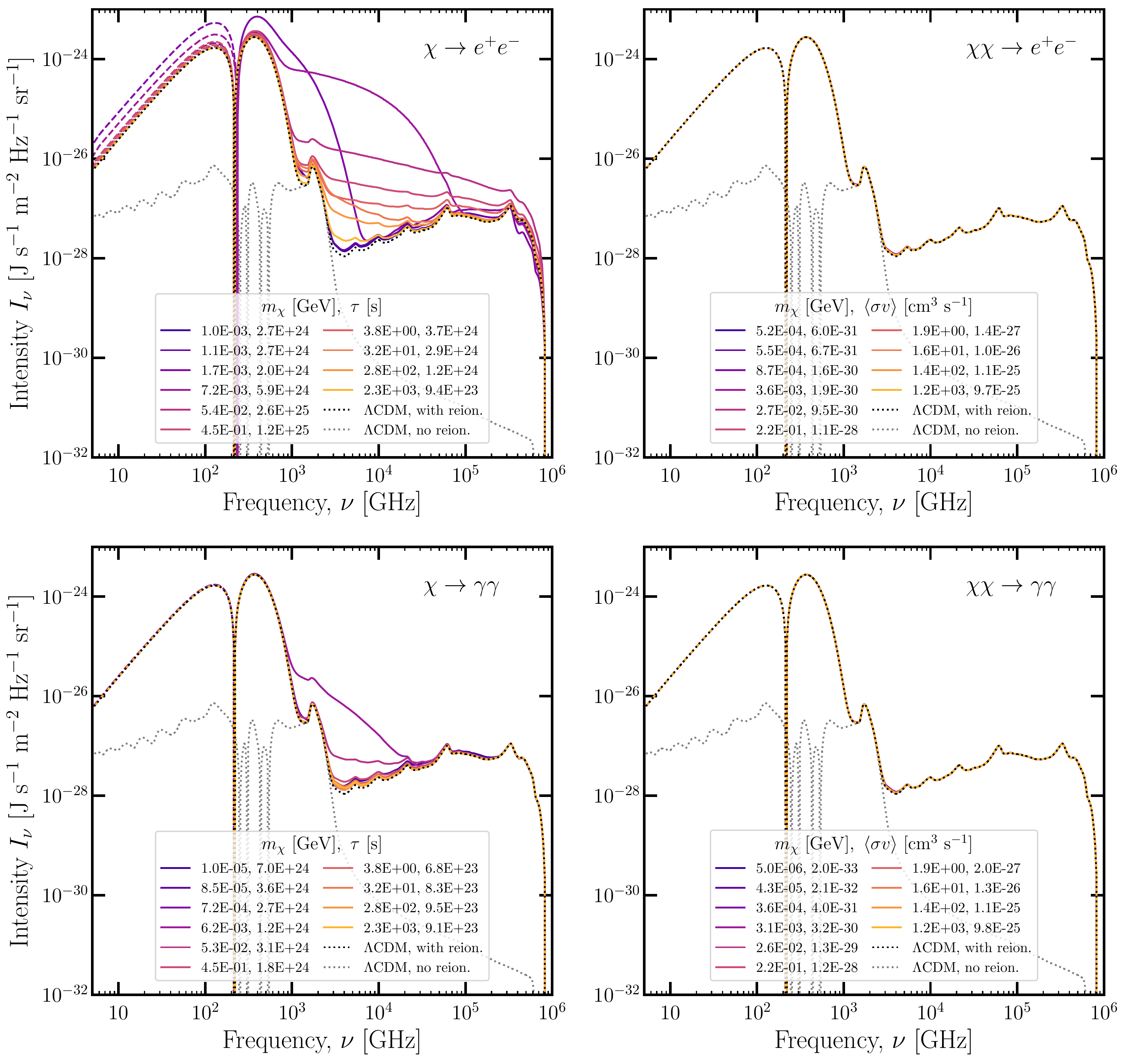}
	\caption{The distortion to the CMB spectrum including different types of DM energy injection; this distortion is relative to the CMB blackbody and hence includes components from $\Lambda$CDM processes.
		For each mass, the lifetime/cross-section is chosen to be at the edge of the CMB constraints given in Refs.~\cite{1610.06933,1506.03811}.
		The left panels show the effect of decaying DM, while the right is from $s$-wave annihilation.
		In the top row, the decay/annihilation products are $e^+ e^-$ pairs, while the bottom row is for photon pairs.
		Negative values for the distortion are represented by colored dashed lines.
		The black dotted line in each panel shows the spectral distortion expected in the absence of exotic energy injection and including a reionization model; the grey dotted line shows the result when we also turn off reionization.}
	\label{fig:dist_grid}
\end{figure*}
\begin{figure*}
	\includegraphics[width=\textwidth]{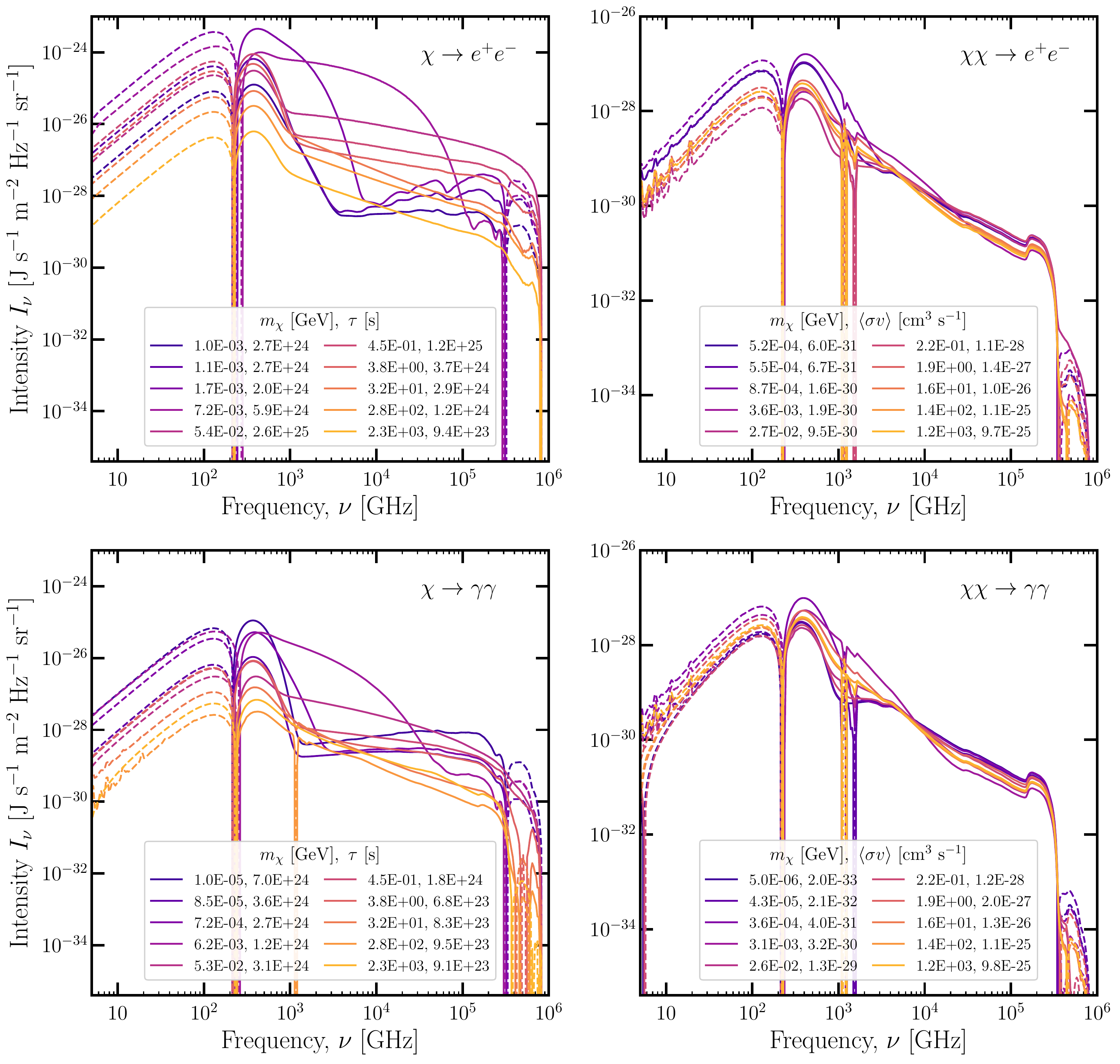}
	\caption{
		Same as Fig.~\ref{fig:dist_grid}, but subtracting the component of the distortion that would be present in the absence of exotic energy injection.
		In other words, here we show the spectral distortion relative to the expected $\Lambda$CDM contribution.
	}
	\label{fig:dist_grid_noLCDM}
\end{figure*}
Fig.~\ref{fig:dist_grid} shows the spectral distortion calculated by \dhis including the effects of DM energy injection for some sample DM models; we emphasize that the public code is capable of repeating these calculations for other final states, and more general exotic injection scenarios. 
We choose scenarios with annihilation or decay purely to photons or $e^+ e^-$ pairs because those are the simplest allowed channels for DM masses below the muon and pion thresholds, which is where these cosmological constraints are most competitive. 
Furthermore, even at higher masses, SM particles heavier than an electron promptly decay, producing electrons, positrons, and photons (as well as neutrinos, which are effectively free-streaming, and (anti)nuclei which typically have a significantly lower branching ratio) so that the results for a general scenario can be estimated by taking linear combinations of the results for photons and $e^+ e^-$.
\dhis includes a submodule than can calculate the electron and photon spectra from the injection of any arbitrary SM particle, based on the \textsc{pppc4dmid} results~\cite{Cirelli:2010xx}.

To be very explicit, the options we set for calculating these distortions are:
\begin{itemize}
	\item  \texttt{backreaction = True},
	
	\item  \texttt{struct\_boost = None},
	
	\item  \texttt{reion\_switch = True},
	
	\item  \texttt{reion\_method = `Puchwein'},
	
	\item  \texttt{distort = True},
	
	\item  \texttt{nmax = 200},
	
	\item  \texttt{iterations = 5},
	
	\item  \texttt{reprocess\_distortion = True}.
\end{itemize}
We conservatively choose not to include structure boost formation in our main results, and discuss the effect of structure formation later in this section.
Regarding \texttt{nmax} and \texttt{iterations}, we choose these values since the ionization level and spectral distortions are converged for \texttt{nmax = 200} and after only one iteration~\cite{paperI}.

For decaying DM, we choose the lifetime at each mass point to be at the minimum lifetime allowed by the CMB anisotropy constraints from Ref.~\cite{1610.06933}; for $s$-wave annihilation, the CMB anisotropy constraints on the cross-section are taken from Ref.~\cite{1506.03811}.
For comparison, we also plot on each panel a black dotted line, which is the spectral distortion calculated by \dhis in the absence of exotic energy injection but including the reionization model; the grey dotted line shows what is left when we also turn off reionization.
The largest distortions come from DM decaying to $e^+ e^-$ pairs at the masses that are least constrained, e.g. around \SIrange{1}{10}{\mega\eV}. 

For the annihilation models, exotic energy injection has a small effect on the CMB blackbody spectrum compared to the distortion that would be expected even in the absence of exotic injections; we call this latter component ``$\Lambda$CDM contributions'' since they are the spectral distortions that would still be present in a $\Lambda$CDM universe.
Hence, the spectral distortions from annihilations are indistinguishable by eye. This indicates that the CMB anisotropy constraints are strong enough to ensure that distortions from annihilation are small relative to the $\Lambda$CDM contribution.
Fig.~\ref{fig:dist_grid_noLCDM} shows the same spectral distortions as in Fig.~\ref{fig:dist_grid}, but with the $\Lambda$CDM contributions subtracted out.
At the lowest frequencies, this difference is very small and subject to numerical instabilities, hence we only show the distortion above a few GHz.
The largest of the distortions from Fig.~\ref{fig:dist_grid_noLCDM} is also plotted on Fig.~\ref{fig:LCDM} for comparison.

\begin{figure}
	\includegraphics[width=\columnwidth]{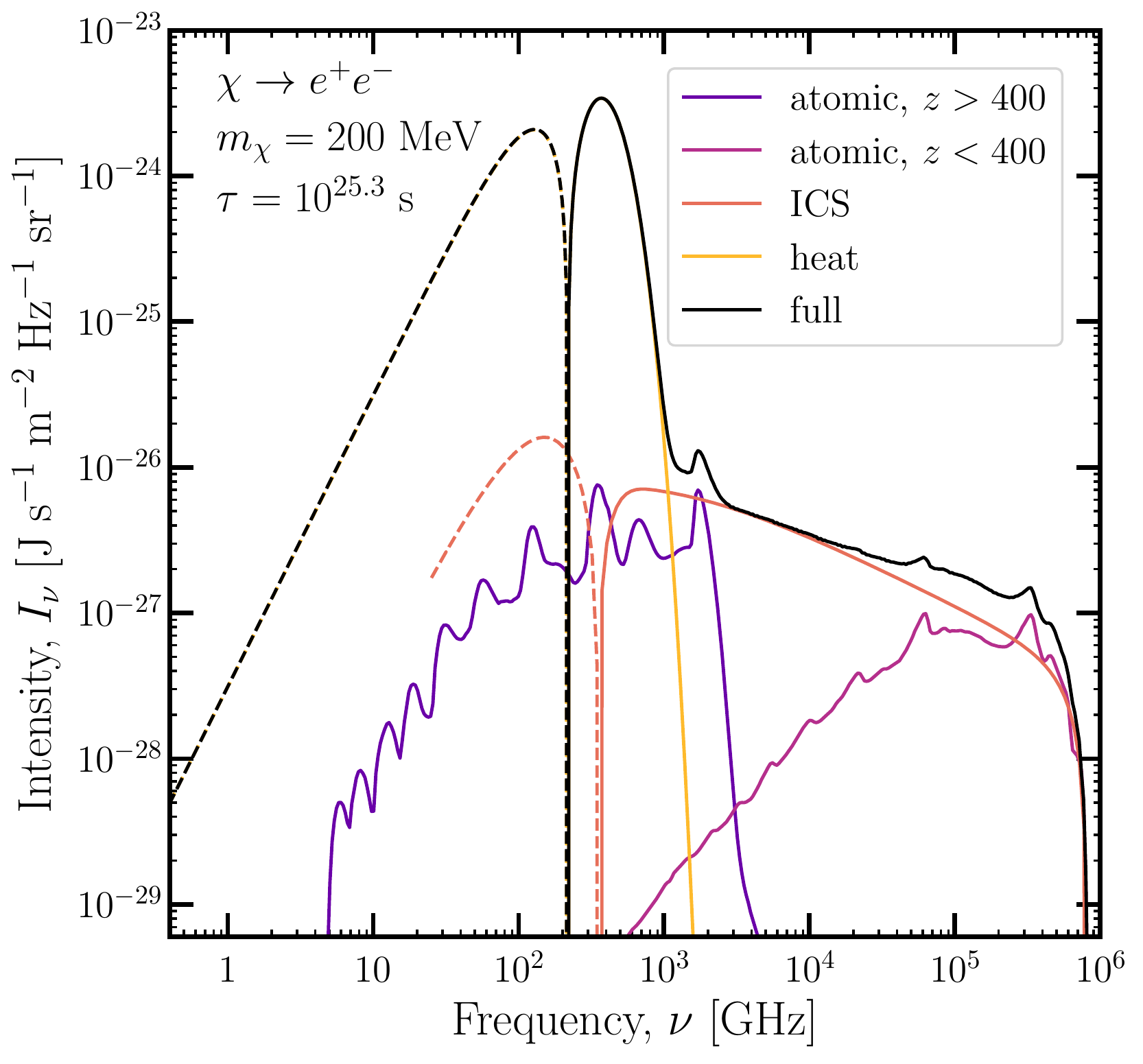}
	\caption{
		The different contributions to the total spectral distortion, including $\Lambda$CDM contributions, due to DM decaying to $e^+ e^-$ pairs, with a mass of $200$ MeV and a lifetime of $10^{25.3}$ s.
		The contributions we show are atomic lines from $z > 400$ which are dominated by the redshifts around recombination (purple), the atomic lines from $z < 400$ which are dominated by the redshifts around reionization (magenta), the photons resulting from ICS (salmon), and the $y$-type distortion resulting from heating of the IGM (yellow).
		The total distortion is shown in black.
	}
	\label{fig:components}
\end{figure}
One can immediately distinguish contributions from various processes and redshifts to the distortions in Fig.~\ref{fig:dist_grid}.
All of the distortions show a trough-peak feature that resembles a $y$-type distortion; this is primarily due to Compton scattering of photons off of a hot IGM heated by DM and reionization.
There is also a narrow peak around 2000 GHz; this is the high energy end of the signal from atomic transitions at recombination, e.g. the magenta line in Fig.~\ref{fig:LCDM}.
Additionally, between about $3000$ and $10^6$ GHz, all the distortions show a spiky shoulder.
This feature contains a smooth component caused by ICS; the spikes are due to the enhancement of atomic line emission around reionization, as neutral hydrogen atoms are raised to excited and ionized states which can subsequently decay and emit line photons.
In Fig.~\ref{fig:components}, we show an example of these different contributions to the distortion resulting from DM decaying to $e^+ e^-$ pairs, with a mass of $200$ MeV and a lifetime of $10^{25.3}$ s.

By comparing the black and grey dotted lines in Fig.~\ref{fig:dist_grid}, we can clearly see that reionization has a large impact on the spectral distortion.
When we set \texttt{reion = True}, we include additional terms in the ionization and temperature evolution, which contribute in a number of different ways: the increase in temperature at late times leads to a large $y$-parameter, and the increasing ionization leads to additional emission of atomic lines.
In principle, our MLA treatment is also affected by the radiation fields that cause reionization in the first place, which could dominate over the late-time spectral distortion at certain frequencies.
One could add a model for these extra radiation fields as a new source of injected photons in \texttt{DarkHistory}; however, for this work we choose only to include the effect of reionization on the spectral distortions through the terms for the ionization and temperature equations (see e.g.\ Eq.~(36) in \citetalias{paperI}).

In addition, since the energy injection rate from $s$-wave annihilation goes as $\rho_\chi^2 \propto (1+z)^6$, where $\rho_\chi$ is the energy density in dark matter, and the same rate for decay goes as $\rho_\chi \propto (1+z)^3$, then decaying DM tends to modify the distortion more at late times compared to annihilation; hence, the $y$-distortion from late time heating is much larger for decay scenarios.
Currently unconstrained decaying DM models will generally give signals large enough to be detectable by future experiments such as PIXIE~\cite{2011JCAP...07..025K, 2016SPIE.9904E..0WK}. 

\begin{figure*}
	\includegraphics[width=\columnwidth]{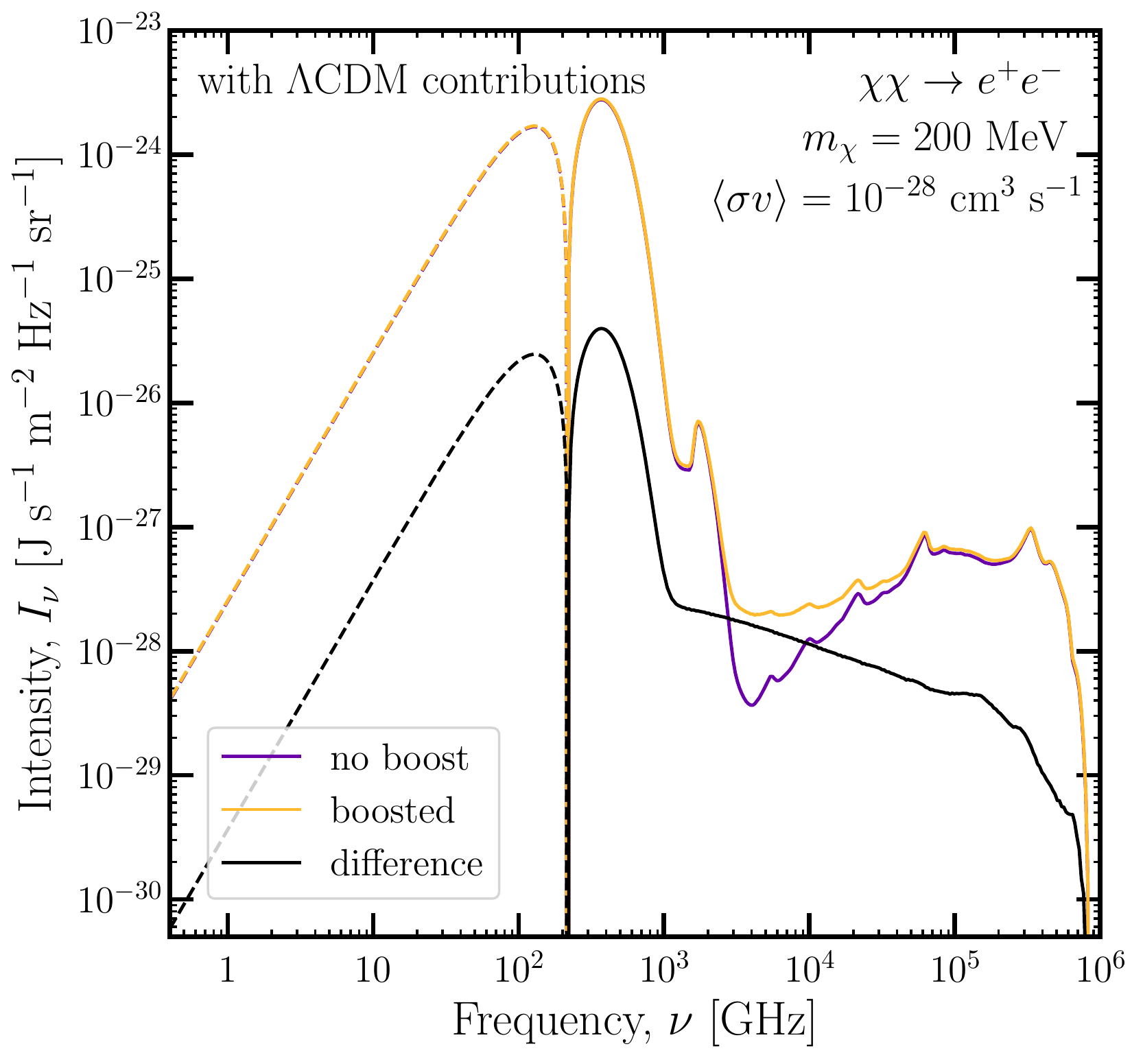}
	\includegraphics[width=\columnwidth]{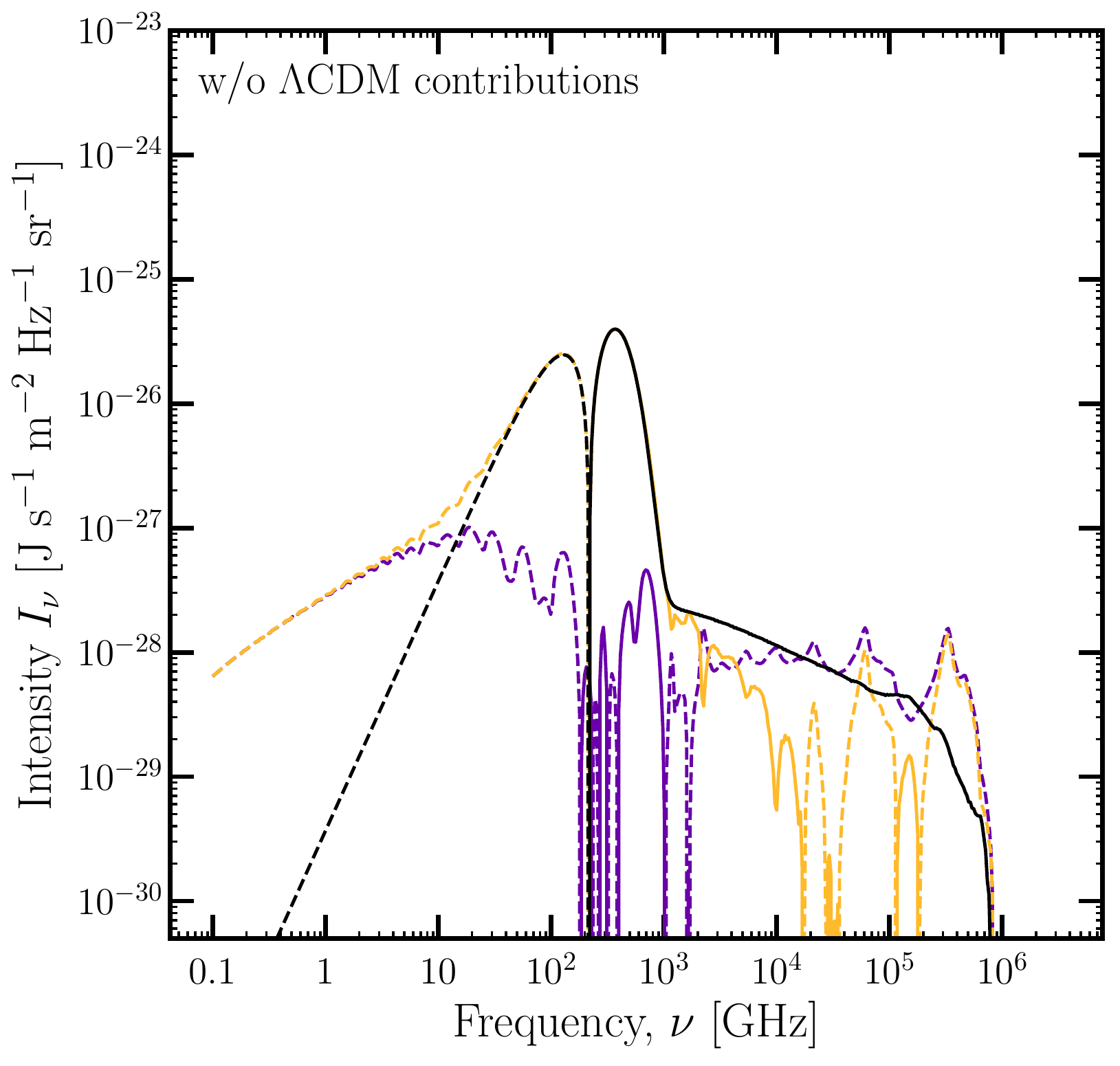}
	\caption{
		The effect of structure formation on the distortion.
		The left panel includes the $\Lambda$CDM and reionization contribution; the right panel has this contribution subtracted out.
		The purple line shows the distortion when we do not include the boost in power due to structure formation; yellow shows the distortion when we use the default model for structure formation in \dhis (see text for details).
		At late times, both the $y$-type distortion from heating and the ICS component of the distortion are enhanced.
		The difference between the two curves is shown by the black line.
	}
	\label{fig:struct_boost}
\end{figure*}
Small scale substructure can amplify spectral distortions from annihilating dark matter at late times; once dark matter halos begin to collapse, the annihilation rate can be significantly enhanced by the factor $\langle \rho_\chi^2 \rangle / \langle \rho_\chi \rangle^2$.
Fig.~\ref{fig:struct_boost} shows the effect of including structure formation; the boost factor is calculated assuming the halos have an Einasto profile and substructure, with properties as discussed in Ref.~\cite{1604.02457}.
The main difference is to enhance the heating $y$-type spectral distortion and the ICS contribution from later redshifts.
At frequencies less than a couple thousand GHz, the contribution of $\Lambda$CDM processes dominates over the spectral distortion solely from annihilations; hence, the only visible change to the spectral distortion in the left panel comes from the ICS component, which raises the high frequency shoulder around 4000 GHz.
For the models we considered, this effect is at most at the level of a few times $10^{-26}$ J s$^{-1}$ m$^{-2}$ Hz$^{-1}$ sr$^{-1}$, which may be just detectable by e.g. PIXIE~\cite{2011JCAP...07..025K, 2016SPIE.9904E..0WK}.
In the right panel, when we subtract out the $\Lambda$CDM contribution to the distortion, we can more clearly see the effect on the heating $y$-type component.

We reiterate that we only show contributions to the spectral distortions from between $1+z=3000$ and $1+z=4$.
We have estimated the contribution from other redshifts and find that it is subdominant to the contributions calculated here for the models we have studied; the degree to which the contribution from $3000 > 1+z > 4$ dominates depends on the channel for energy injection, see Appendix~\ref{app:other_rs} for details.
We also examine the rate of energy deposited into spectral distortions and find that it is consistent with the energy deposited into the continuum channel plus $y$-type distortion contributions, see Appendix~\ref{app:eng_rate} for details.

\section{Ionization Histories from Low Mass Dark Matter}
\label{sec:ionization}

Given the upgrades to the treatment of low-energy electrons described in \citetalias{paperI}, we can now accurately calculate the ionization histories with particles injected with energy less than a few keV. 
In this section, we present these results for the first time.
Moreover, since we have now implemented the capability for \dhis to track many excited states of hydrogen, transitions of these excited states can in principle modify the evolution of the free electron fraction, compared to the results derived by the previous version of \texttt{DarkHistory}.
The modified ionization rate and spectral distortion can affect each other: 
atomic transitions and recombination lead to the absorption or emission of low-energy photons,
while spectral distortions will affect the various atomic rates through the photon phase space density.
These modifications can in principle be large; however, given existing constraints on DM annihilation and decay, we will show that the effects on these processes are not significant enough to alter present constraints.

\subsection{Extending to lower energies}
\label{sec:ion_low_mass}

%
\begin{figure*}
	\includegraphics[width=\textwidth]{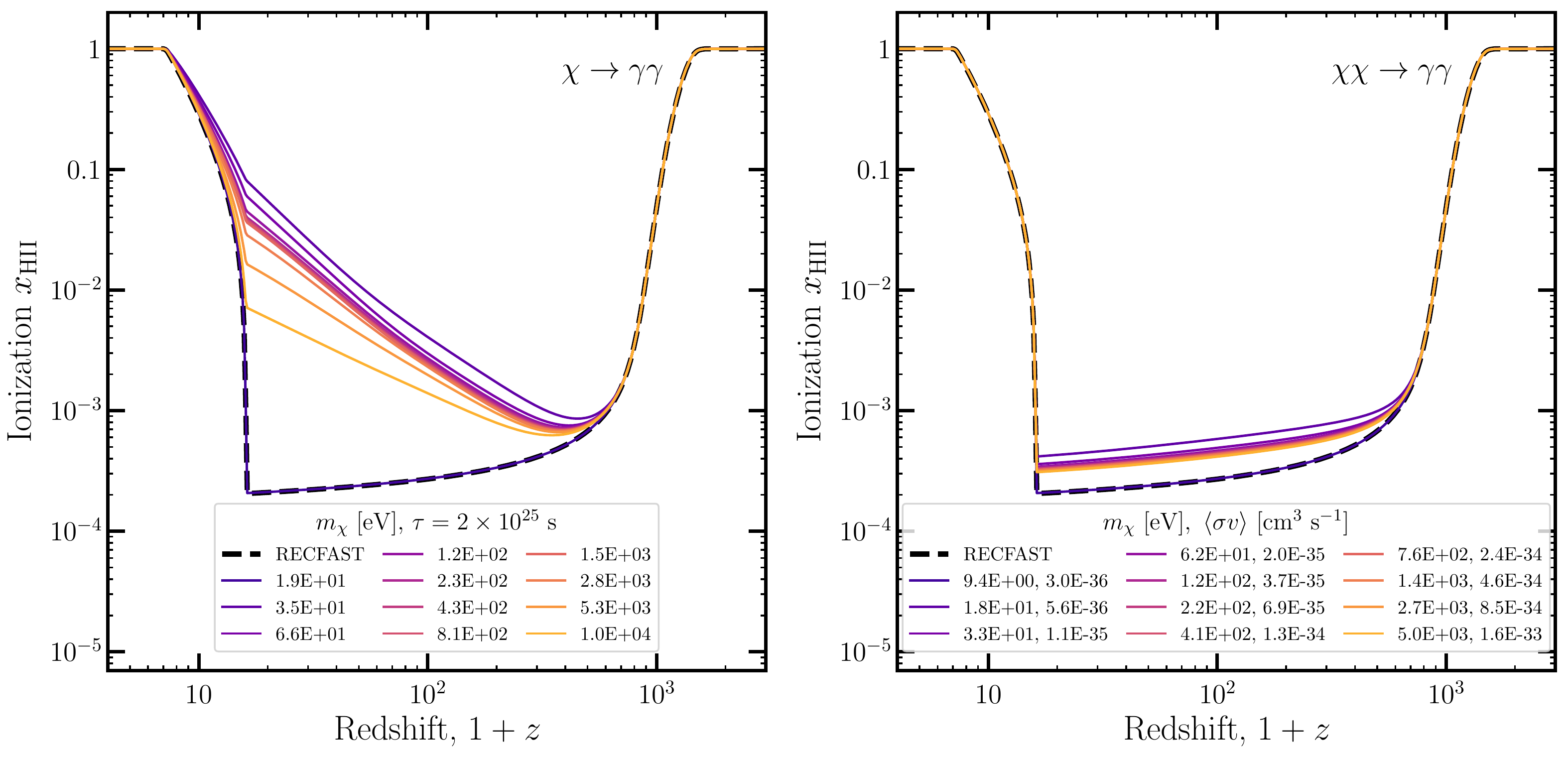}
	\caption{
		Ionization histories for dark matter decaying or annihilating to photons with masses $m_\chi <$ 10 keV. For decays, the lifetime is fixed to $\tau = 2 \times 10^{25}$ s; for annihilations, we choose the lifetime to be at the Planck 2018 constraint assuming 100\% efficiency of energy deposition, $\langle \sigma v \rangle / m_\chi = 3.2 \times 10^{-28}$ cm$^3$ s$^{-1}$ GeV$^{-1}$~\cite{Planck:2018vyg}.
		The black dashed line shows the ionization history calculated with \texttt{Recfast}~\cite{Seager:1999km,Seager_1999} in the absence of exotic energy injection.
	}
	\label{fig:xe_lowm}
\end{figure*}
A novel application of the method described in \citetalias{paperI} is the ability to accurately calculate the ionization histories resulting from dark matter with masses less than a few keV.
At energies lower than this, previous calculations relied on Monte Carlo results based on Ref.~\cite{MEDEAII}, while the applicability of the photoionization cross section used in earlier works as we approach the ionization potential of hydrogen and helium was unclear. 
With our improved low-energy treatment outlined in \citetalias{paperI}, we can now reliably calculate $x_\text{HII} (z)$ for arbitrarily low dark matter masses.

Fig.~\ref{fig:xe_lowm} shows the ionization histories from exotic energy injection.
The left panel shows dark matter decaying to photons, with masses between 19 eV and 10 keV and lifetime fixed at $\tau = 2 \times 10^{25}$ s.
At this lifetime, we see that the presence of exotic energy injection can significantly impact the ionization history prior to reionization; hence, these modifications can also be constrained by their impact on CMB anisotropies (see Section~\ref{sec:anisotropy}).
As $m_\chi$ drops below $2\mathcal{R}$, where $\mathcal{R} \equiv \SI{13.6}{\eV}$ denotes the hydrogen ionization potential, the resulting photons drop below the hydrogen ionization threshold; below this mass, exotic energy injection cannot directly ionize hydrogen and the curve calculated by \dhis almost exactly matches the standard \texttt{Recfast} ionization history; the difference arises from changes to recombination due to the presence of nonthermal photons below \SI{13.6}{\eV}, which we can calculate accurately, and will discuss further below.

The right panel of Fig.~\ref{fig:xe_lowm} is for dark matter annihilating to photons, with cross-section fixed to the Planck 2018 constraint assuming that all injected energy is transferred to the IGM, $\langle \sigma v \rangle / m_\chi = 3.2 \times 10^{-28}$ cm$^3$ s$^{-1}$ GeV$^{-1}$~\cite{Planck:2018vyg}.
We see that at this cross-section, the effect of exotic energy injection on ionization is at a level of less than $\Delta x_e < 10^{-3}$, which is smaller than the effect from decays but still potentially detectable by future experiments.
Again, we see that as the kinetic energy of the primary photons falls below $\mathcal{R}$, the effect on the ionization history is suppressed---although as we will discuss, it does not vanish entirely.

\subsection{Comparison to previous calculations}
\label{sec:old_ion}

%
\begin{figure*}
	\includegraphics[width=\textwidth]{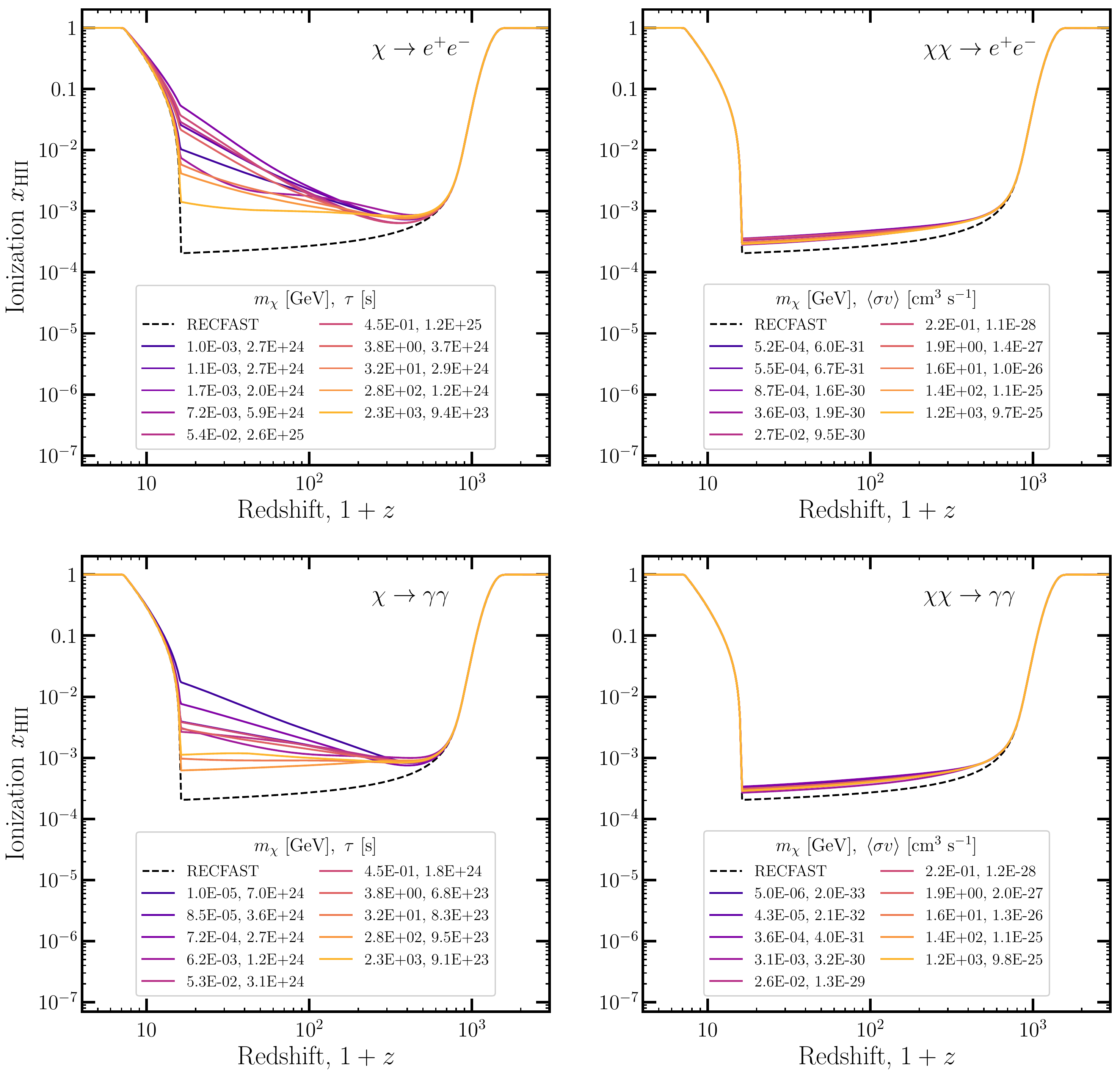}
	\caption{Ionization histories for the same DM models as in Fig.~\ref{fig:dist_grid}.
	The black dashed line shows the ionization history calculated with \texttt{Recfast}~\cite{Seager:1999km,Seager_1999} in the absence of exotic energy injection.
	While all models with exotic energy injection show an increase in the global ionization, the change in ionization is largest for the decaying DM models.}
	\label{fig:xe_grid}
\end{figure*}
\begin{figure*}
	\includegraphics[width=\textwidth]{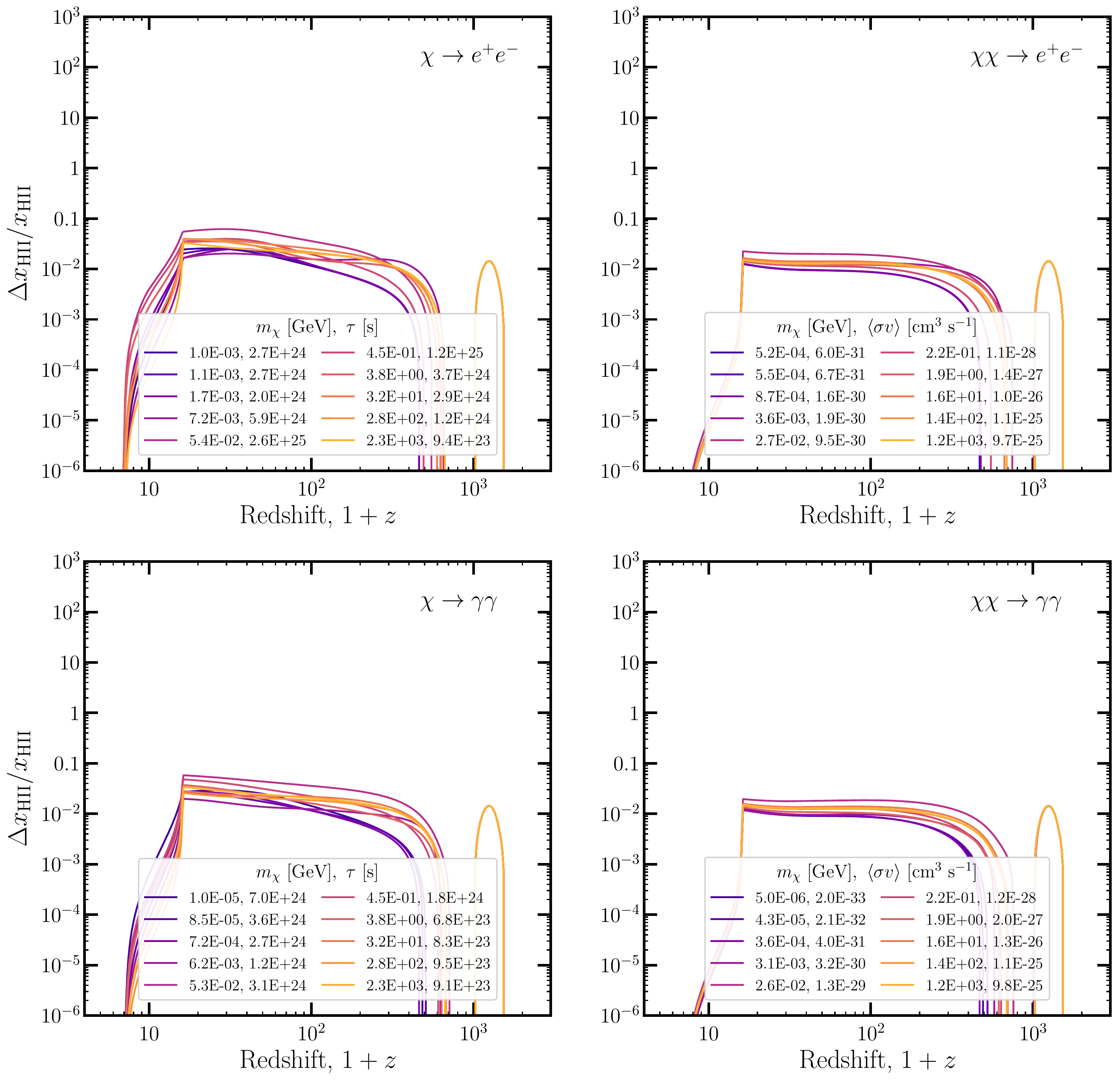}
	\caption{Difference in the ionization history relative to that calculated using \dhis \texttt{v1.0}.
	In other words, this is the difference in ionization between the two methods, divided by the history calculated with \dhis \texttt{v1.0}.
	The differences are at the level of less than 10 percent, which shows that tracking more excited states of hydrogen and the feedback between spectral distortions and ionization does not significantly modify the global ionization history.}
	\label{fig:delta_xe_grid_oldDH}
\end{figure*}
Due to the feedback between modifications to the global ionization history and spectral distortions, one might expect that accurately tracking the spectrum of background photons could change our calculation of the free electron fraction, $x_e$, relative to \dhis \texttt{v1.0} even from higher DM masses.
Here, we examine the size of this effect.

Fig.~\ref{fig:xe_grid} shows the ionization histories derived using the same DM models shown in Fig.~\ref{fig:dist_grid}.
Shown in the black dashed line is the ionization history calculated by \texttt{Recfast} in the absence of DM energy injection (including the hydrogen fudge factor, which we set to 1.125, and the double Gaussian function correction~\cite{Seager:1999km,Seager_1999}).
In all cases, as expected, including energy injections from DM annihilation/decay yields ionization histories that are larger than the \texttt{Recfast} history at all redshifts.
For DM annihilating to photons or $e^+ e^-$ pairs, we see that the spread in ionization histories is quite small.
For decaying DM, the spread in ionization histories is larger, and the histories can deviate more significantly from the \texttt{Recfast} calculation at late times (consistent with e.g. Ref.~\cite{1604.02457}).
We also show in Appendix~\ref{app:recfast_xcheck} the relative difference between the ionization history calculated with our new method versus that derived with \texttt{Recfast}. 

The reason for the difference in variation between annihilation and decay scenarios is due to the redshift dependence of the energy injection and the fact that we are using CMB-derived limits.
For decay, the constraint is mainly controlled by energy injection at redshift $1+z \sim 300$~\cite{1610.06933}, hence we would expect the ionization and difference relative to the \texttt{Recfast} curve at this redshift to be similar in the left panels of Figs.~\ref{fig:xe_grid}.
This can be clearly seen in the $\chi \rightarrow \gamma\gamma$ panels.
For annihilation constraints, the principal component comes from redshift $1+z \sim 600$~\cite{1506.03811}, hence the ionization should be similar at this redshift in the right panels.
Since the power from annihilating DM decreases with redshift as $(1+z)^6$, then energy injection effectively shuts off after $1+z \sim 600$ and we see very little spread to the ionization histories, whereas the power from decaying DM only decreases as $(1+z)^3$ so there is much greater variation in the ionization histories at late times.

In order to determine how much the new machinery to track hydrogen levels and spectral distortions affects the ionization history, in Fig.~\ref{fig:delta_xe_grid_oldDH}, we show the relative difference between the ionization histories calculated using the new and original versions of \texttt{DarkHistory}; in Appendix~\ref{app:recfast_xcheck}, we also show the relative difference between the new \dhis and \texttt{Recfast}.
Again, we see that the spread in the histories is somewhat larger for decay scenarios than annihilation.
The largest differences are at less than the 10 percent level; hence, while feedback from the spectral distortions does modify the ionization history, this difference is not large for unconstrained values of $\tau$.
An important implication of this is that DM decay and annihilation results that depend on the ionization histories calculated by previous versions of this code remain largely unchanged.
This includes the constraints on decaying and annihilating DM derived using the effect of ionization on the CMB anisotropies~\cite{1506.03811,1610.06933}.
We discuss the CMB anisotropy constraints in greater detail in the next section.

Finally, although the validity of the approximations made in \dhis \texttt{v1.0} was not clear for injections at dark matter masses less than 10 keV, we find that the differences between the ionization histories calculated with \dhis \texttt{v1.0} and with the improved treatment are also less than 10\% at all masses and redshifts.
The agreement between \dhis \texttt{v1.0} and the upgraded version shows that these approximations in fact yield accurate results for sub-keV dark matter, at least in the context of calculations of the modified ionization history and constraints that are determined by this modification.

\section{CMB anisotropy}
\label{sec:anisotropy}

Energy deposition into ionization from DM annihilation or decay modifies the process of recombination, which in turn modifies the CMB anisotropy power spectrum~\cite{0906.1197, 0907.3985}. 
Existing limits based on Planck data set strong limits on such processes, especially in the sub-GeV DM mass range~\cite{1506.03811, 1610.06933, 1610.10051}. 
We can parametrize the power deposited into ionization relative to the power injected by DM as
\begin{equation}
	\left( \frac{dE}{dV dt} \right)_\text{H ion}= f_\text{H ion} \left( \frac{dE}{dV dt} \right)_\text{inj}.
\end{equation}
Properly computing the limits on DM interactions therefore relies on an accurate computation of $f_\text{H ion}(z)$ for annihilation or decay into $e^+e^-$ pairs and $\gamma \gamma$ pairs. 
Previous works on these limits only presented constraints down to a DM mass where the kinetic energy of each injected particle in these channel was $\SI{5}{\kilo\eV}$. 
As mentioned in Section~\ref{sec:ion_low_mass}, at lower energies, Monte Carlo results based on Ref.~\cite{MEDEAII} provided $f_\text{H ion}(z)$ for $e^+e^-$ injection, but required interpolation over a limited number of injection energies, and the applicability of cross-sections used in \dhis\texttt{v1.0} near the ionization potential of hydrogen and helium were unclear. 

With the method described in \citetalias{paperI}, we can now confidently extend the CMB power spectrum limits down to much lower kinetic energies of the injected particles. 
This is particularly important for DM particles such as ALPs decaying into photons, since these constraints are among the strongest available for $2 \mathcal{R} < m_\chi \lesssim \SI{10}{\kilo\eV}$, with CMB spectral distortions providing relevant limits for $m_\chi \geq 2 \mathcal{R}$~\cite{Bolliet:2020ofj}. 

We can obtain a good estimate for the constraint on DM decay by noting that the limit on the lifetime $\tau$ is approximately proportional to $f_\text{H ion}(z = 300)$~\cite{1610.06933}, since the impact of energy injection as a function of redshift on the CMB power spectrum is peaked at that redshift~\cite{2012PhRvD..85d3522F}. 
We can therefore estimate limits for decay by simply computing $f_\text{H ion}(z = 300)$ as a function of $m_\chi$ for both channels, and rescaling that shape to match previous constraints on $\tau$. 
The options we use to calculate $f_\text{H ion}(z = 300)$ are:
\begin{itemize}
	\item  \texttt{backreaction = False},
	
	\item  \texttt{struct\_boost = None},
	
	\item  \texttt{reion\_switch = False},
	
	\item  \texttt{distort = True},
	
	\item  \texttt{nmax = 10},
	
	\item  \texttt{iterations = 1}
	
	\item  \texttt{reprocess\_distortion = False}.
\end{itemize}
Since we are only seeking an estimate of the CMB anisotropy constraint, we lower \texttt{nmax} to 10, \texttt{iterations} to 1, and set \texttt{reprocess\_distortion = False} to speed up the calculation.
This estimate is valid down to about $m_\chi = 2 \mathcal{R}$; for masses lower than this, the resulting photons are below the hydrogen ionization threshold; therefore, $f_\text{H ion}(z = 300)=0$.

\begin{figure}
	\includegraphics[width=\columnwidth]{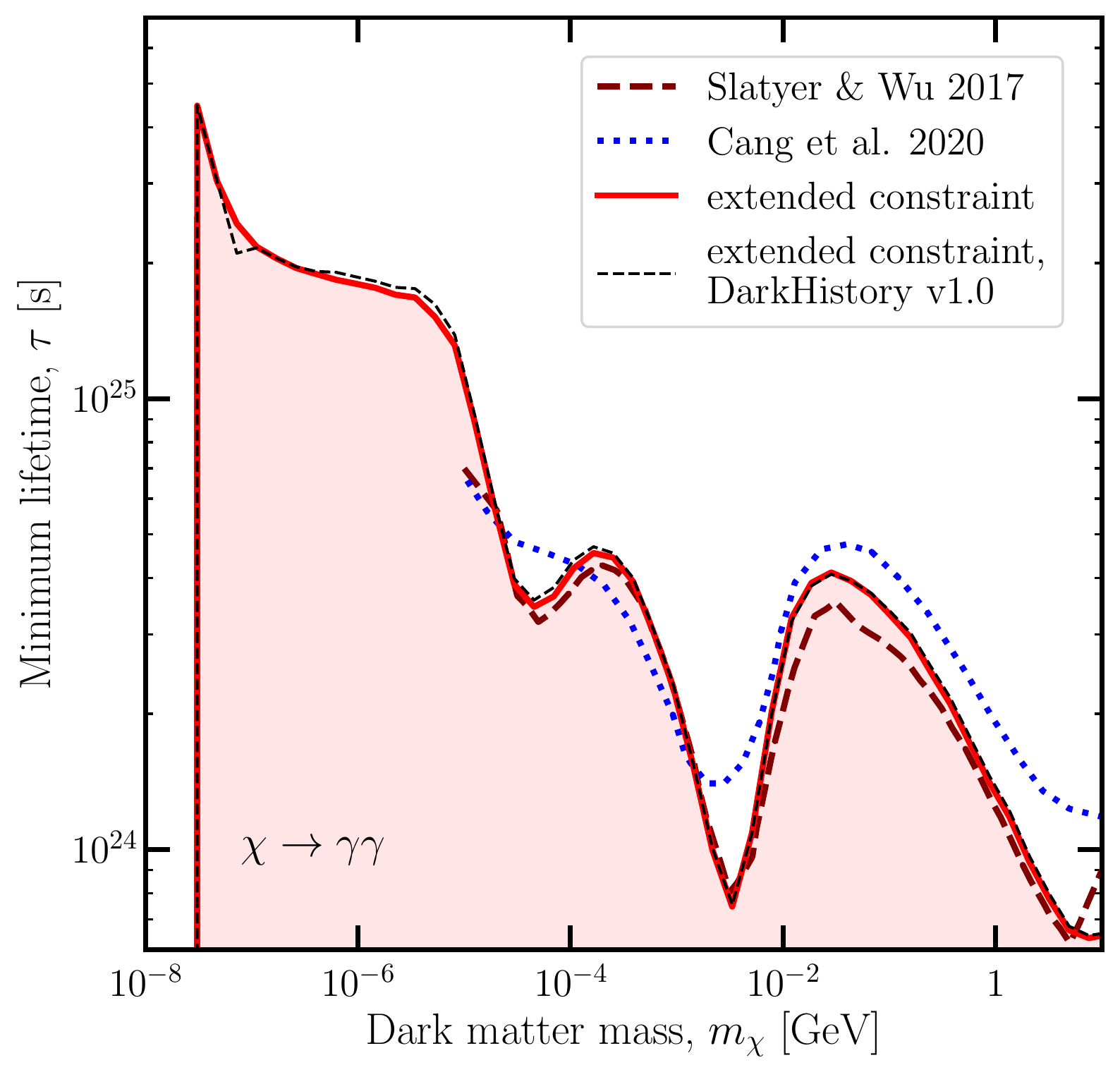}
	\caption{
		Estimated constraints on the lifetime for DM that decays to a pair photons, shown as the red contour. 
		These constraints were derived using the results of Ref.~\cite{1610.06933} which are shown as a dashed dark red line.
		We show the same estimate using \dhis \texttt{v1.0} in the thin dashed black line.
		Both of these results made use of Planck 2015 data.
		We also plot Ref.~\cite{Cang:2020exa}, which used Planck 2018 data, as a dotted blue line for comparison; we choose not to extend these constraints, since it is unclear if our estimation method applies to this data.
	}
	\label{fig:anisotropy_constraint}
\end{figure}
\begin{figure}
	\includegraphics[width=\columnwidth]{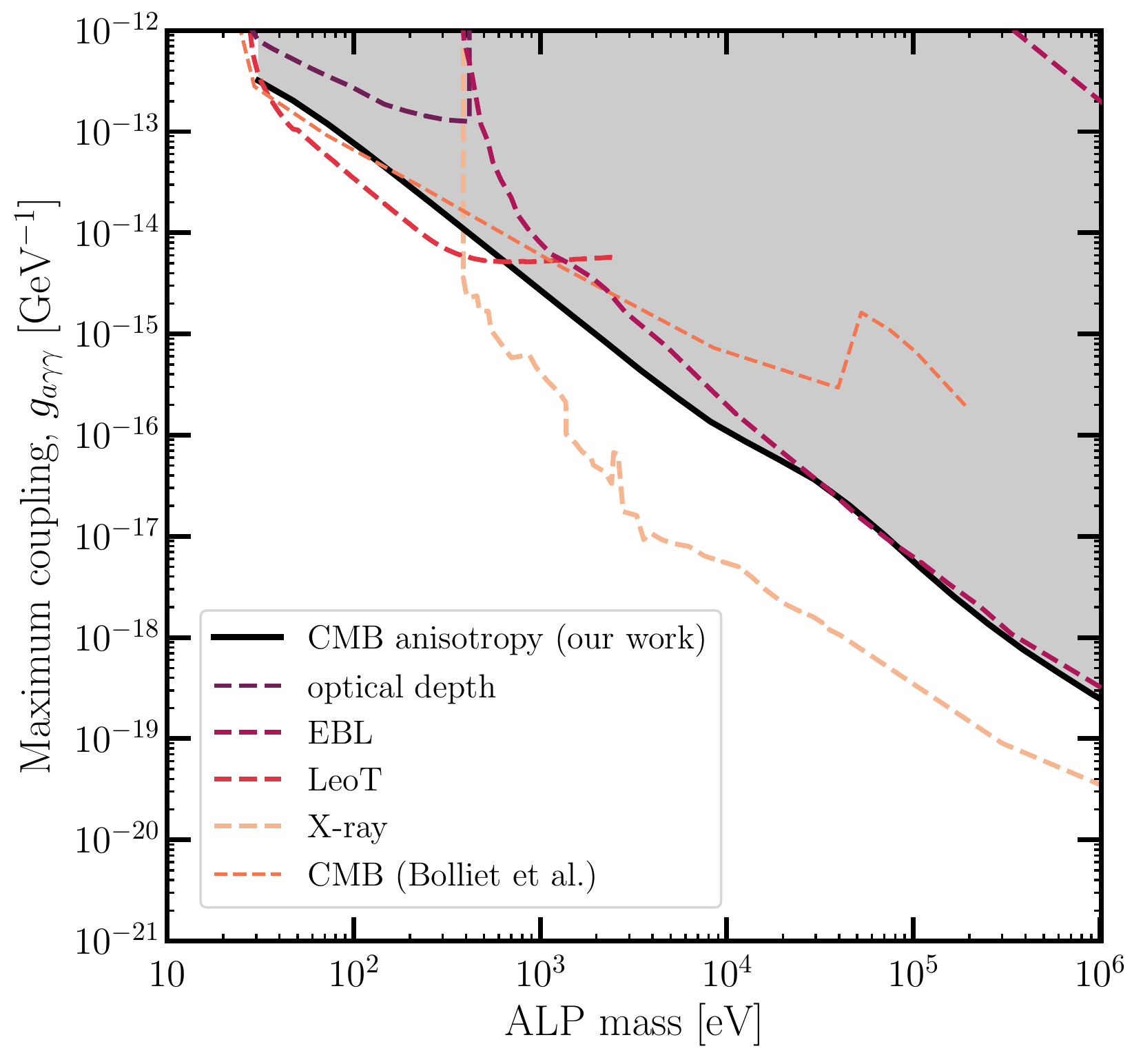}
	\caption{Constraints on the coupling of ALPs to photons.
	Our estimate of the constraint from modifications of CMB anisotropies is shown in the black contour.
	For comparison, we also show constraints from measurements of the ionization history/optical depth ($x_e$)~\cite{Cadamuro:2011fd}, extragalactic background light (EBL), heating in LeoT using the more conservative gas temperature of 7552 K~\cite{Wadekar:2021qae}, CMB spectral distortion and anisotropy constraints from Ref.~\cite{Bolliet:2020ofj}, and X-rays~\cite{Cadamuro:2011fd}.
	The constraints were plotted using the \texttt{AxionLimits} code~\cite{AxionLimits}.}
	\label{fig:ALP_constraint}
\end{figure}
\begin{figure}
	\includegraphics[width=\columnwidth]{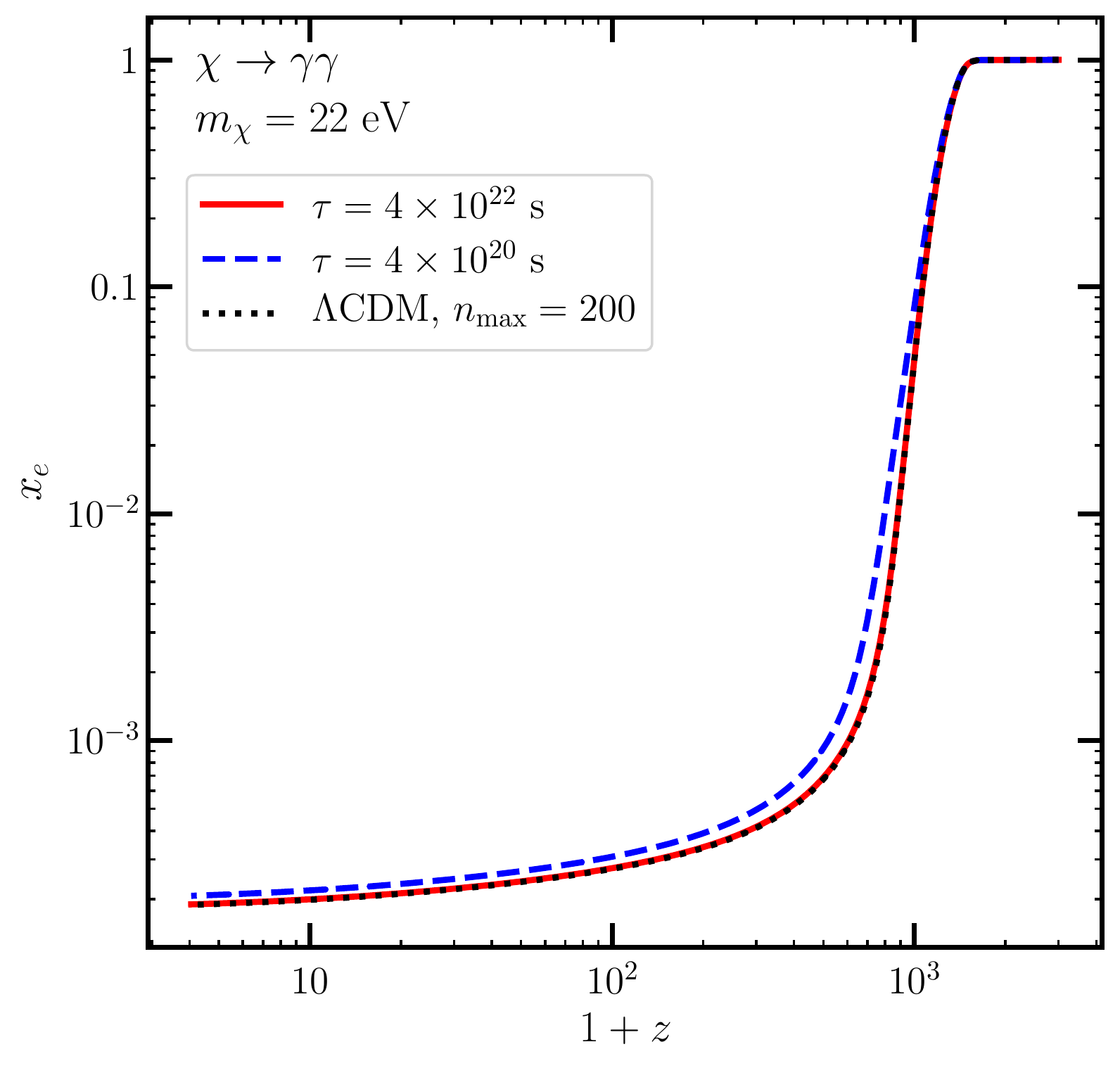}
	\caption{Ionization histories for DM decaying to photons at a mass of $m_\chi =$ 22 eV, i.e. for injection of photons with energies less than $\mathcal{R}$.
	At a DM lifetime of $\tau = 4 \times 10^{22}$ (red), which is taken from constraints based on measurements of the EBL~\cite{Nakayama:2022jza}, the ionization history is indistinguishable from the ionization history with no exotic energy injection (dotted black).
	In order to have an observable change to the ionization, we must decrease the lifetime quite a bit, e.g. to $\tau = 4 \times 10^{20}$ s (dashed blue).
	}
	\label{fig:subRydberg}
\end{figure}

Fig.~\ref{fig:anisotropy_constraint} shows our estimated lifetime constraints on DM decaying into a pair of photons from extending the bounds in Ref.~\cite{1610.06933}, which used the Planck 2015 dataset and likelihoods (as discussed in e.g.~\cite{Planck:2015bpv}). 
For $m_\chi > \SI{10}{\kilo\eV}$, we obtain excellent agreement between our current result and those of Ref.~\cite{1610.06933} as expected, but extend the anisotropy power spectrum limits down to $m_\chi = 2 \mathcal{R}$, where we find $\tau \gtrsim \SI{e25}{\second}$. 
We show the same estimate using \dhis \texttt{v1.0} in the thin dashed black line.
Per the discussion in Section\ref{sec:ionization}, we find that the approximations made in \dhis \texttt{v1.0} are in fact very good for this purpose, such that $f_\text{H ion}(z = 300)$ calculated using \dhis \texttt{v1.0} is remarkably close to the result using our improved low-energy treatment.
Hence, the constraints between the two versions of \dhis are nearly identical, although we are more confident in our control over theoretical uncertainties in the upgraded version. 
As we will discuss below, we can in principle also obtain constraints for $m_\chi < 2 \mathcal{R}$ coming from modifications to recombination due to the presence nonthermal low-energy photons, but these constraints appear to be weaker than existing limits, and not been computed carefully here.

We also show the constraints from Ref.~\cite{Cang:2020exa}, which used Planck 2018 + BAO data.
The shape of these constraints is smoother compared to Ref.~\cite{1610.06933}, since they used a different statistical approach (treating $m_\chi$ as a model parameter in their MCMC rather than performing a separate MCMC for each fixed $m_\chi$ value); hence it is not clear if the same principal component approach can be applied here.
As a result, we choose only to extend the estimated constraints using the results of Ref.~\cite{1610.06933} based on Planck 2015 data.
The most robust constraint would come from repeating the MCMC analysis for the Planck 2018 data using the latest version of \dhis presented here.

The limits on injected photons translate directly to a constraint on the ALP-photon coupling $g_{a \gamma \gamma}$, which we show in Fig.~\ref{fig:ALP_constraint}. 
Comparing to the limits derived using CMB anisotropy data in Ref.~\cite{Bolliet:2020ofj}, our constraints become stronger above $m_a \sim 100$ eV; this may at least in part be explained by simplifications in their treatment that are expected to break down at higher energies~\cite{Bolliet:private_comm}.
In addition, while weaker than limits based on heating of dwarf galaxies like LeoT~\cite{Wadekar:2021qae}, our constraints have completely complementary systematics.  
Note that this result is based on the full information provided by the Planck CMB power spectrum---across all available angular scales and using TT, TE and EE power spectra---instead of simply relying on the inferred optical depth to the surface of last scattering from WMAP7 to set a limit, as was done in Ref.~\cite{Cadamuro:2011fd}.
These results therefore supercede the estimate in Ref.~\cite{Cadamuro:2011fd} (labelled $x_e$ in their Fig. 11). 

For DM with $m_\chi < 2 \mathcal{R}$ decaying to photons, the primary photons cannot ionize anything, but the ionization history can still be affected: for example, injected photons can excite neutral hydrogen, and the resulting excited atoms have a higher probability of being ionized by a CMB photon.
Hence, although $f_\text{H ion} = 0$, there will still be some constraint on energy injection from the overall ionization history.
The current most competitive bound on DM decaying to photons at $m_\chi =$ 22 eV is given by measurements of the extragalactic background light (EBL) and corresponds to a lifetime of about $\tau \approx 4 \times 10^{22}$ s~\cite{Nakayama:2022jza}.
At this lifetime, we find the change to the ionization history is negligible.
In order to get a potentially observable change to the ionization, we would have to lower the lifetime by at least an order of magnitude or two, so any CMB anisotropy constraints for injection of sub-$\mathcal{R}$ photons are likely already ruled out by the EBL.
Fig.~\ref{fig:subRydberg} shows the ionization histories for $\Lambda$CDM and also including dark matter decaying to photons at $m_\chi =$ 22 eV and lifetimes of $\tau \approx 4 \times 10^{22}$ s and $4 \times 10^{20}$ s (to ensure this calculation is accurate, we again set \texttt{nmax = 200} and \texttt{iterations = 5} for this figure).
The curve with $\tau \approx 4 \times 10^{22}$ s is nearly identical to the $\Lambda$CDM curve; hence, it is unlikely that the resulting CMB anisotropy constraints would be competitive with current bounds in this low mass range.

Finally, Ref.~\cite{Langhoff:2022bij} pointed out that for sufficiently large couplings, a relic density of ALPs may freeze in from the thermal bath, forming an irreducible contribution to the DM density.
Although these ALPs may only comprise a small fraction of DM, DM searches can still be used to set very strong constraints on ALP parameter space using this relic abundance.
Ref.~\cite{Langhoff:2022bij} applied this argument to CMB constraints, using the anisotropy costraints in Ref.~\cite{Cang:2020exa} above masses of 10 keV and the spectral distortion information from Ref.~\cite{Bolliet:2020ofj} below 10 keV.
In principle, with the upgrades to \dhis presented here, we can now extend the CMB anisotropy constraints on such a population to arbitrarily low masses; we leave a detailed analysis to future work.

\textit{Note:} In the late stages of preparing this manuscript, we became aware of another work near completion~\cite{Capozzi:2023xie}, which calculates the CMB anisotropy bounds on sub-keV dark matter, performing a full MCMC analysis using Planck 2018 data and \dhis \texttt{v1.0}.
Our results in this respect are complementary; that work provides a complete calculation of the constraints using \dhis \texttt{v1.0} (which appear broadly consistent with our estimates), while our results show that using the new version is unlikely to modify the constraints significantly.

\section{Conclusion}
\label{sec:conclusion}

In \citetalias{paperI}, we described a major upgrade to the capabilities of the \dhis package for modeling the effects of exotic energy injection in the early universe, including an improved treatment for energy deposition by low-energy electrons, tracking the full background spectrum of photons, and taking into account interactions between the photon bath and the hydrogen gas with an arbitrary number of excited states.
This treatment assumes that energy is deposited homogeneously, which may become inaccurate at late times once structure formation is well underway.
In this work, we describe several applications of this new machinery.

First, we demonstrated that we can now compute the CMB spectral distortions resulting from general exotic energy injections in the $z < 3000$ universe; specifically, we studied models of decaying and annihilating DM, with interaction rates at the exclusion limit from CMB anisotropy constraints, and predicted their signals in CMB spectral distortion.
The largest predicted signals come from light decaying DM, yielding spectral distortions up to about one part in $10^6$, and would be observable by future experiments such as PIXIE, which could detect spectral distortions down to about one part in $10^8$~\cite{2011JCAP...07..025K, 2016SPIE.9904E..0WK}. 
Other proposed experiments that may be able to detect this signal include PRISTINE~\cite{Chluba:2019nxa}, BISOU~\cite{Maffei:2021xur}, and FOSSIL~\cite{Chang:2022tzj}.
The signal from injection of high-energy particles can also be distinguished from other expected contributions by the shape of the high-frequency tail of the distortion, although we have not yet examined the impact of foregrounds on the detectability of these distortions.

Secondly, we extended the calculation of the modified ionization history from decaying DM to arbitrarily low masses.
We also showed that even though spectral distortions and modifications to the ionization history are coupled in the evolution equations, including spectral distortions (in models that are not currently excluded) does not significantly change the ionization history, with effects at the few-percent level and consistently below $10\%$.
This means that any constraints on DM annihilation and decay that were derived using the ionization history calculated with \dhis \texttt{v1.0} remain largely unchanged.

With these results in hand, we estimated the resulting CMB anisotropy constraints on low-mass scenarios. 
These results can also be translated into a limit on the coupling between ALPs and photons, which is competitive with other CMB-derived constraints and has complementary systematics to the leading constraints from e.g.~heating of dwarf galaxies.

There are numerous other possible applications of this improved technology, including studying the effects of the photons from exotic energy injection on the 21-cm signal or the EBL.
We leave these directions as topics for future study.

\section*{Acknowledgements}

We thank Sandeep Acharya and Rishi Khatri for the use of their Green's functions data, as well as Trey W. Jensen and Yacine Ali-Ha\"{i}moud for useful discussions. TRS was supported by the Simons Foundation (Grant Number 929255, T.R.S) and by the National Science Foundation under Cooperative Agreement PHY-2019786 (The NSF AI Institute for Artificial Intelligence and Fundamental Interactions, \url{http://iaifi.org/}).
WQ and GWR were supported by the National Science Foundation Graduate Research Fellowship under Grant No. 1745302.
WQ was also supported by National Science Foundation Graduate Research Fellowship under Grant No. 2141064 and GWR by the U.S. Department of Energy, Office of Science, Office of Nuclear Physics under grant Contract Number DE-SC0011090.
WQ, GWR, and \& TRS were supported by the U.S. Department of Energy, Office of Science, Office of High Energy Physics of U.S. Department of Energy under grant Contract Number DE-SC0012567. 
HL is supported by the DOE under Award Number DE-SC0007968, NSF grant PHY2210498, and the Simons Foundation.

This work made use of 
\texttt{Jupyter}~\cite{Kluyver2016JupyterN}, 
\texttt{matplotlib}~\cite{Hunter:2007ouj}, 
\texttt{NumPy}~\cite{Harris:2020xlr}, 
\texttt{SciPy}~\cite{2020NatMe..17..261V}, and 
\texttt{tqdm}~\cite{daCosta-Luis2019},
as well as Webplotdigitizer~\cite{Rohatgi2022}.

\clearpage
\onecolumngrid
\appendix 

\section{Spectral distortions from other redshifts}
\label{app:other_rs}

In Section~\ref{sec:dist_results}, we showed the spectral distortions resulting from energy injection between redshifts of $3000 > 1+z > 4$.
By default, \dhis only treats this redshift range, since $1+z = 3000$ is the highest redshift for which we have photon cooling transfer functions, and at lower redshifts one has to treat helium reionization.
However, energy injection from other redshift ranges can have contributions to the distortion that are comparable in amplitude.

\begin{figure*}
	\includegraphics[width=\textwidth]{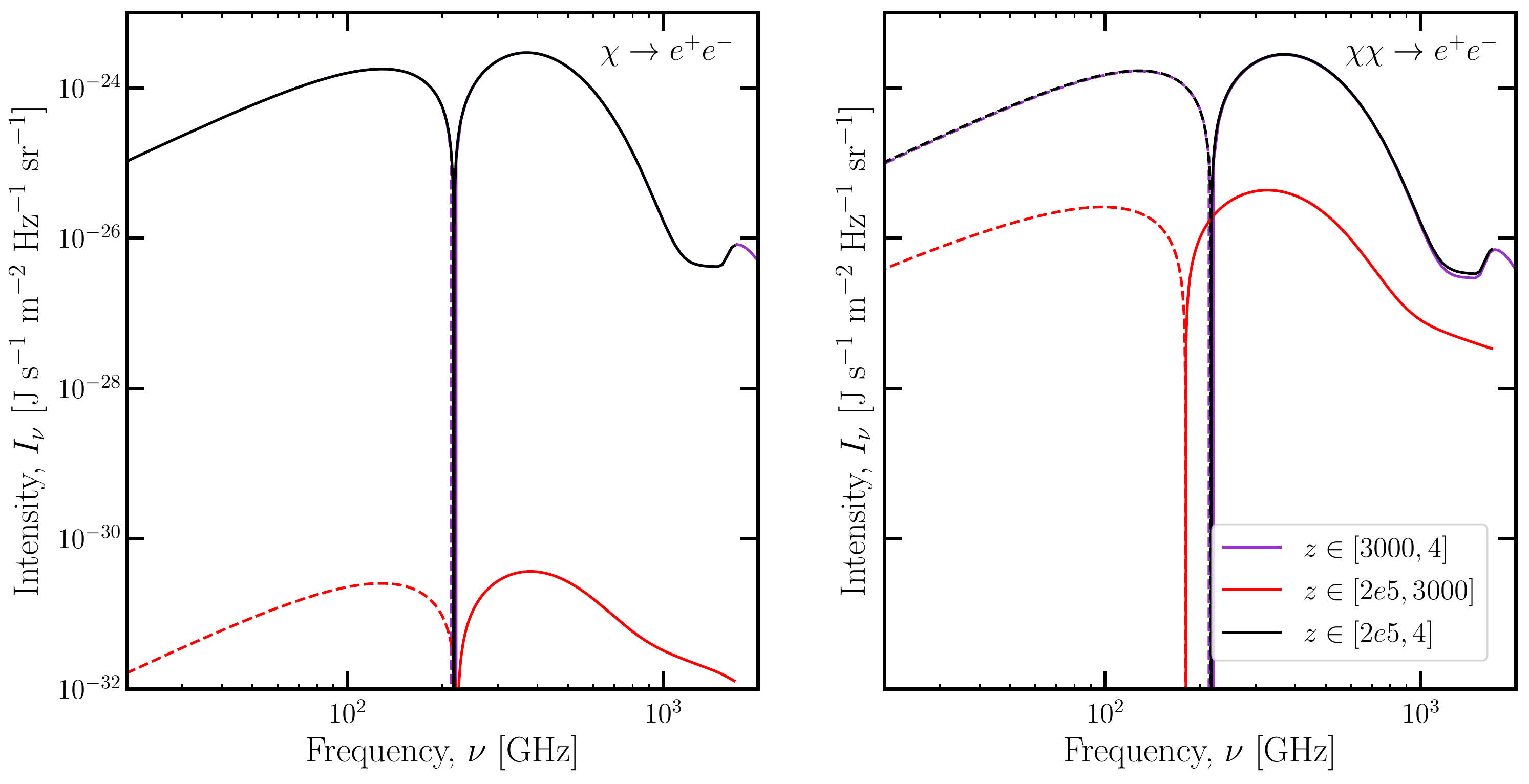}
	\caption{
		Contribution to the spectral distortion from high redshifts. 
		The DM masses are chosen such that the injected electrons and positrons have kinetic energy of 1 GeV; for decaying DM, the lifetime is $10^{25}$ s and for annihilation, the cross-section is $10^{-26}$ cm$^3$ s$^{-1}$.
		Red shows the contribution from redshifts $2 \times 10^5 > 1+z > 3000$ calculated using the Green's functions described in Ref.~\cite{Acharya:2018iwh}.
		Purple is the contribution from \texttt{DarkHistory}, while black shows the sum of the two contributions.
		Note that the $x$-axis range here is much narrower than for other figures.
	}
	\label{fig:high_rs}
\end{figure*}

For redshifts between $2 \times 10^5 > 1+z > 3000$, one can determine the spectral distortion contribution using the Green's functions calculated in Ref.~\cite{Acharya:2018iwh}.
We find that the early redshift contribution is subdominant for the masses we have tested, as shown in Fig.~\ref{fig:high_rs}; the left panel shows decaying DM and the right shows annihilating DM.
The mass of the DM particle is chosen such that the injected electrons and positrons have a kinetic energy of 1 GeV, so that we can use the Green's functions presented in Ref.~\cite{Acharya:2018iwh}.
The red line shows the early contribution and the purple line shows the contribution calculated by \texttt{DarkHistory}; black shows the sum of the two contributions, i.e. the total spectral distortion contributed by energy injection between $2 \times 10^5 > 1+z > 4$.

The relative size of the contributions can be roughly understood by considering the redshift dependence of the energy injection rate.
For decays, the energy injected per log redshift interval as a fraction of the CMB energy density is proportional to $\rho_\chi / H / u_\text{CMB} (T)$, where $H$ is the Hubble parameter and $u_\text{CMB}(T)$ is the blackbody energy density with temperature $T$.
Hence, this quantity depends on redshift as $(1+z)^{-2.5}$ during matter domination and $(1+z)^{-3}$ during radiation domination.
Then integrating this over the appropriate log redshift intervals, we estimate that the contribution from $3000 > 1+z >4$ should be larger than that of $2 \times 10^5 > 1+z > 3000$ by about seven orders of magnitude, which is roughly consistent with Fig.~\ref{fig:high_rs}.
Similarly for annihilation, the energy injected per Hubble time as a fraction of the CMB energy density is $\rho_\chi^2 / H / u_\text{CMB} (T)$, which goes as $(1+z)^{0.5}$ during matter domination and is constant during radiation domination.
Integrating this over the two redshift ranges, we find that two contributions are nearly equal, with the later contribution being smaller by a factor of only about 2.
However, as we found in \citetalias{paperI}, when the universe is fully ionized, i.e. prior to $z \sim 1100$, the conversion of energy injection to spectral distortions is less efficient; the reason is there are no photoionizations, and hence no secondary electrons, which can efficiently contribute to heating.
Hence, this suppresses the distortion from earlier redshifts by a couple orders of magnitude.
Note that these estimates are independent of dark matter mass, so we expect it to be generally true that the contribution from $3000 > 1+z > 4$ dominates over the contribution from higher redshifts.

\begin{figure*}
	\includegraphics[width=\textwidth]{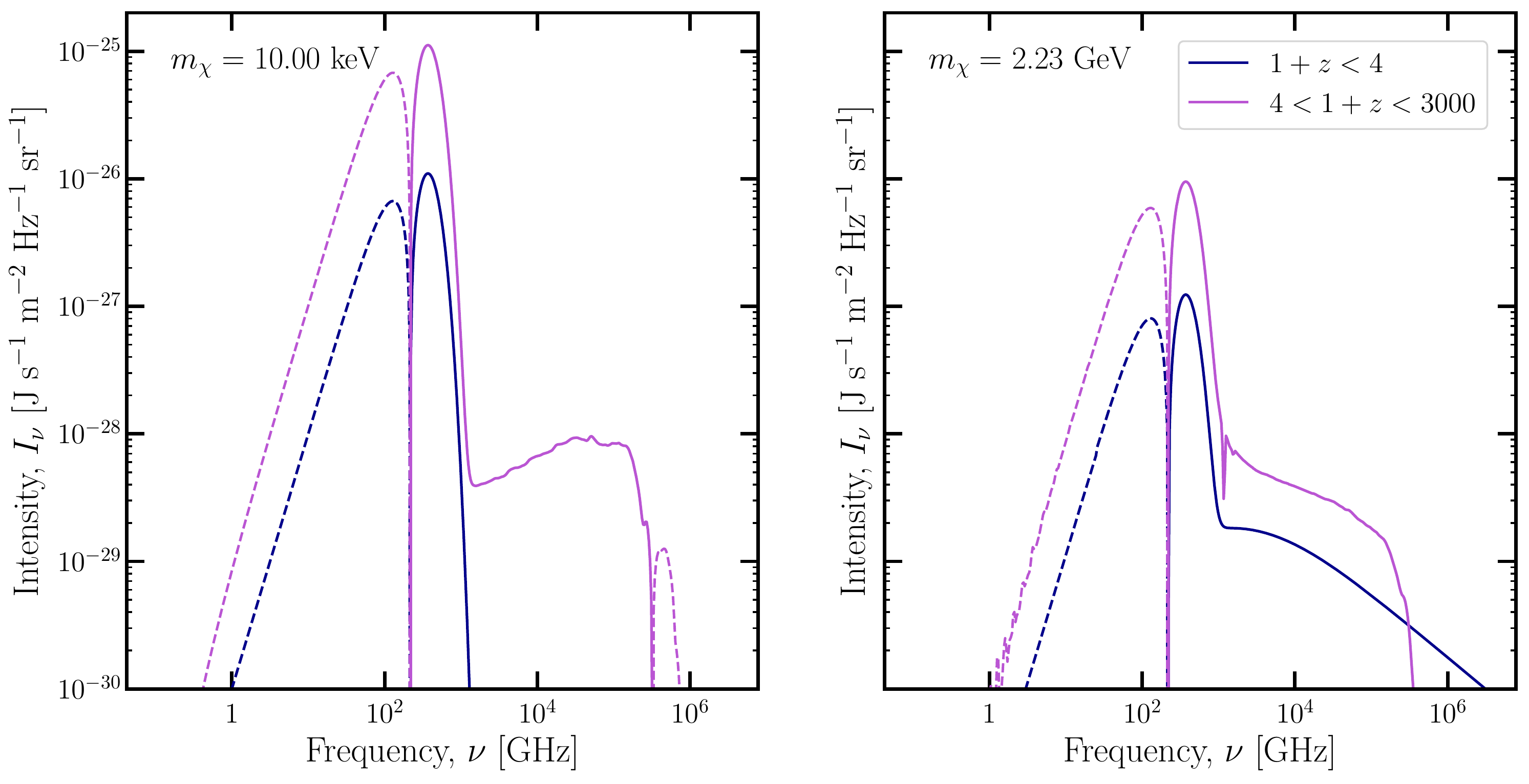}
	\caption{
		Estimate of the contribution to the spectral distortions from low redshifts.
		Here, we assume DM decays to photons with the lifetime taken to be at CMB constraints~\cite{1610.06933}.
		Purple shows the early contribution from \texttt{DarkHistory}, while blue shows the late time contribution.
	}
	\label{fig:low_rs}
\end{figure*}
We can also estimate the contribution to the distortion during the latest redshifts, $1+z < 4$.
At these redshifts, we can approximate the universe as completely ionized, hence the only cooling mechanisms available to photons are Compton scattering, pair production, and redshifting; the particle cascades resulting from these processes will eventually contribute to a spectral distortion through heating or ICS.
For photon energies much smaller than the electron mass $m_e$, the photons lose energy to electrons at the rate~\cite{1990ApJ...349..415S,Acharya:2018iwh}.
\begin{equation}
\frac{d \omega}{dt} = n_e \sigma_T \frac{\omega^2}{m_e} ,
\end{equation}
where $n_e$ is the number density of electrons and $\sigma_T$ is the Thomson cross-section.
Then, based on the timescales for electron cooling processes given in Ref~\cite{2010MNRAS.404.1869F}, the dominant cooling process for electrons at energies less than a few MeV is heating, as opposed to ICS.
At energies much larger than this, we can no longer use the previous energy loss formula and we must also take ICS into account.
Hence, we can estimate the low-redshift spectral distortion in this regime using three steps:
\begin{enumerate}
	\item We calculate the rate at which electrons are upscattered by Compton scattering.
	
	\item We assume the electrons gain most of the injected photon's energy; this can be seen by averaging the Compton scattering formula
	\begin{equation}
		\omega' = \frac{\omega}{1 + \frac{\omega}{m_e} (1-\cos\theta)}
	\end{equation}
	over the angle $\theta$, where $\omega'$ is the energy of the photon after scattering.
	We find that
	\begin{equation}
		\frac{\langle \omega' \rangle}{\omega} = \frac{m_e}{\omega} \log \left( 1 + \frac{\omega}{m_e} \right).
	\end{equation}
	In the limit that $\omega \gg m_e$, this asymptotes to zero, hence the photon loses most of its energy to the electron upon scattering.
	
	\item We process the electrons through the electron cooling module of \dhis to determine the spectrum of secondary photons from ICS, as well as how much energy was deposited into heating.
\end{enumerate}

Fig.~\ref{fig:low_rs} compares the spectral distortion calculated by \texttt{DarkHistory}, shown in purple, and the missing contribution from late times, shown in blue.
We assume DM decays to photons with a mass of 10 keV or 2.23 GeV and lifetime taken to be at the CMB bounds~\cite{1610.06933}; these two models are representative of results for energy deposition by low and high energy particles.
The amplitude of the late-time distortion is smaller than the contribution from $3000 > 1+z > 4$ by about an order of magnitude in the models we consider. 
Again, we can estimate the relative size of these contributions using the argument that we outlined above for high redshifts.
Using the redshift dependence for energy injection by decaying dark matter, one might guess that the late time contributions from $1+z < 4$ should be larger in amplitude than that of $3000 > 1+z > 4$ by a factor of about 10. However, since the universe is also fully ionized at these late times, the amplitude of the resulting spectral distortion is suppressed and turns out to be smaller than the distortion from $3000 > 1+z > 4$ by an order of magnitude for the two masses that we have tested.

Moreover, if the late time spectral distortion does not have significant ICS or atomic line contributions, then it is degenerate with the distortions from reionization and structure formation, since all of these are $y$-type distortions sourced by heating.
If ICS is significant, then the distortion will have a high energy tail that can be distinguished from other sources of spectral distortions, as shown in the right panel of Fig.~\ref{fig:low_rs}.

\begin{figure*}
	\includegraphics[width=\textwidth]{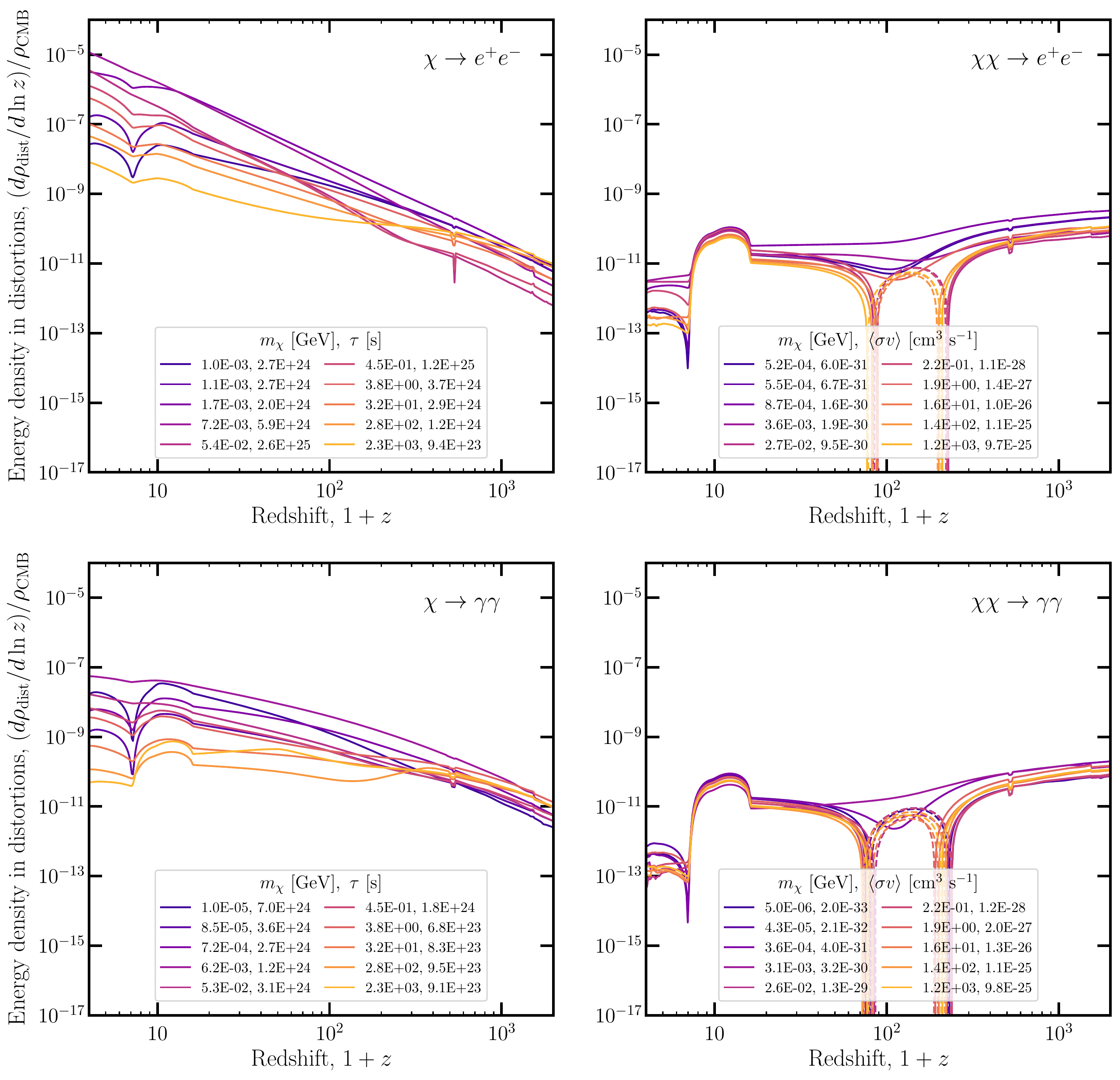}
	\caption{
		Rate of change in the energy of the spectral distortion relative to $\Lambda$CDM for the same models as in  Fig.~\ref{fig:dist_grid}.
		In other words, this figure shows how the energy in the spectra shown in Fig.~\ref{fig:dist_grid_noLCDM} evolves with time.
	}
	\label{fig:eng_rate_grid_noLCDM}
\end{figure*}
\begin{figure}
	\includegraphics[width=0.5\columnwidth]{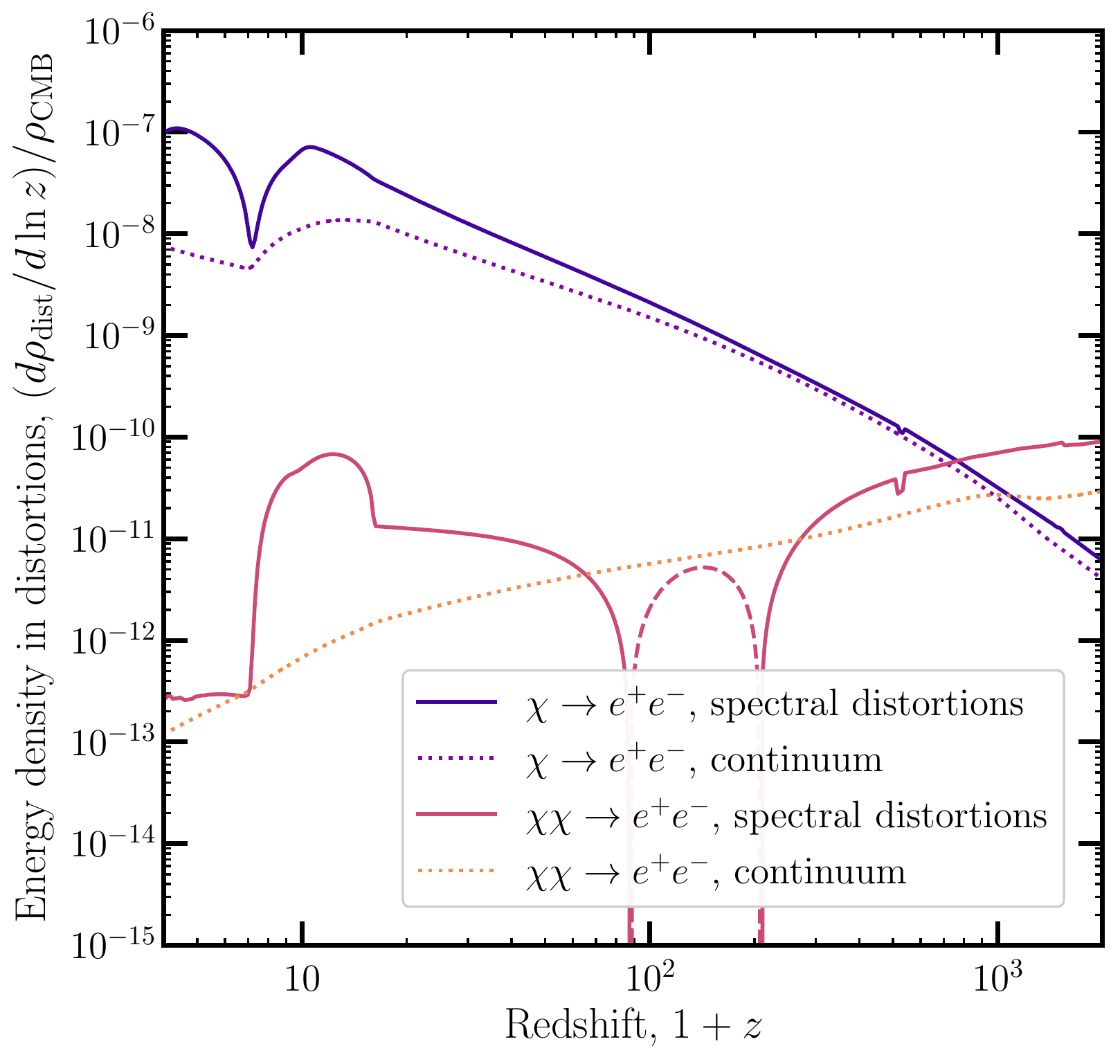}
	\caption{
		Comparison of the rate of change in the energy of the spectral distortion to the rate of energy deposited in the continuum channel for two of the models shown in Fig.~\ref{fig:eng_rate_grid_noLCDM}.
	}
	\label{fig:dist_vs_cont}
\end{figure}
%

\section{Energy density in spectral distortions}
\label{app:eng_rate}
%
Here, we validate our results by examining how much energy is deposited into the spectral distortions as a function of redshift.
Fig.~\ref{fig:eng_rate_grid_noLCDM} shows the rate of change in the energy of the spectral distortion, not including the contributions to the distortion from $\Lambda$CDM processes.
In other words, this figure shows how the energy in the spectra shown in Fig.~\ref{fig:dist_grid_noLCDM} evolves with time.

For comparison, in Fig.~\ref{fig:dist_vs_cont}, we show two of the models in Fig.~\ref{fig:dist_grid_noLCDM}, one for decay to $e^+ e^-$ and one for annihilation to $e^+ e^-$, and also show an estimate for the rate at which energy is deposited into the continuum for those same models, computed as discussed in \citetalias{paperI}. 
This calculation largely tracks the $f_\text{cont}$ calculation in the original version of \texttt{DarkHistory}, albeit with improvements to the treatment of low-energy electrons; it was intended as a calorimetric estimate of the total power into spectral distortions. 
The full spectral distortion calculation includes effects that the continuum calculation does not---in particular, $y$-type distortions from heating are not included in the contributions to the continuum channel---and provides the spectrum itself rather than simply an integrated quantity. 
This comparison can be viewed as a cross-check on our previous estimate of the integrated energy, and we do find that the shape and normalization of the curves is generally similar; however, detailed sensitivity estimates should use the spectral distortion results.

We see that the general shape of all the curves is consistent with the scaling for the energy injection rate discussed in App.~\ref{app:other_rs}: for decays, this rate scales as $(1+z)^{-2.5}$ during matter domination and $(1+z)^{-3}$ during radiation domination and for annihilations, the rate goes as $(1+z)^{0.5}$ during matter domination and is constant during radiation domination.
There are obvious exceptions to this trend.
For example, when reionization begins, there is an increase to the rate of energy deposited in spectral distortions due to photoheating and emission from excited hydrogen atoms.
There is also a small artifact around $1+z \sim 500$ due to a change in the way we calculate the $y$ parameter after this redshift, see \citetalias{paperI} for details.

In addition, for $s$-wave annihilation, there is a negative feature around $1+z \sim 100$.
The main contributions to spectral distortions at this time are atomic line emission and a negative amplitude $y$-type distortion. 
This $y$-distortion is present because the matter is always slightly colder than the CMB; the matter temperature would cool faster than the CMB if their temperatures were not coupled at early times.
The atomic lines are a positive contribution to the energy density in spectral distortions, whereas the $y$-distortion is a negative contribution.
Both contributions are enhanced in the presence of exotic energy injection due to the larger residual ionization at these times; hence the contribution to the energy density caused by exotic energy injection can be negative when the enhancement of the $y$-distortion dominates.

\section{Comparison to \texttt{Recfast}}
\label{app:recfast_xcheck}

%
\begin{figure*}
	\includegraphics[width=\textwidth]{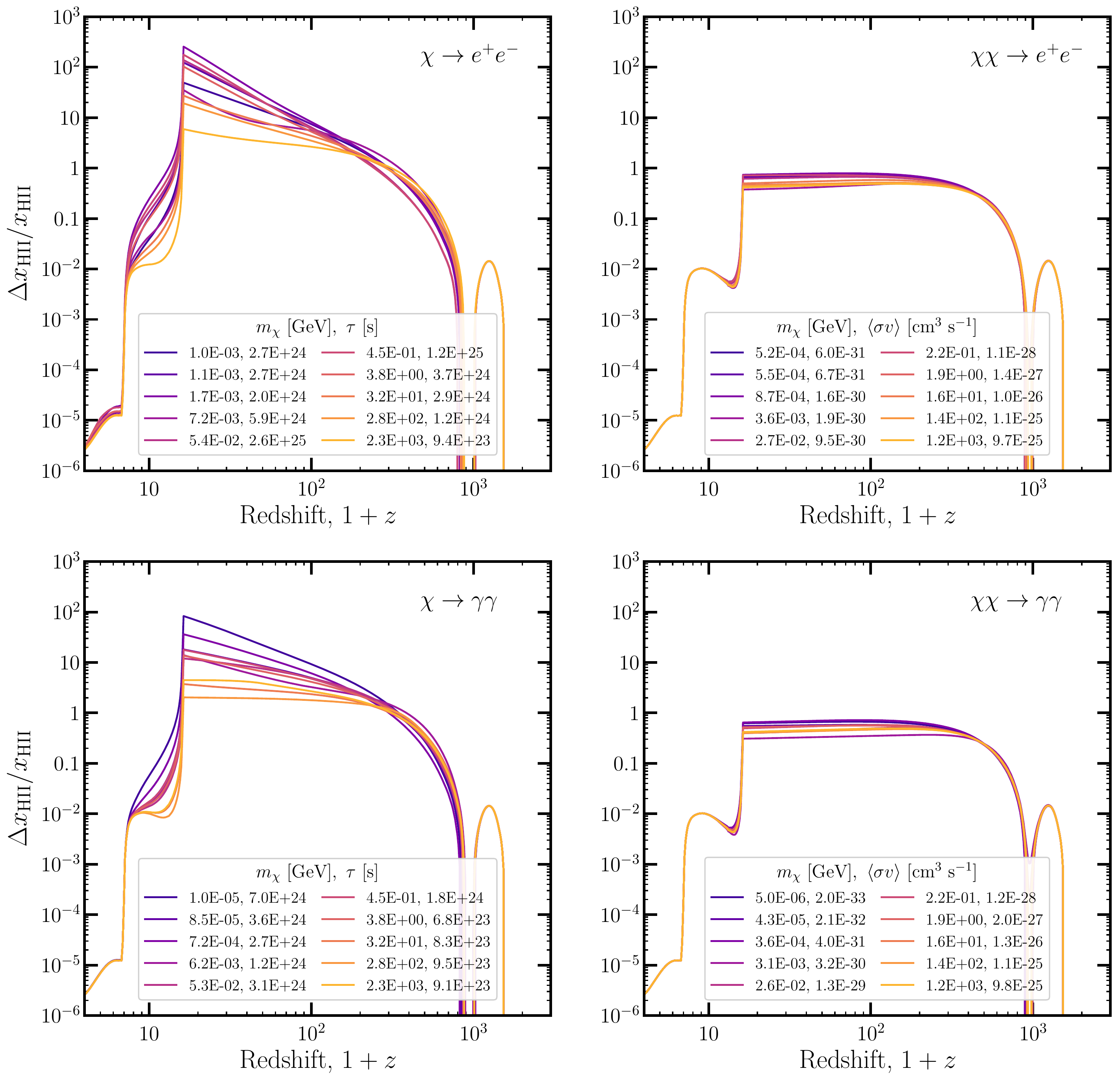}
	\caption{Difference in the ionization history relative to that calculated using \texttt{Recfast}.
		In other words, this is the difference in ionization between the two methods, divided by the history calculated with  \texttt{Recfast}.
		While the differences in ionization between histories with/without exotic energy injection is $\mathcal{O}(1)$ for annihilating DM, the relative change can be as large as a factor of a few hundred for decaying models.}
	\label{fig:delta_xe_grid_recfast}
\end{figure*}

In Section~\ref{sec:old_ion}, we showed that the difference in the ionization histories calculated using the method described in \citetalias{paperI} and \dhis \texttt{v1.0} were not significant.
Hence, a comparison of the ionization histories calculated with the new method against \texttt{Recfast} is not technically new, but we show the results here for completion.

Fig.\ref{fig:delta_xe_grid_recfast} shows the relative difference in $x_\text{HII} (z)$ calculated using the upgrades from \citetalias{paperI} and \texttt{Recfast}; in other words, Fig.\ref{fig:delta_xe_grid_recfast} shows the change in the ionization when we include the effect of exotic energy injection.
The signal is largest for decaying dark matter models, where the ionization history can be enhanced by as much as $\mathcal{O} (100)$ around $1+z \sim 20$.

\bibliography{bib}

\begin{thebibliography}{103}%
\makeatletter
\providecommand \@ifxundefined [1]{%
 \@ifx{#1\undefined}
}%
\providecommand \@ifnum [1]{%
 \ifnum #1\expandafter \@firstoftwo
 \else \expandafter \@secondoftwo
 \fi
}%
\providecommand \@ifx [1]{%
 \ifx #1\expandafter \@firstoftwo
 \else \expandafter \@secondoftwo
 \fi
}%
\providecommand \natexlab [1]{#1}%
\providecommand \enquote  [1]{``#1''}%
\providecommand \bibnamefont  [1]{#1}%
\providecommand \bibfnamefont [1]{#1}%
\providecommand \citenamefont [1]{#1}%
\providecommand \href@noop [0]{\@secondoftwo}%
\providecommand \href [0]{\begingroup \@sanitize@url \@href}%
\providecommand \@href[1]{\@@startlink{#1}\@@href}%
\providecommand \@@href[1]{\endgroup#1\@@endlink}%
\providecommand \@sanitize@url [0]{\catcode `\\12\catcode `\$12\catcode
  `\&12\catcode `\#12\catcode `\^12\catcode `\_12\catcode `\%12\relax}%
\providecommand \@@startlink[1]{}%
\providecommand \@@endlink[0]{}%
\providecommand \url  [0]{\begingroup\@sanitize@url \@url }%
\providecommand \@url [1]{\endgroup\@href {#1}{\urlprefix }}%
\providecommand \urlprefix  [0]{URL }%
\providecommand \Eprint [0]{\href }%
\providecommand \doibase [0]{http://dx.doi.org/}%
\providecommand \selectlanguage [0]{\@gobble}%
\providecommand \bibinfo  [0]{\@secondoftwo}%
\providecommand \bibfield  [0]{\@secondoftwo}%
\providecommand \translation [1]{[#1]}%
\providecommand \BibitemOpen [0]{}%
\providecommand \bibitemStop [0]{}%
\providecommand \bibitemNoStop [0]{.\EOS\space}%
\providecommand \EOS [0]{\spacefactor3000\relax}%
\providecommand \BibitemShut  [1]{\csname bibitem#1\endcsname}%
\let\auto@bib@innerbib\@empty
\bibitem [{\citenamefont {Slatyer}\ \emph {et~al.}(2009)\citenamefont
  {Slatyer}, \citenamefont {Padmanabhan},\ and\ \citenamefont
  {Finkbeiner}}]{0906.1197}%
  \BibitemOpen
  \bibfield  {author} {\bibinfo {author} {\bibfnamefont {Tracy~R.}\
  \bibnamefont {Slatyer}}, \bibinfo {author} {\bibfnamefont {Nikhil}\
  \bibnamefont {Padmanabhan}}, \ and\ \bibinfo {author} {\bibfnamefont
  {Douglas~P.}\ \bibnamefont {Finkbeiner}},\ }\bibfield  {title} {\enquote
  {\bibinfo {title} {{CMB Constraints on WIMP Annihilation: Energy Absorption
  During the Recombination Epoch}},}\ }\href {\doibase
  10.1103/PhysRevD.80.043526} {\bibfield  {journal} {\bibinfo  {journal} {Phys.
  Rev. D}\ }\textbf {\bibinfo {volume} {80}},\ \bibinfo {pages} {043526}
  (\bibinfo {year} {2009})},\ \Eprint {http://arxiv.org/abs/0906.1197}
  {arXiv:0906.1197 [astro-ph.CO]} \BibitemShut {NoStop}%
\bibitem [{\citenamefont {Cirelli}\ \emph {et~al.}(2009)\citenamefont
  {Cirelli}, \citenamefont {Iocco},\ and\ \citenamefont {Panci}}]{0907.0719}%
  \BibitemOpen
  \bibfield  {author} {\bibinfo {author} {\bibfnamefont {Marco}\ \bibnamefont
  {Cirelli}}, \bibinfo {author} {\bibfnamefont {Fabio}\ \bibnamefont {Iocco}},
  \ and\ \bibinfo {author} {\bibfnamefont {Paolo}\ \bibnamefont {Panci}},\
  }\bibfield  {title} {\enquote {\bibinfo {title} {{Constraints on Dark Matter
  annihilations from reionization and heating of the intergalactic gas}},}\
  }\href {\doibase 10.1088/1475-7516/2009/10/009} {\bibfield  {journal}
  {\bibinfo  {journal} {JCAP}\ }\textbf {\bibinfo {volume} {10}},\ \bibinfo
  {pages} {009} (\bibinfo {year} {2009})},\ \Eprint
  {http://arxiv.org/abs/0907.0719} {arXiv:0907.0719 [astro-ph.CO]} \BibitemShut
  {NoStop}%
\bibitem [{\citenamefont {Kanzaki}\ \emph {et~al.}(2010)\citenamefont
  {Kanzaki}, \citenamefont {Kawasaki},\ and\ \citenamefont
  {Nakayama}}]{0907.3985}%
  \BibitemOpen
  \bibfield  {author} {\bibinfo {author} {\bibfnamefont {Toru}\ \bibnamefont
  {Kanzaki}}, \bibinfo {author} {\bibfnamefont {Masahiro}\ \bibnamefont
  {Kawasaki}}, \ and\ \bibinfo {author} {\bibfnamefont {Kazunori}\ \bibnamefont
  {Nakayama}},\ }\bibfield  {title} {\enquote {\bibinfo {title} {{Effects of
  Dark Matter Annihilation on the Cosmic Microwave Background}},}\ }\href
  {\doibase 10.1143/PTP.123.853} {\bibfield  {journal} {\bibinfo  {journal}
  {Prog. Theor. Phys.}\ }\textbf {\bibinfo {volume} {123}},\ \bibinfo {pages}
  {853--865} (\bibinfo {year} {2010})},\ \Eprint
  {http://arxiv.org/abs/0907.3985} {arXiv:0907.3985 [astro-ph.CO]} \BibitemShut
  {NoStop}%
\bibitem [{\citenamefont {Diamanti}\ \emph {et~al.}(2014)\citenamefont
  {Diamanti}, \citenamefont {Lopez-Honorez}, \citenamefont {Mena},
  \citenamefont {Palomares-Ruiz},\ and\ \citenamefont {Vincent}}]{1308.2578}%
  \BibitemOpen
  \bibfield  {author} {\bibinfo {author} {\bibfnamefont {Roberta}\ \bibnamefont
  {Diamanti}}, \bibinfo {author} {\bibfnamefont {Laura}\ \bibnamefont
  {Lopez-Honorez}}, \bibinfo {author} {\bibfnamefont {Olga}\ \bibnamefont
  {Mena}}, \bibinfo {author} {\bibfnamefont {Sergio}\ \bibnamefont
  {Palomares-Ruiz}}, \ and\ \bibinfo {author} {\bibfnamefont {Aaron~C.}\
  \bibnamefont {Vincent}},\ }\bibfield  {title} {\enquote {\bibinfo {title}
  {{Constraining Dark Matter Late-Time Energy Injection: Decays and P-Wave
  Annihilations}},}\ }\href {\doibase 10.1088/1475-7516/2014/02/017} {\bibfield
   {journal} {\bibinfo  {journal} {JCAP}\ }\textbf {\bibinfo {volume} {02}},\
  \bibinfo {pages} {017} (\bibinfo {year} {2014})},\ \Eprint
  {http://arxiv.org/abs/1308.2578} {arXiv:1308.2578 [astro-ph.CO]} \BibitemShut
  {NoStop}%
\bibitem [{\citenamefont {Evoli}\ \emph {et~al.}(2014)\citenamefont {Evoli},
  \citenamefont {Mesinger},\ and\ \citenamefont {Ferrara}}]{1408.1109}%
  \BibitemOpen
  \bibfield  {author} {\bibinfo {author} {\bibfnamefont {Carmelo}\ \bibnamefont
  {Evoli}}, \bibinfo {author} {\bibfnamefont {Andrei}\ \bibnamefont
  {Mesinger}}, \ and\ \bibinfo {author} {\bibfnamefont {Andrea}\ \bibnamefont
  {Ferrara}},\ }\bibfield  {title} {\enquote {\bibinfo {title} {{Unveiling the
  nature of dark matter with high redshift 21 cm line experiments}},}\ }\href
  {\doibase 10.1088/1475-7516/2014/11/024} {\bibfield  {journal} {\bibinfo
  {journal} {JCAP}\ }\textbf {\bibinfo {volume} {11}},\ \bibinfo {pages} {024}
  (\bibinfo {year} {2014})},\ \Eprint {http://arxiv.org/abs/1408.1109}
  {arXiv:1408.1109 [astro-ph.HE]} \BibitemShut {NoStop}%
\bibitem [{\citenamefont {Slatyer}(2016{\natexlab{a}})}]{1506.03811}%
  \BibitemOpen
  \bibfield  {author} {\bibinfo {author} {\bibfnamefont {Tracy~R.}\
  \bibnamefont {Slatyer}},\ }\bibfield  {title} {\enquote {\bibinfo {title}
  {{Indirect dark matter signatures in the cosmic dark ages. I. Generalizing
  the bound on s-wave dark matter annihilation from Planck results}},}\ }\href
  {\doibase 10.1103/PhysRevD.93.023527} {\bibfield  {journal} {\bibinfo
  {journal} {Phys. Rev. D}\ }\textbf {\bibinfo {volume} {93}},\ \bibinfo
  {pages} {023527} (\bibinfo {year} {2016}{\natexlab{a}})},\ \Eprint
  {http://arxiv.org/abs/1506.03811} {arXiv:1506.03811 [hep-ph]} \BibitemShut
  {NoStop}%
\bibitem [{\citenamefont {Lopez-Honorez}\ \emph {et~al.}(2016)\citenamefont
  {Lopez-Honorez}, \citenamefont {Mena}, \citenamefont {Molin\'e},
  \citenamefont {Palomares-Ruiz},\ and\ \citenamefont {Vincent}}]{1603.06795}%
  \BibitemOpen
  \bibfield  {author} {\bibinfo {author} {\bibfnamefont {Laura}\ \bibnamefont
  {Lopez-Honorez}}, \bibinfo {author} {\bibfnamefont {Olga}\ \bibnamefont
  {Mena}}, \bibinfo {author} {\bibfnamefont {\'Angeles}\ \bibnamefont
  {Molin\'e}}, \bibinfo {author} {\bibfnamefont {Sergio}\ \bibnamefont
  {Palomares-Ruiz}}, \ and\ \bibinfo {author} {\bibfnamefont {Aaron~C.}\
  \bibnamefont {Vincent}},\ }\bibfield  {title} {\enquote {\bibinfo {title}
  {{The 21 cm signal and the interplay between dark matter annihilations and
  astrophysical processes}},}\ }\href {\doibase 10.1088/1475-7516/2016/08/004}
  {\bibfield  {journal} {\bibinfo  {journal} {JCAP}\ }\textbf {\bibinfo
  {volume} {08}},\ \bibinfo {pages} {004} (\bibinfo {year} {2016})},\ \Eprint
  {http://arxiv.org/abs/1603.06795} {arXiv:1603.06795 [astro-ph.CO]}
  \BibitemShut {NoStop}%
\bibitem [{\citenamefont {Liu}\ \emph {et~al.}(2016)\citenamefont {Liu},
  \citenamefont {Slatyer},\ and\ \citenamefont {Zavala}}]{1604.02457}%
  \BibitemOpen
  \bibfield  {author} {\bibinfo {author} {\bibfnamefont {Hongwan}\ \bibnamefont
  {Liu}}, \bibinfo {author} {\bibfnamefont {Tracy~R.}\ \bibnamefont {Slatyer}},
  \ and\ \bibinfo {author} {\bibfnamefont {Jes\'us}\ \bibnamefont {Zavala}},\
  }\bibfield  {title} {\enquote {\bibinfo {title} {{Contributions to cosmic
  reionization from dark matter annihilation and decay}},}\ }\href {\doibase
  10.1103/PhysRevD.94.063507} {\bibfield  {journal} {\bibinfo  {journal} {Phys.
  Rev. D}\ }\textbf {\bibinfo {volume} {94}},\ \bibinfo {pages} {063507}
  (\bibinfo {year} {2016})},\ \Eprint {http://arxiv.org/abs/1604.02457}
  {arXiv:1604.02457 [astro-ph.CO]} \BibitemShut {NoStop}%
\bibitem [{\citenamefont {Slatyer}\ and\ \citenamefont
  {Wu}(2017)}]{1610.06933}%
  \BibitemOpen
  \bibfield  {author} {\bibinfo {author} {\bibfnamefont {Tracy~R.}\
  \bibnamefont {Slatyer}}\ and\ \bibinfo {author} {\bibfnamefont {Chih-Liang}\
  \bibnamefont {Wu}},\ }\bibfield  {title} {\enquote {\bibinfo {title}
  {{General Constraints on Dark Matter Decay from the Cosmic Microwave
  Background}},}\ }\href {\doibase 10.1103/PhysRevD.95.023010} {\bibfield
  {journal} {\bibinfo  {journal} {Phys. Rev. D}\ }\textbf {\bibinfo {volume}
  {95}},\ \bibinfo {pages} {023010} (\bibinfo {year} {2017})},\ \Eprint
  {http://arxiv.org/abs/1610.06933} {arXiv:1610.06933 [astro-ph.CO]}
  \BibitemShut {NoStop}%
\bibitem [{\citenamefont {Poulin}\ \emph {et~al.}(2017)\citenamefont {Poulin},
  \citenamefont {Lesgourgues},\ and\ \citenamefont {Serpico}}]{1610.10051}%
  \BibitemOpen
  \bibfield  {author} {\bibinfo {author} {\bibfnamefont {Vivian}\ \bibnamefont
  {Poulin}}, \bibinfo {author} {\bibfnamefont {Julien}\ \bibnamefont
  {Lesgourgues}}, \ and\ \bibinfo {author} {\bibfnamefont {Pasquale~D.}\
  \bibnamefont {Serpico}},\ }\bibfield  {title} {\enquote {\bibinfo {title}
  {{Cosmological constraints on exotic injection of electromagnetic energy}},}\
  }\href {\doibase 10.1088/1475-7516/2017/03/043} {\bibfield  {journal}
  {\bibinfo  {journal} {JCAP}\ }\textbf {\bibinfo {volume} {03}},\ \bibinfo
  {pages} {043} (\bibinfo {year} {2017})},\ \Eprint
  {http://arxiv.org/abs/1610.10051} {arXiv:1610.10051 [astro-ph.CO]}
  \BibitemShut {NoStop}%
\bibitem [{\citenamefont {Hiss}\ \emph {et~al.}(2018)\citenamefont {Hiss},
  \citenamefont {Walther}, \citenamefont {Hennawi}, \citenamefont {O\~norbe},
  \citenamefont {O'Meara}, \citenamefont {Rorai},\ and\ \citenamefont
  {Luki\'c}}]{1710.00700}%
  \BibitemOpen
  \bibfield  {author} {\bibinfo {author} {\bibfnamefont {Hector}\ \bibnamefont
  {Hiss}}, \bibinfo {author} {\bibfnamefont {Michael}\ \bibnamefont {Walther}},
  \bibinfo {author} {\bibfnamefont {Joseph~F.}\ \bibnamefont {Hennawi}},
  \bibinfo {author} {\bibfnamefont {Jos\'e}\ \bibnamefont {O\~norbe}}, \bibinfo
  {author} {\bibfnamefont {John~M.}\ \bibnamefont {O'Meara}}, \bibinfo {author}
  {\bibfnamefont {Alberto}\ \bibnamefont {Rorai}}, \ and\ \bibinfo {author}
  {\bibfnamefont {Zarija}\ \bibnamefont {Luki\'c}},\ }\bibfield  {title}
  {\enquote {\bibinfo {title} {{A New Measurement of the
  Temperature\textendash{}density Relation of the IGM from Voigt Profile
  Fitting}},}\ }\href {\doibase 10.3847/1538-4357/aada86} {\bibfield  {journal}
  {\bibinfo  {journal} {Astrophys. J.}\ }\textbf {\bibinfo {volume} {865}},\
  \bibinfo {pages} {42} (\bibinfo {year} {2018})},\ \Eprint
  {http://arxiv.org/abs/1710.00700} {arXiv:1710.00700 [astro-ph.CO]}
  \BibitemShut {NoStop}%
\bibitem [{\citenamefont {D'Amico}\ \emph {et~al.}(2018)\citenamefont
  {D'Amico}, \citenamefont {Panci},\ and\ \citenamefont
  {Strumia}}]{1803.03629}%
  \BibitemOpen
  \bibfield  {author} {\bibinfo {author} {\bibfnamefont {Guido}\ \bibnamefont
  {D'Amico}}, \bibinfo {author} {\bibfnamefont {Paolo}\ \bibnamefont {Panci}},
  \ and\ \bibinfo {author} {\bibfnamefont {Alessandro}\ \bibnamefont
  {Strumia}},\ }\bibfield  {title} {\enquote {\bibinfo {title} {{Bounds on Dark
  Matter annihilations from 21 cm data}},}\ }\href {\doibase
  10.1103/PhysRevLett.121.011103} {\bibfield  {journal} {\bibinfo  {journal}
  {Phys. Rev. Lett.}\ }\textbf {\bibinfo {volume} {121}},\ \bibinfo {pages}
  {011103} (\bibinfo {year} {2018})},\ \Eprint
  {http://arxiv.org/abs/1803.03629} {arXiv:1803.03629 [astro-ph.CO]}
  \BibitemShut {NoStop}%
\bibitem [{\citenamefont {Liu}\ and\ \citenamefont
  {Slatyer}(2018)}]{1803.09739}%
  \BibitemOpen
  \bibfield  {author} {\bibinfo {author} {\bibfnamefont {Hongwan}\ \bibnamefont
  {Liu}}\ and\ \bibinfo {author} {\bibfnamefont {Tracy~R.}\ \bibnamefont
  {Slatyer}},\ }\bibfield  {title} {\enquote {\bibinfo {title} {{Implications
  of a 21-cm signal for dark matter annihilation and decay}},}\ }\href
  {\doibase 10.1103/PhysRevD.98.023501} {\bibfield  {journal} {\bibinfo
  {journal} {Phys. Rev. D}\ }\textbf {\bibinfo {volume} {98}},\ \bibinfo
  {pages} {023501} (\bibinfo {year} {2018})},\ \Eprint
  {http://arxiv.org/abs/1803.09739} {arXiv:1803.09739 [astro-ph.CO]}
  \BibitemShut {NoStop}%
\bibitem [{\citenamefont {Cheung}\ \emph {et~al.}(2019)\citenamefont {Cheung},
  \citenamefont {Kuo}, \citenamefont {Ng},\ and\ \citenamefont
  {Tsai}}]{1803.09398}%
  \BibitemOpen
  \bibfield  {author} {\bibinfo {author} {\bibfnamefont {Kingman}\ \bibnamefont
  {Cheung}}, \bibinfo {author} {\bibfnamefont {Jui-Lin}\ \bibnamefont {Kuo}},
  \bibinfo {author} {\bibfnamefont {Kin-Wang}\ \bibnamefont {Ng}}, \ and\
  \bibinfo {author} {\bibfnamefont {Yue-Lin~Sming}\ \bibnamefont {Tsai}},\
  }\bibfield  {title} {\enquote {\bibinfo {title} {{The impact of EDGES 21-cm
  data on dark matter interactions}},}\ }\href {\doibase
  10.1016/j.physletb.2018.11.058} {\bibfield  {journal} {\bibinfo  {journal}
  {Phys. Lett. B}\ }\textbf {\bibinfo {volume} {789}},\ \bibinfo {pages}
  {137--144} (\bibinfo {year} {2019})},\ \Eprint
  {http://arxiv.org/abs/1803.09398} {arXiv:1803.09398 [astro-ph.CO]}
  \BibitemShut {NoStop}%
\bibitem [{\citenamefont {Mitridate}\ and\ \citenamefont
  {Podo}(2018)}]{1803.11169}%
  \BibitemOpen
  \bibfield  {author} {\bibinfo {author} {\bibfnamefont {Andrea}\ \bibnamefont
  {Mitridate}}\ and\ \bibinfo {author} {\bibfnamefont {Alessandro}\
  \bibnamefont {Podo}},\ }\bibfield  {title} {\enquote {\bibinfo {title}
  {{Bounds on Dark Matter decay from 21 cm line}},}\ }\href {\doibase
  10.1088/1475-7516/2018/05/069} {\bibfield  {journal} {\bibinfo  {journal}
  {JCAP}\ }\textbf {\bibinfo {volume} {05}},\ \bibinfo {pages} {069} (\bibinfo
  {year} {2018})},\ \Eprint {http://arxiv.org/abs/1803.11169} {arXiv:1803.11169
  [hep-ph]} \BibitemShut {NoStop}%
\bibitem [{\citenamefont {Clark}\ \emph {et~al.}(2018)\citenamefont {Clark},
  \citenamefont {Dutta}, \citenamefont {Gao}, \citenamefont {Ma},\ and\
  \citenamefont {Strigari}}]{1803.09390}%
  \BibitemOpen
  \bibfield  {author} {\bibinfo {author} {\bibfnamefont {Steven}\ \bibnamefont
  {Clark}}, \bibinfo {author} {\bibfnamefont {Bhaskar}\ \bibnamefont {Dutta}},
  \bibinfo {author} {\bibfnamefont {Yu}~\bibnamefont {Gao}}, \bibinfo {author}
  {\bibfnamefont {Yin-Zhe}\ \bibnamefont {Ma}}, \ and\ \bibinfo {author}
  {\bibfnamefont {Louis~E.}\ \bibnamefont {Strigari}},\ }\bibfield  {title}
  {\enquote {\bibinfo {title} {{21 cm limits on decaying dark matter and
  primordial black holes}},}\ }\href {\doibase 10.1103/PhysRevD.98.043006}
  {\bibfield  {journal} {\bibinfo  {journal} {Phys. Rev. D}\ }\textbf {\bibinfo
  {volume} {98}},\ \bibinfo {pages} {043006} (\bibinfo {year} {2018})},\
  \Eprint {http://arxiv.org/abs/1803.09390} {arXiv:1803.09390 [astro-ph.HE]}
  \BibitemShut {NoStop}%
\bibitem [{\citenamefont {Walther}\ \emph {et~al.}(2019)\citenamefont
  {Walther}, \citenamefont {O\~norbe}, \citenamefont {Hennawi},\ and\
  \citenamefont {Luki\'c}}]{1808.04367}%
  \BibitemOpen
  \bibfield  {author} {\bibinfo {author} {\bibfnamefont {Michael}\ \bibnamefont
  {Walther}}, \bibinfo {author} {\bibfnamefont {Jose}\ \bibnamefont
  {O\~norbe}}, \bibinfo {author} {\bibfnamefont {Joseph~F.}\ \bibnamefont
  {Hennawi}}, \ and\ \bibinfo {author} {\bibfnamefont {Zarija}\ \bibnamefont
  {Luki\'c}},\ }\bibfield  {title} {\enquote {\bibinfo {title} {{New
  Constraints on IGM Thermal Evolution from the Ly\ensuremath{\alpha} Forest
  Power Spectrum}},}\ }\href {\doibase 10.3847/1538-4357/aafad1} {\bibfield
  {journal} {\bibinfo  {journal} {Astrophys. J.}\ }\textbf {\bibinfo {volume}
  {872}},\ \bibinfo {pages} {13} (\bibinfo {year} {2019})},\ \Eprint
  {http://arxiv.org/abs/1808.04367} {arXiv:1808.04367 [astro-ph.CO]}
  \BibitemShut {NoStop}%
\bibitem [{\citenamefont {McDermott}\ and\ \citenamefont
  {Witte}(2020)}]{1911.05086}%
  \BibitemOpen
  \bibfield  {author} {\bibinfo {author} {\bibfnamefont {Samuel~D.}\
  \bibnamefont {McDermott}}\ and\ \bibinfo {author} {\bibfnamefont {Samuel~J.}\
  \bibnamefont {Witte}},\ }\bibfield  {title} {\enquote {\bibinfo {title}
  {{Cosmological evolution of light dark photon dark matter}},}\ }\href
  {\doibase 10.1103/PhysRevD.101.063030} {\bibfield  {journal} {\bibinfo
  {journal} {Phys. Rev. D}\ }\textbf {\bibinfo {volume} {101}},\ \bibinfo
  {pages} {063030} (\bibinfo {year} {2020})},\ \Eprint
  {http://arxiv.org/abs/1911.05086} {arXiv:1911.05086 [hep-ph]} \BibitemShut
  {NoStop}%
\bibitem [{\citenamefont {Gaikwad}\ \emph {et~al.}(2020)\citenamefont {Gaikwad}
  \emph {et~al.}}]{2001.10018}%
  \BibitemOpen
  \bibfield  {author} {\bibinfo {author} {\bibfnamefont {Prakash}\ \bibnamefont
  {Gaikwad}} \emph {et~al.},\ }\bibfield  {title} {\enquote {\bibinfo {title}
  {{Probing the thermal state of the intergalactic medium at z \ensuremath{>} 5
  with the transmission spikes in high-resolution Ly \ensuremath{\alpha} forest
  spectra}},}\ }\href {\doibase 10.1093/mnras/staa907} {\bibfield  {journal}
  {\bibinfo  {journal} {Mon. Not. Roy. Astron. Soc.}\ }\textbf {\bibinfo
  {volume} {494}},\ \bibinfo {pages} {5091--5109} (\bibinfo {year} {2020})},\
  \Eprint {http://arxiv.org/abs/2001.10018} {arXiv:2001.10018 [astro-ph.CO]}
  \BibitemShut {NoStop}%
\bibitem [{\citenamefont {Caputo}\ \emph {et~al.}(2020)\citenamefont {Caputo},
  \citenamefont {Liu}, \citenamefont {Mishra-Sharma},\ and\ \citenamefont
  {Ruderman}}]{2002.05165}%
  \BibitemOpen
  \bibfield  {author} {\bibinfo {author} {\bibfnamefont {Andrea}\ \bibnamefont
  {Caputo}}, \bibinfo {author} {\bibfnamefont {Hongwan}\ \bibnamefont {Liu}},
  \bibinfo {author} {\bibfnamefont {Siddharth}\ \bibnamefont {Mishra-Sharma}},
  \ and\ \bibinfo {author} {\bibfnamefont {Joshua~T.}\ \bibnamefont
  {Ruderman}},\ }\bibfield  {title} {\enquote {\bibinfo {title} {{Dark Photon
  Oscillations in Our Inhomogeneous Universe}},}\ }\href {\doibase
  10.1103/PhysRevLett.125.221303} {\bibfield  {journal} {\bibinfo  {journal}
  {Phys. Rev. Lett.}\ }\textbf {\bibinfo {volume} {125}},\ \bibinfo {pages}
  {221303} (\bibinfo {year} {2020})},\ \Eprint
  {http://arxiv.org/abs/2002.05165} {arXiv:2002.05165 [astro-ph.CO]}
  \BibitemShut {NoStop}%
\bibitem [{\citenamefont {Witte}\ \emph {et~al.}(2020)\citenamefont {Witte},
  \citenamefont {Rosauro-Alcaraz}, \citenamefont {McDermott},\ and\
  \citenamefont {Poulin}}]{2003.13698}%
  \BibitemOpen
  \bibfield  {author} {\bibinfo {author} {\bibfnamefont {Samuel~J.}\
  \bibnamefont {Witte}}, \bibinfo {author} {\bibfnamefont {Salvador}\
  \bibnamefont {Rosauro-Alcaraz}}, \bibinfo {author} {\bibfnamefont
  {Samuel~D.}\ \bibnamefont {McDermott}}, \ and\ \bibinfo {author}
  {\bibfnamefont {Vivian}\ \bibnamefont {Poulin}},\ }\bibfield  {title}
  {\enquote {\bibinfo {title} {{Dark photon dark matter in the presence of
  inhomogeneous structure}},}\ }\href {\doibase 10.1007/JHEP06(2020)132}
  {\bibfield  {journal} {\bibinfo  {journal} {JHEP}\ }\textbf {\bibinfo
  {volume} {06}},\ \bibinfo {pages} {132} (\bibinfo {year} {2020})},\ \Eprint
  {http://arxiv.org/abs/2003.13698} {arXiv:2003.13698 [astro-ph.CO]}
  \BibitemShut {NoStop}%
\bibitem [{\citenamefont {Liu}\ \emph {et~al.}(2021)\citenamefont {Liu},
  \citenamefont {Qin}, \citenamefont {Ridgway},\ and\ \citenamefont
  {Slatyer}}]{2008.01084}%
  \BibitemOpen
  \bibfield  {author} {\bibinfo {author} {\bibfnamefont {Hongwan}\ \bibnamefont
  {Liu}}, \bibinfo {author} {\bibfnamefont {Wenzer}\ \bibnamefont {Qin}},
  \bibinfo {author} {\bibfnamefont {Gregory~W.}\ \bibnamefont {Ridgway}}, \
  and\ \bibinfo {author} {\bibfnamefont {Tracy~R.}\ \bibnamefont {Slatyer}},\
  }\bibfield  {title} {\enquote {\bibinfo {title} {{Lyman-\ensuremath{\alpha}
  constraints on cosmic heating from dark matter annihilation and decay}},}\
  }\href {\doibase 10.1103/PhysRevD.104.043514} {\bibfield  {journal} {\bibinfo
   {journal} {Phys. Rev. D}\ }\textbf {\bibinfo {volume} {104}},\ \bibinfo
  {pages} {043514} (\bibinfo {year} {2021})},\ \Eprint
  {http://arxiv.org/abs/2008.01084} {arXiv:2008.01084 [astro-ph.CO]}
  \BibitemShut {NoStop}%
\bibitem [{\citenamefont {Gaikwad}\ \emph {et~al.}(2021)\citenamefont
  {Gaikwad}, \citenamefont {Srianand}, \citenamefont {Haehnelt},\ and\
  \citenamefont {Choudhury}}]{2009.00016}%
  \BibitemOpen
  \bibfield  {author} {\bibinfo {author} {\bibfnamefont {Prakash}\ \bibnamefont
  {Gaikwad}}, \bibinfo {author} {\bibfnamefont {Raghunathan}\ \bibnamefont
  {Srianand}}, \bibinfo {author} {\bibfnamefont {Martin~G.}\ \bibnamefont
  {Haehnelt}}, \ and\ \bibinfo {author} {\bibfnamefont {Tirthankar~Roy}\
  \bibnamefont {Choudhury}},\ }\bibfield  {title} {\enquote {\bibinfo {title}
  {{A consistent and robust measurement of the thermal state of the IGM at 2
  \ensuremath{\leq} z \ensuremath{\leq} 4 from a large sample of
  \,Ly\,\ensuremath{\alpha} forest spectra: evidence for late and rapid He\,ii
  reionization}},}\ }\href {\doibase 10.1093/mnras/stab2017} {\bibfield
  {journal} {\bibinfo  {journal} {Mon. Not. Roy. Astron. Soc.}\ }\textbf
  {\bibinfo {volume} {506}},\ \bibinfo {pages} {4389--4412} (\bibinfo {year}
  {2021})},\ \Eprint {http://arxiv.org/abs/2009.00016} {arXiv:2009.00016
  [astro-ph.CO]} \BibitemShut {NoStop}%
\bibitem [{\citenamefont {Bolton}\ \emph {et~al.}(2022)\citenamefont {Bolton},
  \citenamefont {Caputo}, \citenamefont {Liu},\ and\ \citenamefont
  {Viel}}]{2206.13520}%
  \BibitemOpen
  \bibfield  {author} {\bibinfo {author} {\bibfnamefont {James~S.}\
  \bibnamefont {Bolton}}, \bibinfo {author} {\bibfnamefont {Andrea}\
  \bibnamefont {Caputo}}, \bibinfo {author} {\bibfnamefont {Hongwan}\
  \bibnamefont {Liu}}, \ and\ \bibinfo {author} {\bibfnamefont {Matteo}\
  \bibnamefont {Viel}},\ }\bibfield  {title} {\enquote {\bibinfo {title}
  {{Hints of dark photon dark matter from observations and hydrodynamical
  simulations of the low-redshift Lyman-$\alpha$ forest}},}\ }\href@noop {} {\
  (\bibinfo {year} {2022})},\ \Eprint {http://arxiv.org/abs/2206.13520}
  {arXiv:2206.13520 [hep-ph]} \BibitemShut {NoStop}%
\bibitem [{\citenamefont {Lauer}\ \emph {et~al.}(2022)\citenamefont {Lauer}
  \emph {et~al.}}]{Lauer:2022fgc}%
  \BibitemOpen
  \bibfield  {author} {\bibinfo {author} {\bibfnamefont {Tod~R.}\ \bibnamefont
  {Lauer}} \emph {et~al.},\ }\bibfield  {title} {\enquote {\bibinfo {title}
  {{Anomalous Flux in the Cosmic Optical Background Detected with New Horizons
  Observations}},}\ }\href {\doibase 10.3847/2041-8213/ac573d} {\bibfield
  {journal} {\bibinfo  {journal} {Astrophys. J. Lett.}\ }\textbf {\bibinfo
  {volume} {927}},\ \bibinfo {pages} {L8} (\bibinfo {year} {2022})},\ \Eprint
  {http://arxiv.org/abs/2202.04273} {arXiv:2202.04273 [astro-ph.GA]}
  \BibitemShut {NoStop}%
\bibitem [{\citenamefont {Bernal}\ \emph {et~al.}(2022)\citenamefont {Bernal},
  \citenamefont {Sato-Polito},\ and\ \citenamefont
  {Kamionkowski}}]{Bernal:2022wsu}%
  \BibitemOpen
  \bibfield  {author} {\bibinfo {author} {\bibfnamefont {Jos\'e~Luis}\
  \bibnamefont {Bernal}}, \bibinfo {author} {\bibfnamefont {Gabriela}\
  \bibnamefont {Sato-Polito}}, \ and\ \bibinfo {author} {\bibfnamefont {Marc}\
  \bibnamefont {Kamionkowski}},\ }\bibfield  {title} {\enquote {\bibinfo
  {title} {{The cosmic optical background excess, dark matter, and
  line-intensity mapping}},}\ }\href@noop {} {\  (\bibinfo {year} {2022})},\
  \Eprint {http://arxiv.org/abs/2203.11236} {arXiv:2203.11236 [astro-ph.CO]}
  \BibitemShut {NoStop}%
\bibitem [{\citenamefont {Hirata}(2006)}]{Hirata:2005mz}%
  \BibitemOpen
  \bibfield  {author} {\bibinfo {author} {\bibfnamefont {Christopher~M.}\
  \bibnamefont {Hirata}},\ }\bibfield  {title} {\enquote {\bibinfo {title}
  {{Wouthuysen-Field coupling strength and application to high-redshift 21 cm
  radiation}},}\ }\href {\doibase 10.1111/j.1365-2966.2005.09949.x} {\bibfield
  {journal} {\bibinfo  {journal} {Mon. Not. Roy. Astron. Soc.}\ }\textbf
  {\bibinfo {volume} {367}},\ \bibinfo {pages} {259--274} (\bibinfo {year}
  {2006})},\ \Eprint {http://arxiv.org/abs/astro-ph/0507102}
  {arXiv:astro-ph/0507102} \BibitemShut {NoStop}%
\bibitem [{\citenamefont {Hirata}\ and\ \citenamefont
  {Padmanabhan}(2006)}]{Hirata:2006bt}%
  \BibitemOpen
  \bibfield  {author} {\bibinfo {author} {\bibfnamefont {Christopher~M.}\
  \bibnamefont {Hirata}}\ and\ \bibinfo {author} {\bibfnamefont {Nikhil}\
  \bibnamefont {Padmanabhan}},\ }\bibfield  {title} {\enquote {\bibinfo {title}
  {{Cosmological production of H(2) before the formation of the first
  galaxies}},}\ }\href {\doibase 10.1111/j.1365-2966.2006.10924.x} {\bibfield
  {journal} {\bibinfo  {journal} {Mon. Not. Roy. Astron. Soc.}\ }\textbf
  {\bibinfo {volume} {372}},\ \bibinfo {pages} {1175--1186} (\bibinfo {year}
  {2006})},\ \Eprint {http://arxiv.org/abs/astro-ph/0606437}
  {arXiv:astro-ph/0606437} \BibitemShut {NoStop}%
\bibitem [{\citenamefont {{Stecher}}\ and\ \citenamefont
  {{Williams}}(1967)}]{1967ApJ...149L..29S}%
  \BibitemOpen
  \bibfield  {author} {\bibinfo {author} {\bibfnamefont {T.~P.}\ \bibnamefont
  {{Stecher}}}\ and\ \bibinfo {author} {\bibfnamefont {D.~A.}\ \bibnamefont
  {{Williams}}},\ }\bibfield  {title} {\enquote {\bibinfo {title}
  {Photodestruction of hydrogen molecules in hi regions},}\ }\href {\doibase
  10.1086/180047} {\bibfield  {journal} {\bibinfo  {journal} {ApJ}\ }\textbf
  {\bibinfo {volume} {149}},\ \bibinfo {pages} {L29} (\bibinfo {year}
  {1967})}\BibitemShut {NoStop}%
\bibitem [{\citenamefont {{Abgrall}}\ \emph {et~al.}(1992)\citenamefont
  {{Abgrall}}, \citenamefont {{Le Bourlot}}, \citenamefont {{Pineau Des
  Forets}}, \citenamefont {{Roueff}}, \citenamefont {{Flower}},\ and\
  \citenamefont {{Heck}}}]{1992A&A...253..525A}%
  \BibitemOpen
  \bibfield  {author} {\bibinfo {author} {\bibfnamefont {H.}~\bibnamefont
  {{Abgrall}}}, \bibinfo {author} {\bibfnamefont {J.}~\bibnamefont {{Le
  Bourlot}}}, \bibinfo {author} {\bibfnamefont {G.}~\bibnamefont {{Pineau Des
  Forets}}}, \bibinfo {author} {\bibfnamefont {E.}~\bibnamefont {{Roueff}}},
  \bibinfo {author} {\bibfnamefont {D.~R.}\ \bibnamefont {{Flower}}}, \ and\
  \bibinfo {author} {\bibfnamefont {L.}~\bibnamefont {{Heck}}},\ }\bibfield
  {title} {\enquote {\bibinfo {title} {Photodissociation of h$_2$ and the
  h/h$_2$ transition in interstellar clouds},}\ }\href@noop {} {\bibfield
  {journal} {\bibinfo  {journal} {A\&A}\ }\textbf {\bibinfo {volume} {253}},\
  \bibinfo {pages} {525--536} (\bibinfo {year} {1992})}\BibitemShut {NoStop}%
\bibitem [{\citenamefont {{Haiman}}\ \emph {et~al.}(1996)\citenamefont
  {{Haiman}}, \citenamefont {{Rees}},\ and\ \citenamefont
  {{Loeb}}}]{1996ApJ...467..522H}%
  \BibitemOpen
  \bibfield  {author} {\bibinfo {author} {\bibfnamefont {Zoltan}\ \bibnamefont
  {{Haiman}}}, \bibinfo {author} {\bibfnamefont {Martin~J.}\ \bibnamefont
  {{Rees}}}, \ and\ \bibinfo {author} {\bibfnamefont {Abraham}\ \bibnamefont
  {{Loeb}}},\ }\bibfield  {title} {\enquote {\bibinfo {title} {H$_2$ cooling of
  primordial gas triggered by uv irradiation},}\ }\href {\doibase
  10.1086/177628} {\bibfield  {journal} {\bibinfo  {journal} {ApJ}\ }\textbf
  {\bibinfo {volume} {467}},\ \bibinfo {pages} {522} (\bibinfo {year}
  {1996})},\ \Eprint {http://arxiv.org/abs/astro-ph/9511126}
  {arXiv:astro-ph/9511126 [astro-ph]} \BibitemShut {NoStop}%
\bibitem [{\citenamefont {{Mather}}\ \emph {et~al.}(1994)\citenamefont
  {{Mather}}, \citenamefont {{Cheng}}, \citenamefont {{Cottingham}},
  \citenamefont {{Eplee}}, \citenamefont {{Fixsen}}, \citenamefont
  {{Hewagama}}, \citenamefont {{Isaacman}}, \citenamefont {{Jensen}},
  \citenamefont {{Meyer}}, \citenamefont {{Noerdlinger}}, \citenamefont
  {{Read}}, \citenamefont {{Rosen}}, \citenamefont {{Shafer}}, \citenamefont
  {{Wright}}, \citenamefont {{Bennett}}, \citenamefont {{Boggess}},
  \citenamefont {{Hauser}}, \citenamefont {{Kelsall}}, \citenamefont
  {{Moseley}}, \citenamefont {{Silverberg}}, \citenamefont {{Smoot}},
  \citenamefont {{Weiss}},\ and\ \citenamefont
  {{Wilkinson}}}]{COBE_FIRAS_Mather}%
  \BibitemOpen
  \bibfield  {author} {\bibinfo {author} {\bibfnamefont {J.~C.}\ \bibnamefont
  {{Mather}}}, \bibinfo {author} {\bibfnamefont {E.~S.}\ \bibnamefont
  {{Cheng}}}, \bibinfo {author} {\bibfnamefont {D.~A.}\ \bibnamefont
  {{Cottingham}}}, \bibinfo {author} {\bibfnamefont {Jr.}\ \bibnamefont
  {{Eplee}}, \bibfnamefont {R.~E.}}, \bibinfo {author} {\bibfnamefont {D.~J.}\
  \bibnamefont {{Fixsen}}}, \bibinfo {author} {\bibfnamefont {T.}~\bibnamefont
  {{Hewagama}}}, \bibinfo {author} {\bibfnamefont {R.~B.}\ \bibnamefont
  {{Isaacman}}}, \bibinfo {author} {\bibfnamefont {K.~A.}\ \bibnamefont
  {{Jensen}}}, \bibinfo {author} {\bibfnamefont {S.~S.}\ \bibnamefont
  {{Meyer}}}, \bibinfo {author} {\bibfnamefont {P.~D.}\ \bibnamefont
  {{Noerdlinger}}}, \bibinfo {author} {\bibfnamefont {S.~M.}\ \bibnamefont
  {{Read}}}, \bibinfo {author} {\bibfnamefont {L.~P.}\ \bibnamefont {{Rosen}}},
  \bibinfo {author} {\bibfnamefont {R.~A.}\ \bibnamefont {{Shafer}}}, \bibinfo
  {author} {\bibfnamefont {E.~L.}\ \bibnamefont {{Wright}}}, \bibinfo {author}
  {\bibfnamefont {C.~L.}\ \bibnamefont {{Bennett}}}, \bibinfo {author}
  {\bibfnamefont {N.~W.}\ \bibnamefont {{Boggess}}}, \bibinfo {author}
  {\bibfnamefont {M.~G.}\ \bibnamefont {{Hauser}}}, \bibinfo {author}
  {\bibfnamefont {T.}~\bibnamefont {{Kelsall}}}, \bibinfo {author}
  {\bibfnamefont {Jr.}\ \bibnamefont {{Moseley}}, \bibfnamefont {S.~H.}},
  \bibinfo {author} {\bibfnamefont {R.~F.}\ \bibnamefont {{Silverberg}}},
  \bibinfo {author} {\bibfnamefont {G.~F.}\ \bibnamefont {{Smoot}}}, \bibinfo
  {author} {\bibfnamefont {R.}~\bibnamefont {{Weiss}}}, \ and\ \bibinfo
  {author} {\bibfnamefont {D.~T.}\ \bibnamefont {{Wilkinson}}},\ }\bibfield
  {title} {\enquote {\bibinfo {title} {{Measurement of the Cosmic Microwave
  Background Spectrum by the COBE FIRAS Instrument}},}\ }\href {\doibase
  10.1086/173574} {\bibfield  {journal} {\bibinfo  {journal} {ApJ}\ }\textbf
  {\bibinfo {volume} {420}},\ \bibinfo {pages} {439} (\bibinfo {year}
  {1994})}\BibitemShut {NoStop}%
\bibitem [{\citenamefont {{Fixsen}}\ \emph {et~al.}(1996)\citenamefont
  {{Fixsen}}, \citenamefont {{Cheng}}, \citenamefont {{Gales}}, \citenamefont
  {{Mather}}, \citenamefont {{Shafer}},\ and\ \citenamefont
  {{Wright}}}]{COBE_FIRAS_Fixsen}%
  \BibitemOpen
  \bibfield  {author} {\bibinfo {author} {\bibfnamefont {D.~J.}\ \bibnamefont
  {{Fixsen}}}, \bibinfo {author} {\bibfnamefont {E.~S.}\ \bibnamefont
  {{Cheng}}}, \bibinfo {author} {\bibfnamefont {J.~M.}\ \bibnamefont
  {{Gales}}}, \bibinfo {author} {\bibfnamefont {J.~C.}\ \bibnamefont
  {{Mather}}}, \bibinfo {author} {\bibfnamefont {R.~A.}\ \bibnamefont
  {{Shafer}}}, \ and\ \bibinfo {author} {\bibfnamefont {E.~L.}\ \bibnamefont
  {{Wright}}},\ }\bibfield  {title} {\enquote {\bibinfo {title} {{The Cosmic
  Microwave Background Spectrum from the Full COBE FIRAS Data Set}},}\ }\href
  {\doibase 10.1086/178173} {\bibfield  {journal} {\bibinfo  {journal} {ApJ}\
  }\textbf {\bibinfo {volume} {473}},\ \bibinfo {pages} {576} (\bibinfo {year}
  {1996})},\ \Eprint {http://arxiv.org/abs/astro-ph/9605054}
  {arXiv:astro-ph/9605054 [astro-ph]} \BibitemShut {NoStop}%
\bibitem [{\citenamefont {{Fixsen}}\ \emph {et~al.}(2011)\citenamefont
  {{Fixsen}}, \citenamefont {{Kogut}}, \citenamefont {{Levin}}, \citenamefont
  {{Limon}}, \citenamefont {{Lubin}}, \citenamefont {{Mirel}}, \citenamefont
  {{Seiffert}}, \citenamefont {{Singal}}, \citenamefont {{Wollack}},
  \citenamefont {{Villela}},\ and\ \citenamefont
  {{Wuensche}}}]{2011ApJ...734....5F}%
  \BibitemOpen
  \bibfield  {author} {\bibinfo {author} {\bibfnamefont {D.~J.}\ \bibnamefont
  {{Fixsen}}}, \bibinfo {author} {\bibfnamefont {A.}~\bibnamefont {{Kogut}}},
  \bibinfo {author} {\bibfnamefont {S.}~\bibnamefont {{Levin}}}, \bibinfo
  {author} {\bibfnamefont {M.}~\bibnamefont {{Limon}}}, \bibinfo {author}
  {\bibfnamefont {P.}~\bibnamefont {{Lubin}}}, \bibinfo {author} {\bibfnamefont
  {P.}~\bibnamefont {{Mirel}}}, \bibinfo {author} {\bibfnamefont
  {M.}~\bibnamefont {{Seiffert}}}, \bibinfo {author} {\bibfnamefont
  {J.}~\bibnamefont {{Singal}}}, \bibinfo {author} {\bibfnamefont
  {E.}~\bibnamefont {{Wollack}}}, \bibinfo {author} {\bibfnamefont
  {T.}~\bibnamefont {{Villela}}}, \ and\ \bibinfo {author} {\bibfnamefont
  {C.~A.}\ \bibnamefont {{Wuensche}}},\ }\bibfield  {title} {\enquote {\bibinfo
  {title} {{ARCADE 2 Measurement of the Absolute Sky Brightness at 3-90
  GHz}},}\ }\href {\doibase 10.1088/0004-637X/734/1/5} {\bibfield  {journal}
  {\bibinfo  {journal} {ApJ}\ }\textbf {\bibinfo {volume} {734}},\ \bibinfo
  {eid} {5} (\bibinfo {year} {2011})},\ \Eprint
  {http://arxiv.org/abs/0901.0555} {arXiv:0901.0555 [astro-ph.CO]} \BibitemShut
  {NoStop}%
\bibitem [{\citenamefont {{Seiffert}}\ \emph {et~al.}(2011)\citenamefont
  {{Seiffert}}, \citenamefont {{Fixsen}}, \citenamefont {{Kogut}},
  \citenamefont {{Levin}}, \citenamefont {{Limon}}, \citenamefont {{Lubin}},
  \citenamefont {{Mirel}}, \citenamefont {{Singal}}, \citenamefont {{Villela}},
  \citenamefont {{Wollack}},\ and\ \citenamefont
  {{Wuensche}}}]{2011ApJ...734....6S}%
  \BibitemOpen
  \bibfield  {author} {\bibinfo {author} {\bibfnamefont {M.}~\bibnamefont
  {{Seiffert}}}, \bibinfo {author} {\bibfnamefont {D.~J.}\ \bibnamefont
  {{Fixsen}}}, \bibinfo {author} {\bibfnamefont {A.}~\bibnamefont {{Kogut}}},
  \bibinfo {author} {\bibfnamefont {S.~M.}\ \bibnamefont {{Levin}}}, \bibinfo
  {author} {\bibfnamefont {M.}~\bibnamefont {{Limon}}}, \bibinfo {author}
  {\bibfnamefont {P.~M.}\ \bibnamefont {{Lubin}}}, \bibinfo {author}
  {\bibfnamefont {P.}~\bibnamefont {{Mirel}}}, \bibinfo {author} {\bibfnamefont
  {J.}~\bibnamefont {{Singal}}}, \bibinfo {author} {\bibfnamefont
  {T.}~\bibnamefont {{Villela}}}, \bibinfo {author} {\bibfnamefont
  {E.}~\bibnamefont {{Wollack}}}, \ and\ \bibinfo {author} {\bibfnamefont
  {C.~A.}\ \bibnamefont {{Wuensche}}},\ }\bibfield  {title} {\enquote {\bibinfo
  {title} {{Interpretation of the ARCADE 2 Absolute Sky Brightness
  Measurement}},}\ }\href {\doibase 10.1088/0004-637X/734/1/6} {\bibfield
  {journal} {\bibinfo  {journal} {ApJ}\ }\textbf {\bibinfo {volume} {734}},\
  \bibinfo {eid} {6} (\bibinfo {year} {2011})}\BibitemShut {NoStop}%
\bibitem [{\citenamefont {{Kogut}}\ \emph {et~al.}(2011)\citenamefont
  {{Kogut}}, \citenamefont {{Fixsen}}, \citenamefont {{Chuss}}, \citenamefont
  {{Dotson}}, \citenamefont {{Dwek}}, \citenamefont {{Halpern}}, \citenamefont
  {{Hinshaw}}, \citenamefont {{Meyer}}, \citenamefont {{Moseley}},
  \citenamefont {{Seiffert}}, \citenamefont {{Spergel}},\ and\ \citenamefont
  {{Wollack}}}]{2011JCAP...07..025K}%
  \BibitemOpen
  \bibfield  {author} {\bibinfo {author} {\bibfnamefont {A.}~\bibnamefont
  {{Kogut}}}, \bibinfo {author} {\bibfnamefont {D.~J.}\ \bibnamefont
  {{Fixsen}}}, \bibinfo {author} {\bibfnamefont {D.~T.}\ \bibnamefont
  {{Chuss}}}, \bibinfo {author} {\bibfnamefont {J.}~\bibnamefont {{Dotson}}},
  \bibinfo {author} {\bibfnamefont {E.}~\bibnamefont {{Dwek}}}, \bibinfo
  {author} {\bibfnamefont {M.}~\bibnamefont {{Halpern}}}, \bibinfo {author}
  {\bibfnamefont {G.~F.}\ \bibnamefont {{Hinshaw}}}, \bibinfo {author}
  {\bibfnamefont {S.~M.}\ \bibnamefont {{Meyer}}}, \bibinfo {author}
  {\bibfnamefont {S.~H.}\ \bibnamefont {{Moseley}}}, \bibinfo {author}
  {\bibfnamefont {M.~D.}\ \bibnamefont {{Seiffert}}}, \bibinfo {author}
  {\bibfnamefont {D.~N.}\ \bibnamefont {{Spergel}}}, \ and\ \bibinfo {author}
  {\bibfnamefont {E.~J.}\ \bibnamefont {{Wollack}}},\ }\bibfield  {title}
  {\enquote {\bibinfo {title} {{The Primordial Inflation Explorer (PIXIE): a
  nulling polarimeter for cosmic microwave background observations}},}\ }\href
  {\doibase 10.1088/1475-7516/2011/07/025} {\bibfield  {journal} {\bibinfo
  {journal} {JCAP}\ }\textbf {\bibinfo {volume} {2011}},\ \bibinfo {eid} {025}
  (\bibinfo {year} {2011})},\ \Eprint {http://arxiv.org/abs/1105.2044}
  {arXiv:1105.2044 [astro-ph.CO]} \BibitemShut {NoStop}%
\bibitem [{\citenamefont {Chluba}\ \emph {et~al.}(2021)\citenamefont {Chluba}
  \emph {et~al.}}]{Chluba:2019nxa}%
  \BibitemOpen
  \bibfield  {author} {\bibinfo {author} {\bibfnamefont {J.}~\bibnamefont
  {Chluba}} \emph {et~al.},\ }\bibfield  {title} {\enquote {\bibinfo {title}
  {{New horizons in cosmology with spectral distortions of the cosmic microwave
  background}},}\ }\href {\doibase 10.1007/s10686-021-09729-5} {\bibfield
  {journal} {\bibinfo  {journal} {Exper. Astron.}\ }\textbf {\bibinfo {volume}
  {51}},\ \bibinfo {pages} {1515--1554} (\bibinfo {year} {2021})},\ \Eprint
  {http://arxiv.org/abs/1909.01593} {arXiv:1909.01593 [astro-ph.CO]}
  \BibitemShut {NoStop}%
\bibitem [{\citenamefont {Kogut}\ \emph {et~al.}(2019)\citenamefont {Kogut},
  \citenamefont {Abitbol}, \citenamefont {Chluba}, \citenamefont
  {Delabrouille}, \citenamefont {Fixsen}, \citenamefont {Hill}, \citenamefont
  {Patil},\ and\ \citenamefont {Rotti}}]{Kogut:2019vqh}%
  \BibitemOpen
  \bibfield  {author} {\bibinfo {author} {\bibfnamefont {A.}~\bibnamefont
  {Kogut}}, \bibinfo {author} {\bibfnamefont {M.~H.}\ \bibnamefont {Abitbol}},
  \bibinfo {author} {\bibfnamefont {J.}~\bibnamefont {Chluba}}, \bibinfo
  {author} {\bibfnamefont {J.}~\bibnamefont {Delabrouille}}, \bibinfo {author}
  {\bibfnamefont {D.}~\bibnamefont {Fixsen}}, \bibinfo {author} {\bibfnamefont
  {J.~C.}\ \bibnamefont {Hill}}, \bibinfo {author} {\bibfnamefont {S.~P.}\
  \bibnamefont {Patil}}, \ and\ \bibinfo {author} {\bibfnamefont
  {A.}~\bibnamefont {Rotti}},\ }\bibfield  {title} {\enquote {\bibinfo {title}
  {{CMB Spectral Distortions: Status and Prospects}},}\ }\href@noop {} {\
  (\bibinfo {year} {2019})},\ \Eprint {http://arxiv.org/abs/1907.13195}
  {arXiv:1907.13195 [astro-ph.CO]} \BibitemShut {NoStop}%
\bibitem [{\citenamefont {Maffei}\ \emph {et~al.}(2021)\citenamefont {Maffei}
  \emph {et~al.}}]{Maffei:2021xur}%
  \BibitemOpen
  \bibfield  {author} {\bibinfo {author} {\bibfnamefont {B.}~\bibnamefont
  {Maffei}} \emph {et~al.},\ }\bibfield  {title} {\enquote {\bibinfo {title}
  {{BISOU: A balloon project to measure the CMB spectral distortions}},}\ }in\
  \href {\doibase 10.1142/9789811269776_0129} {\emph {\bibinfo {booktitle}
  {{16th Marcel Grossmann Meeting on~Recent Developments in Theoretical and
  Experimental General Relativity, Astrophysics and Relativistic Field
  Theories}}}}\ (\bibinfo {year} {2021})\ \Eprint
  {http://arxiv.org/abs/2111.00246} {arXiv:2111.00246 [astro-ph.IM]}
  \BibitemShut {NoStop}%
\bibitem [{\citenamefont {Chang}\ \emph {et~al.}(2022)\citenamefont {Chang}
  \emph {et~al.}}]{Chang:2022tzj}%
  \BibitemOpen
  \bibfield  {author} {\bibinfo {author} {\bibfnamefont {Clarence~L.}\
  \bibnamefont {Chang}} \emph {et~al.},\ }\bibfield  {title} {\enquote
  {\bibinfo {title} {{Snowmass2021 Cosmic Frontier: Cosmic Microwave Background
  Measurements White Paper}},}\ }\href@noop {} {\  (\bibinfo {year} {2022})},\
  \Eprint {http://arxiv.org/abs/2203.07638} {arXiv:2203.07638 [astro-ph.CO]}
  \BibitemShut {NoStop}%
\bibitem [{\citenamefont {Chluba}\ \emph {et~al.}(2019)\citenamefont {Chluba}
  \emph {et~al.}}]{Chluba:2019kpb}%
  \BibitemOpen
  \bibfield  {author} {\bibinfo {author} {\bibfnamefont {J.}~\bibnamefont
  {Chluba}} \emph {et~al.},\ }\bibfield  {title} {\enquote {\bibinfo {title}
  {{Spectral Distortions of the CMB as a Probe of Inflation, Recombination,
  Structure Formation and Particle Physics}: {Astro2020 Science White
  Paper}},}\ }\href@noop {} {\bibfield  {journal} {\bibinfo  {journal} {Bull.
  Am. Astron. Soc.}\ }\textbf {\bibinfo {volume} {51}},\ \bibinfo {pages} {184}
  (\bibinfo {year} {2019})},\ \Eprint {http://arxiv.org/abs/1903.04218}
  {arXiv:1903.04218 [astro-ph.CO]} \BibitemShut {NoStop}%
\bibitem [{\citenamefont {{Illarionov}}\ and\ \citenamefont
  {{Siuniaev}}(1975)}]{Illarionov_Sunyaev}%
  \BibitemOpen
  \bibfield  {author} {\bibinfo {author} {\bibfnamefont {A.~F.}\ \bibnamefont
  {{Illarionov}}}\ and\ \bibinfo {author} {\bibfnamefont {R.~A.}\ \bibnamefont
  {{Siuniaev}}},\ }\bibfield  {title} {\enquote {\bibinfo {title}
  {{Comptonization, characteristic radiation spectra, and thermal balance of
  low-density plasma}},}\ }\href@noop {} {\bibfield  {journal} {\bibinfo
  {journal} {Soviet Astronomy}\ }\textbf {\bibinfo {volume} {18}},\ \bibinfo
  {pages} {413--419} (\bibinfo {year} {1975})}\BibitemShut {NoStop}%
\bibitem [{\citenamefont {{Zeldovich}}\ and\ \citenamefont
  {{Sunyaev}}(1969)}]{Zeldovich_Sunyaev}%
  \BibitemOpen
  \bibfield  {author} {\bibinfo {author} {\bibfnamefont {Ya.~B.}\ \bibnamefont
  {{Zeldovich}}}\ and\ \bibinfo {author} {\bibfnamefont {R.~A.}\ \bibnamefont
  {{Sunyaev}}},\ }\bibfield  {title} {\enquote {\bibinfo {title} {{The
  Interaction of Matter and Radiation in a Hot-Model Universe}},}\ }\href
  {\doibase 10.1007/BF00661821} {\bibfield  {journal} {\bibinfo  {journal}
  {Astrophysics and Space Science}\ }\textbf {\bibinfo {volume} {4}},\ \bibinfo
  {pages} {301--316} (\bibinfo {year} {1969})}\BibitemShut {NoStop}%
\bibitem [{\citenamefont {Chluba}(2020)}]{Chluba:2018cww}%
  \BibitemOpen
  \bibfield  {author} {\bibinfo {author} {\bibfnamefont {Jens}\ \bibnamefont
  {Chluba}},\ }\bibfield  {title} {\enquote {\bibinfo {title} {{Future Steps in
  Cosmology using Spectral Distortions of the Cosmic Microwave Background}},}\
  }\href {\doibase 10.3254/ENFI200012} {\bibfield  {journal} {\bibinfo
  {journal} {Proc. Int. Sch. Phys. Fermi}\ }\textbf {\bibinfo {volume} {200}},\
  \bibinfo {pages} {265--309} (\bibinfo {year} {2020})},\ \Eprint
  {http://arxiv.org/abs/1806.02915} {arXiv:1806.02915 [astro-ph.CO]}
  \BibitemShut {NoStop}%
\bibitem [{\citenamefont {Chluba}(2016)}]{Chluba:2016bvg}%
  \BibitemOpen
  \bibfield  {author} {\bibinfo {author} {\bibfnamefont {Jens}\ \bibnamefont
  {Chluba}},\ }\bibfield  {title} {\enquote {\bibinfo {title} {{Which spectral
  distortions does $\Lambda$CDM actually predict?}}}\ }\href {\doibase
  10.1093/mnras/stw945} {\bibfield  {journal} {\bibinfo  {journal} {Mon. Not.
  Roy. Astron. Soc.}\ }\textbf {\bibinfo {volume} {460}},\ \bibinfo {pages}
  {227--239} (\bibinfo {year} {2016})},\ \Eprint
  {http://arxiv.org/abs/1603.02496} {arXiv:1603.02496 [astro-ph.CO]}
  \BibitemShut {NoStop}%
\bibitem [{\citenamefont {{Rybicki}}\ and\ \citenamefont
  {{dell'Antonio}}(1993)}]{1993ASPC...51..548R}%
  \BibitemOpen
  \bibfield  {author} {\bibinfo {author} {\bibfnamefont {G.~B.}\ \bibnamefont
  {{Rybicki}}}\ and\ \bibinfo {author} {\bibfnamefont {I.~P.}\ \bibnamefont
  {{dell'Antonio}}},\ }\bibfield  {title} {\enquote {\bibinfo {title}
  {{Spectral Distortions in the CMB from Recombination.}}}\ }in\ \href@noop {}
  {\emph {\bibinfo {booktitle} {Observational Cosmology}}},\ \bibinfo {series}
  {Astronomical Society of the Pacific Conference Series}, Vol.~\bibinfo
  {volume} {51},\ \bibinfo {editor} {edited by\ \bibinfo {editor}
  {\bibfnamefont {Guido~L.}\ \bibnamefont {{Chincarini}}}, \bibinfo {editor}
  {\bibfnamefont {Angela}\ \bibnamefont {{Iovino}}}, \bibinfo {editor}
  {\bibfnamefont {Tommaso}\ \bibnamefont {{Maccacaro}}}, \ and\ \bibinfo
  {editor} {\bibfnamefont {Dario}\ \bibnamefont {{Maccagni}}}}\ (\bibinfo
  {year} {1993})\ p.\ \bibinfo {pages} {548}\BibitemShut {NoStop}%
\bibitem [{\citenamefont {Rubino-Martin}\ \emph {et~al.}(2006)\citenamefont
  {Rubino-Martin}, \citenamefont {Chluba},\ and\ \citenamefont
  {Sunyaev}}]{Jens2006}%
  \BibitemOpen
  \bibfield  {author} {\bibinfo {author} {\bibfnamefont {J.~A.}\ \bibnamefont
  {Rubino-Martin}}, \bibinfo {author} {\bibfnamefont {J.}~\bibnamefont
  {Chluba}}, \ and\ \bibinfo {author} {\bibfnamefont {R.~A.}\ \bibnamefont
  {Sunyaev}},\ }\bibfield  {title} {\enquote {\bibinfo {title} {{Lines in the
  Cosmic Microwave Background Spectrum from the Epoch of Cosmological Hydrogen
  Recombination}},}\ }\href {\doibase 10.1111/j.1365-2966.2006.10839.x}
  {\bibfield  {journal} {\bibinfo  {journal} {Mon. Not. Roy. Astron. Soc.}\
  }\textbf {\bibinfo {volume} {371}},\ \bibinfo {pages} {1939--1952} (\bibinfo
  {year} {2006})},\ \Eprint {http://arxiv.org/abs/astro-ph/0607373}
  {arXiv:astro-ph/0607373} \BibitemShut {NoStop}%
\bibitem [{\citenamefont {Chluba}\ \emph {et~al.}(2007)\citenamefont {Chluba},
  \citenamefont {Rubino-Martin},\ and\ \citenamefont
  {Sunyaev}}]{Chluba:2006bc}%
  \BibitemOpen
  \bibfield  {author} {\bibinfo {author} {\bibfnamefont {Jens}\ \bibnamefont
  {Chluba}}, \bibinfo {author} {\bibfnamefont {J.~A.}\ \bibnamefont
  {Rubino-Martin}}, \ and\ \bibinfo {author} {\bibfnamefont {R.~A.}\
  \bibnamefont {Sunyaev}},\ }\bibfield  {title} {\enquote {\bibinfo {title}
  {{Cosmological hydrogen recombination: Populations of the high level
  sub-states}},}\ }\href {\doibase 10.1111/j.1365-2966.2006.11239.x} {\bibfield
   {journal} {\bibinfo  {journal} {Mon. Not. Roy. Astron. Soc.}\ }\textbf
  {\bibinfo {volume} {374}},\ \bibinfo {pages} {1310--1320} (\bibinfo {year}
  {2007})},\ \Eprint {http://arxiv.org/abs/astro-ph/0608242}
  {arXiv:astro-ph/0608242} \BibitemShut {NoStop}%
\bibitem [{\citenamefont {Ali-Ha\"\i{}moud}\ \emph {et~al.}(2015)\citenamefont
  {Ali-Ha\"\i{}moud}, \citenamefont {Chluba},\ and\ \citenamefont
  {Kamionkowski}}]{Ali-Haimoud:2015pwa}%
  \BibitemOpen
  \bibfield  {author} {\bibinfo {author} {\bibfnamefont {Yacine}\ \bibnamefont
  {Ali-Ha\"\i{}moud}}, \bibinfo {author} {\bibfnamefont {Jens}\ \bibnamefont
  {Chluba}}, \ and\ \bibinfo {author} {\bibfnamefont {Marc}\ \bibnamefont
  {Kamionkowski}},\ }\bibfield  {title} {\enquote {\bibinfo {title}
  {{Constraints on Dark Matter Interactions with Standard Model Particles from
  Cosmic Microwave Background Spectral Distortions}},}\ }\href {\doibase
  10.1103/PhysRevLett.115.071304} {\bibfield  {journal} {\bibinfo  {journal}
  {Phys. Rev. Lett.}\ }\textbf {\bibinfo {volume} {115}},\ \bibinfo {pages}
  {071304} (\bibinfo {year} {2015})},\ \Eprint
  {http://arxiv.org/abs/1506.04745} {arXiv:1506.04745 [astro-ph.CO]}
  \BibitemShut {NoStop}%
\bibitem [{\citenamefont {Bolliet}\ \emph {et~al.}(2021)\citenamefont
  {Bolliet}, \citenamefont {Chluba},\ and\ \citenamefont
  {Battye}}]{Bolliet:2020ofj}%
  \BibitemOpen
  \bibfield  {author} {\bibinfo {author} {\bibfnamefont {Boris}\ \bibnamefont
  {Bolliet}}, \bibinfo {author} {\bibfnamefont {Jens}\ \bibnamefont {Chluba}},
  \ and\ \bibinfo {author} {\bibfnamefont {Richard}\ \bibnamefont {Battye}},\
  }\bibfield  {title} {\enquote {\bibinfo {title} {{Spectral distortion
  constraints on photon injection from low-mass decaying particles}},}\ }\href
  {\doibase 10.1093/mnras/stab1997} {\bibfield  {journal} {\bibinfo  {journal}
  {Mon. Not. Roy. Astron. Soc.}\ }\textbf {\bibinfo {volume} {507}},\ \bibinfo
  {pages} {3148--3178} (\bibinfo {year} {2021})},\ \Eprint
  {http://arxiv.org/abs/2012.07292} {arXiv:2012.07292 [astro-ph.CO]}
  \BibitemShut {NoStop}%
\bibitem [{\citenamefont {Chluba}(2013)}]{Chluba:2013vsa}%
  \BibitemOpen
  \bibfield  {author} {\bibinfo {author} {\bibfnamefont {Jens}\ \bibnamefont
  {Chluba}},\ }\bibfield  {title} {\enquote {\bibinfo {title} {{Green's
  function of the cosmological thermalization problem}},}\ }\href {\doibase
  10.1093/mnras/stt1025} {\bibfield  {journal} {\bibinfo  {journal} {Mon. Not.
  Roy. Astron. Soc.}\ }\textbf {\bibinfo {volume} {434}},\ \bibinfo {pages}
  {352} (\bibinfo {year} {2013})},\ \Eprint {http://arxiv.org/abs/1304.6120}
  {arXiv:1304.6120 [astro-ph.CO]} \BibitemShut {NoStop}%
\bibitem [{\citenamefont {Chluba}(2015)}]{Chluba:2015hma}%
  \BibitemOpen
  \bibfield  {author} {\bibinfo {author} {\bibfnamefont {Jens}\ \bibnamefont
  {Chluba}},\ }\bibfield  {title} {\enquote {\bibinfo {title} {{Green's
  function of the cosmological thermalization problem \textendash{} II. Effect
  of photon injection and constraints}},}\ }\href {\doibase
  10.1093/mnras/stv2243} {\bibfield  {journal} {\bibinfo  {journal} {Mon. Not.
  Roy. Astron. Soc.}\ }\textbf {\bibinfo {volume} {454}},\ \bibinfo {pages}
  {4182--4196} (\bibinfo {year} {2015})},\ \Eprint
  {http://arxiv.org/abs/1506.06582} {arXiv:1506.06582 [astro-ph.CO]}
  \BibitemShut {NoStop}%
\bibitem [{\citenamefont {Acharya}\ and\ \citenamefont
  {Khatri}(2019{\natexlab{a}})}]{Acharya:2018iwh}%
  \BibitemOpen
  \bibfield  {author} {\bibinfo {author} {\bibfnamefont {Sandeep~Kumar}\
  \bibnamefont {Acharya}}\ and\ \bibinfo {author} {\bibfnamefont {Rishi}\
  \bibnamefont {Khatri}},\ }\bibfield  {title} {\enquote {\bibinfo {title}
  {{Rich structure of non-thermal relativistic CMB spectral distortions from
  high energy particle cascades at redshifts $z\lesssim 2\times 10^5$}},}\
  }\href {\doibase 10.1103/PhysRevD.99.043520} {\bibfield  {journal} {\bibinfo
  {journal} {Phys. Rev. D}\ }\textbf {\bibinfo {volume} {99}},\ \bibinfo
  {pages} {043520} (\bibinfo {year} {2019}{\natexlab{a}})},\ \Eprint
  {http://arxiv.org/abs/1808.02897} {arXiv:1808.02897 [astro-ph.CO]}
  \BibitemShut {NoStop}%
\bibitem [{\citenamefont {Adams}\ \emph {et~al.}(1998)\citenamefont {Adams},
  \citenamefont {Sarkar},\ and\ \citenamefont {Sciama}}]{Adams:1998nr}%
  \BibitemOpen
  \bibfield  {author} {\bibinfo {author} {\bibfnamefont {Jennifer~A.}\
  \bibnamefont {Adams}}, \bibinfo {author} {\bibfnamefont {Subir}\ \bibnamefont
  {Sarkar}}, \ and\ \bibinfo {author} {\bibfnamefont {D.W.}\ \bibnamefont
  {Sciama}},\ }\bibfield  {title} {\enquote {\bibinfo {title} {{CMB anisotropy
  in the decaying neutrino cosmology}},}\ }\href {\doibase
  10.1046/j.1365-8711.1998.02017.x} {\bibfield  {journal} {\bibinfo  {journal}
  {Mon. Not. Roy. Astron. Soc.}\ }\textbf {\bibinfo {volume} {301}},\ \bibinfo
  {pages} {210--214} (\bibinfo {year} {1998})},\ \Eprint
  {http://arxiv.org/abs/astro-ph/9805108} {arXiv:astro-ph/9805108} \BibitemShut
  {NoStop}%
\bibitem [{\citenamefont {Chen}\ and\ \citenamefont
  {Kamionkowski}(2004)}]{Chen:2003gz}%
  \BibitemOpen
  \bibfield  {author} {\bibinfo {author} {\bibfnamefont {Xue-Lei}\ \bibnamefont
  {Chen}}\ and\ \bibinfo {author} {\bibfnamefont {Marc}\ \bibnamefont
  {Kamionkowski}},\ }\bibfield  {title} {\enquote {\bibinfo {title} {{Particle
  decays during the cosmic dark ages}},}\ }\href {\doibase
  10.1103/PhysRevD.70.043502} {\bibfield  {journal} {\bibinfo  {journal} {Phys.
  Rev. D}\ }\textbf {\bibinfo {volume} {70}},\ \bibinfo {pages} {043502}
  (\bibinfo {year} {2004})},\ \Eprint {http://arxiv.org/abs/astro-ph/0310473}
  {arXiv:astro-ph/0310473} \BibitemShut {NoStop}%
\bibitem [{\citenamefont {Padmanabhan}\ and\ \citenamefont
  {Finkbeiner}(2005)}]{Padmanabhan:2005es}%
  \BibitemOpen
  \bibfield  {author} {\bibinfo {author} {\bibfnamefont {Nikhil}\ \bibnamefont
  {Padmanabhan}}\ and\ \bibinfo {author} {\bibfnamefont {Douglas~P.}\
  \bibnamefont {Finkbeiner}},\ }\bibfield  {title} {\enquote {\bibinfo {title}
  {{Detecting dark matter annihilation with CMB polarization: Signatures and
  experimental prospects}},}\ }\href {\doibase 10.1103/PhysRevD.72.023508}
  {\bibfield  {journal} {\bibinfo  {journal} {Phys. Rev. D}\ }\textbf {\bibinfo
  {volume} {72}},\ \bibinfo {pages} {023508} (\bibinfo {year} {2005})},\
  \Eprint {http://arxiv.org/abs/astro-ph/0503486} {arXiv:astro-ph/0503486}
  \BibitemShut {NoStop}%
\bibitem [{\citenamefont {Zhang}\ \emph {et~al.}(2007)\citenamefont {Zhang},
  \citenamefont {Chen}, \citenamefont {Kamionkowski}, \citenamefont {Si},\ and\
  \citenamefont {Zheng}}]{Zhang:2007zzh}%
  \BibitemOpen
  \bibfield  {author} {\bibinfo {author} {\bibfnamefont {Le}~\bibnamefont
  {Zhang}}, \bibinfo {author} {\bibfnamefont {Xuelei}\ \bibnamefont {Chen}},
  \bibinfo {author} {\bibfnamefont {Marc}\ \bibnamefont {Kamionkowski}},
  \bibinfo {author} {\bibfnamefont {Zong-guo}\ \bibnamefont {Si}}, \ and\
  \bibinfo {author} {\bibfnamefont {Zheng}\ \bibnamefont {Zheng}},\ }\bibfield
  {title} {\enquote {\bibinfo {title} {{Constraints on radiative dark-matter
  decay from the cosmic microwave background}},}\ }\href {\doibase
  10.1103/PhysRevD.76.061301} {\bibfield  {journal} {\bibinfo  {journal} {Phys.
  Rev. D}\ }\textbf {\bibinfo {volume} {76}},\ \bibinfo {pages} {061301}
  (\bibinfo {year} {2007})},\ \Eprint {http://arxiv.org/abs/0704.2444}
  {arXiv:0704.2444 [astro-ph]} \BibitemShut {NoStop}%
\bibitem [{\citenamefont {Acharya}\ and\ \citenamefont
  {Khatri}(2019{\natexlab{b}})}]{Acharya:2019uba}%
  \BibitemOpen
  \bibfield  {author} {\bibinfo {author} {\bibfnamefont {Sandeep~Kumar}\
  \bibnamefont {Acharya}}\ and\ \bibinfo {author} {\bibfnamefont {Rishi}\
  \bibnamefont {Khatri}},\ }\bibfield  {title} {\enquote {\bibinfo {title}
  {{CMB anisotropy and BBN constraints on pre-recombination decay of dark
  matter to visible particles}},}\ }\href {\doibase
  10.1088/1475-7516/2019/12/046} {\bibfield  {journal} {\bibinfo  {journal}
  {JCAP}\ }\textbf {\bibinfo {volume} {12}},\ \bibinfo {pages} {046} (\bibinfo
  {year} {2019}{\natexlab{b}})},\ \Eprint {http://arxiv.org/abs/1910.06272}
  {arXiv:1910.06272 [astro-ph.CO]} \BibitemShut {NoStop}%
\bibitem [{\citenamefont {Cang}\ \emph {et~al.}(2020)\citenamefont {Cang},
  \citenamefont {Gao},\ and\ \citenamefont {Ma}}]{Cang:2020exa}%
  \BibitemOpen
  \bibfield  {author} {\bibinfo {author} {\bibfnamefont {Junsong}\ \bibnamefont
  {Cang}}, \bibinfo {author} {\bibfnamefont {Yu}~\bibnamefont {Gao}}, \ and\
  \bibinfo {author} {\bibfnamefont {Yin-Zhe}\ \bibnamefont {Ma}},\ }\bibfield
  {title} {\enquote {\bibinfo {title} {{Probing dark matter with future CMB
  measurements}},}\ }\href {\doibase 10.1103/PhysRevD.102.103005} {\bibfield
  {journal} {\bibinfo  {journal} {Phys. Rev. D}\ }\textbf {\bibinfo {volume}
  {102}},\ \bibinfo {pages} {103005} (\bibinfo {year} {2020})},\ \Eprint
  {http://arxiv.org/abs/2002.03380} {arXiv:2002.03380 [astro-ph.CO]}
  \BibitemShut {NoStop}%
\bibitem [{\citenamefont {Galli}\ \emph {et~al.}(2009)\citenamefont {Galli},
  \citenamefont {Iocco}, \citenamefont {Bertone},\ and\ \citenamefont
  {Melchiorri}}]{Galli:2009zc}%
  \BibitemOpen
  \bibfield  {author} {\bibinfo {author} {\bibfnamefont {Silvia}\ \bibnamefont
  {Galli}}, \bibinfo {author} {\bibfnamefont {Fabio}\ \bibnamefont {Iocco}},
  \bibinfo {author} {\bibfnamefont {Gianfranco}\ \bibnamefont {Bertone}}, \
  and\ \bibinfo {author} {\bibfnamefont {Alessandro}\ \bibnamefont
  {Melchiorri}},\ }\bibfield  {title} {\enquote {\bibinfo {title} {{CMB
  constraints on Dark Matter models with large annihilation cross-section}},}\
  }\href {\doibase 10.1103/PhysRevD.80.023505} {\bibfield  {journal} {\bibinfo
  {journal} {Phys. Rev. D}\ }\textbf {\bibinfo {volume} {80}},\ \bibinfo
  {pages} {023505} (\bibinfo {year} {2009})},\ \Eprint
  {http://arxiv.org/abs/0905.0003} {arXiv:0905.0003 [astro-ph.CO]} \BibitemShut
  {NoStop}%
\bibitem [{\citenamefont {Hisano}\ \emph {et~al.}(2011)\citenamefont {Hisano},
  \citenamefont {Kawasaki}, \citenamefont {Kohri}, \citenamefont {Moroi},
  \citenamefont {Nakayama},\ and\ \citenamefont {Sekiguchi}}]{Hisano:2011dc}%
  \BibitemOpen
  \bibfield  {author} {\bibinfo {author} {\bibfnamefont {Junji}\ \bibnamefont
  {Hisano}}, \bibinfo {author} {\bibfnamefont {Masahiro}\ \bibnamefont
  {Kawasaki}}, \bibinfo {author} {\bibfnamefont {Kazunori}\ \bibnamefont
  {Kohri}}, \bibinfo {author} {\bibfnamefont {Takeo}\ \bibnamefont {Moroi}},
  \bibinfo {author} {\bibfnamefont {Kazunori}\ \bibnamefont {Nakayama}}, \ and\
  \bibinfo {author} {\bibfnamefont {Toyokazu}\ \bibnamefont {Sekiguchi}},\
  }\bibfield  {title} {\enquote {\bibinfo {title} {{Cosmological constraints on
  dark matter models with velocity-dependent annihilation cross section}},}\
  }\href {\doibase 10.1103/PhysRevD.83.123511} {\bibfield  {journal} {\bibinfo
  {journal} {Phys. Rev. D}\ }\textbf {\bibinfo {volume} {83}},\ \bibinfo
  {pages} {123511} (\bibinfo {year} {2011})},\ \Eprint
  {http://arxiv.org/abs/1102.4658} {arXiv:1102.4658 [hep-ph]} \BibitemShut
  {NoStop}%
\bibitem [{\citenamefont {Hutsi}\ \emph {et~al.}(2011)\citenamefont {Hutsi},
  \citenamefont {Chluba}, \citenamefont {Hektor},\ and\ \citenamefont
  {Raidal}}]{Hutsi:2011vx}%
  \BibitemOpen
  \bibfield  {author} {\bibinfo {author} {\bibfnamefont {Gert}\ \bibnamefont
  {Hutsi}}, \bibinfo {author} {\bibfnamefont {Jens}\ \bibnamefont {Chluba}},
  \bibinfo {author} {\bibfnamefont {Andi}\ \bibnamefont {Hektor}}, \ and\
  \bibinfo {author} {\bibfnamefont {Martti}\ \bibnamefont {Raidal}},\
  }\bibfield  {title} {\enquote {\bibinfo {title} {{WMAP7 and future CMB
  constraints on annihilating dark matter: implications on GeV-scale WIMPs}},}\
  }\href {\doibase 10.1051/0004-6361/201116914} {\bibfield  {journal} {\bibinfo
   {journal} {Astron. Astrophys.}\ }\textbf {\bibinfo {volume} {535}},\
  \bibinfo {pages} {A26} (\bibinfo {year} {2011})},\ \Eprint
  {http://arxiv.org/abs/1103.2766} {arXiv:1103.2766 [astro-ph.CO]} \BibitemShut
  {NoStop}%
\bibitem [{\citenamefont {Galli}\ \emph {et~al.}(2011)\citenamefont {Galli},
  \citenamefont {Iocco}, \citenamefont {Bertone},\ and\ \citenamefont
  {Melchiorri}}]{Galli:2011rz}%
  \BibitemOpen
  \bibfield  {author} {\bibinfo {author} {\bibfnamefont {Silvia}\ \bibnamefont
  {Galli}}, \bibinfo {author} {\bibfnamefont {Fabio}\ \bibnamefont {Iocco}},
  \bibinfo {author} {\bibfnamefont {Gianfranco}\ \bibnamefont {Bertone}}, \
  and\ \bibinfo {author} {\bibfnamefont {Alessandro}\ \bibnamefont
  {Melchiorri}},\ }\bibfield  {title} {\enquote {\bibinfo {title} {{Updated CMB
  constraints on Dark Matter annihilation cross-sections}},}\ }\href {\doibase
  10.1103/PhysRevD.84.027302} {\bibfield  {journal} {\bibinfo  {journal} {Phys.
  Rev. D}\ }\textbf {\bibinfo {volume} {84}},\ \bibinfo {pages} {027302}
  (\bibinfo {year} {2011})},\ \Eprint {http://arxiv.org/abs/1106.1528}
  {arXiv:1106.1528 [astro-ph.CO]} \BibitemShut {NoStop}%
\bibitem [{\citenamefont {{Finkbeiner}}\ \emph {et~al.}(2012)\citenamefont
  {{Finkbeiner}}, \citenamefont {{Galli}}, \citenamefont {{Lin}},\ and\
  \citenamefont {{Slatyer}}}]{2012PhRvD..85d3522F}%
  \BibitemOpen
  \bibfield  {author} {\bibinfo {author} {\bibfnamefont {Douglas~P.}\
  \bibnamefont {{Finkbeiner}}}, \bibinfo {author} {\bibfnamefont {Silvia}\
  \bibnamefont {{Galli}}}, \bibinfo {author} {\bibfnamefont {Tongyan}\
  \bibnamefont {{Lin}}}, \ and\ \bibinfo {author} {\bibfnamefont {Tracy~R.}\
  \bibnamefont {{Slatyer}}},\ }\bibfield  {title} {\enquote {\bibinfo {title}
  {{Searching for dark matter in the CMB: A compact parametrization of energy
  injection from new physics}},}\ }\href {\doibase 10.1103/PhysRevD.85.043522}
  {\bibfield  {journal} {\bibinfo  {journal} {\prd}\ }\textbf {\bibinfo
  {volume} {85}},\ \bibinfo {eid} {043522} (\bibinfo {year} {2012})},\ \Eprint
  {http://arxiv.org/abs/1109.6322} {arXiv:1109.6322 [astro-ph.CO]} \BibitemShut
  {NoStop}%
\bibitem [{\citenamefont {Slatyer}(2013)}]{Slatyer:2012yq}%
  \BibitemOpen
  \bibfield  {author} {\bibinfo {author} {\bibfnamefont {Tracy~R.}\
  \bibnamefont {Slatyer}},\ }\bibfield  {title} {\enquote {\bibinfo {title}
  {{Energy Injection And Absorption In The Cosmic Dark Ages}},}\ }\href
  {\doibase 10.1103/PhysRevD.87.123513} {\bibfield  {journal} {\bibinfo
  {journal} {Phys. Rev. D}\ }\textbf {\bibinfo {volume} {87}},\ \bibinfo
  {pages} {123513} (\bibinfo {year} {2013})},\ \Eprint
  {http://arxiv.org/abs/1211.0283} {arXiv:1211.0283 [astro-ph.CO]} \BibitemShut
  {NoStop}%
\bibitem [{\citenamefont {Galli}\ \emph {et~al.}(2013)\citenamefont {Galli},
  \citenamefont {Slatyer}, \citenamefont {Valdes},\ and\ \citenamefont
  {Iocco}}]{Galli:2013dna}%
  \BibitemOpen
  \bibfield  {author} {\bibinfo {author} {\bibfnamefont {Silvia}\ \bibnamefont
  {Galli}}, \bibinfo {author} {\bibfnamefont {Tracy~R.}\ \bibnamefont
  {Slatyer}}, \bibinfo {author} {\bibfnamefont {Marcos}\ \bibnamefont
  {Valdes}}, \ and\ \bibinfo {author} {\bibfnamefont {Fabio}\ \bibnamefont
  {Iocco}},\ }\bibfield  {title} {\enquote {\bibinfo {title} {{Systematic
  Uncertainties In Constraining Dark Matter Annihilation From The Cosmic
  Microwave Background}},}\ }\href {\doibase 10.1103/PhysRevD.88.063502}
  {\bibfield  {journal} {\bibinfo  {journal} {Phys. Rev. D}\ }\textbf {\bibinfo
  {volume} {88}},\ \bibinfo {pages} {063502} (\bibinfo {year} {2013})},\
  \Eprint {http://arxiv.org/abs/1306.0563} {arXiv:1306.0563 [astro-ph.CO]}
  \BibitemShut {NoStop}%
\bibitem [{\citenamefont {Madhavacheril}\ \emph {et~al.}(2014)\citenamefont
  {Madhavacheril}, \citenamefont {Sehgal},\ and\ \citenamefont
  {Slatyer}}]{Madhavacheril:2013cna}%
  \BibitemOpen
  \bibfield  {author} {\bibinfo {author} {\bibfnamefont {Mathew~S.}\
  \bibnamefont {Madhavacheril}}, \bibinfo {author} {\bibfnamefont {Neelima}\
  \bibnamefont {Sehgal}}, \ and\ \bibinfo {author} {\bibfnamefont {Tracy~R.}\
  \bibnamefont {Slatyer}},\ }\bibfield  {title} {\enquote {\bibinfo {title}
  {{Current Dark Matter Annihilation Constraints from CMB and Low-Redshift
  Data}},}\ }\href {\doibase 10.1103/PhysRevD.89.103508} {\bibfield  {journal}
  {\bibinfo  {journal} {Phys. Rev. D}\ }\textbf {\bibinfo {volume} {89}},\
  \bibinfo {pages} {103508} (\bibinfo {year} {2014})},\ \Eprint
  {http://arxiv.org/abs/1310.3815} {arXiv:1310.3815 [astro-ph.CO]} \BibitemShut
  {NoStop}%
\bibitem [{\citenamefont {Slatyer}(2016{\natexlab{b}})}]{1506.03812}%
  \BibitemOpen
  \bibfield  {author} {\bibinfo {author} {\bibfnamefont {Tracy~R.}\
  \bibnamefont {Slatyer}},\ }\bibfield  {title} {\enquote {\bibinfo {title}
  {{Indirect Dark Matter Signatures in the Cosmic Dark Ages II. Ionization,
  Heating and Photon Production from Arbitrary Energy Injections}},}\ }\href
  {\doibase 10.1103/PhysRevD.93.023521} {\bibfield  {journal} {\bibinfo
  {journal} {Phys. Rev.}\ }\textbf {\bibinfo {volume} {D93}},\ \bibinfo {pages}
  {023521} (\bibinfo {year} {2016}{\natexlab{b}})},\ \Eprint
  {http://arxiv.org/abs/1506.03812} {arXiv:1506.03812 [astro-ph.CO]}
  \BibitemShut {NoStop}%
\bibitem [{\citenamefont {Liu}\ \emph {et~al.}(2023)\citenamefont {Liu},
  \citenamefont {Qin}, \citenamefont {Ridgway},\ and\ \citenamefont
  {Slatyer}}]{paperI}%
  \BibitemOpen
  \bibfield  {author} {\bibinfo {author} {\bibfnamefont {Hongwan}\ \bibnamefont
  {Liu}}, \bibinfo {author} {\bibfnamefont {Wenzer}\ \bibnamefont {Qin}},
  \bibinfo {author} {\bibfnamefont {Gregory~W.}\ \bibnamefont {Ridgway}}, \
  and\ \bibinfo {author} {\bibfnamefont {Tracy~R.}\ \bibnamefont {Slatyer}},\
  }\bibfield  {title} {\enquote {\bibinfo {title} {{Exotic energy injection in
  the early universe I: a novel treatment for low-energy electrons and
  photons}},}\ }\href@noop {} {\  (\bibinfo {year} {2023})},\ \Eprint
  {http://arxiv.org/abs/2303.07366} {arXiv:2303.07366 [astro-ph.CO]}
  \BibitemShut {NoStop}%
\bibitem [{\citenamefont {Liu}\ \emph {et~al.}(2020)\citenamefont {Liu},
  \citenamefont {Ridgway},\ and\ \citenamefont {Slatyer}}]{DH}%
  \BibitemOpen
  \bibfield  {author} {\bibinfo {author} {\bibfnamefont {Hongwan}\ \bibnamefont
  {Liu}}, \bibinfo {author} {\bibfnamefont {Gregory~W.}\ \bibnamefont
  {Ridgway}}, \ and\ \bibinfo {author} {\bibfnamefont {Tracy~R.}\ \bibnamefont
  {Slatyer}},\ }\bibfield  {title} {\enquote {\bibinfo {title} {{Code package
  for calculating modified cosmic ionization and thermal histories with dark
  matter and other exotic energy injections}},}\ }\href {\doibase
  10.1103/PhysRevD.101.023530} {\bibfield  {journal} {\bibinfo  {journal}
  {Phys. Rev. D}\ }\textbf {\bibinfo {volume} {101}},\ \bibinfo {pages}
  {023530} (\bibinfo {year} {2020})},\ \Eprint
  {http://arxiv.org/abs/1904.09296} {arXiv:1904.09296 [astro-ph.CO]}
  \BibitemShut {NoStop}%
\bibitem [{\citenamefont {{Kogut}}\ \emph {et~al.}(2016)\citenamefont
  {{Kogut}}, \citenamefont {{Chluba}}, \citenamefont {{Fixsen}}, \citenamefont
  {{Meyer}},\ and\ \citenamefont {{Spergel}}}]{2016SPIE.9904E..0WK}%
  \BibitemOpen
  \bibfield  {author} {\bibinfo {author} {\bibfnamefont {Alan}\ \bibnamefont
  {{Kogut}}}, \bibinfo {author} {\bibfnamefont {Jens}\ \bibnamefont
  {{Chluba}}}, \bibinfo {author} {\bibfnamefont {Dale~J.}\ \bibnamefont
  {{Fixsen}}}, \bibinfo {author} {\bibfnamefont {Stephan}\ \bibnamefont
  {{Meyer}}}, \ and\ \bibinfo {author} {\bibfnamefont {David}\ \bibnamefont
  {{Spergel}}},\ }\bibfield  {title} {\enquote {\bibinfo {title} {{The
  Primordial Inflation Explorer (PIXIE)}},}\ }in\ \href {\doibase
  10.1117/12.2231090} {\emph {\bibinfo {booktitle} {Space Telescopes and
  Instrumentation 2016: Optical, Infrared, and Millimeter Wave}}},\ \bibinfo
  {series} {Society of Photo-Optical Instrumentation Engineers (SPIE)
  Conference Series}, Vol.\ \bibinfo {volume} {9904},\ \bibinfo {editor}
  {edited by\ \bibinfo {editor} {\bibfnamefont {Howard~A.}\ \bibnamefont
  {{MacEwen}}}, \bibinfo {editor} {\bibfnamefont {Giovanni~G.}\ \bibnamefont
  {{Fazio}}}, \bibinfo {editor} {\bibfnamefont {Makenzie}\ \bibnamefont
  {{Lystrup}}}, \bibinfo {editor} {\bibfnamefont {Natalie}\ \bibnamefont
  {{Batalha}}}, \bibinfo {editor} {\bibfnamefont {Nicholas}\ \bibnamefont
  {{Siegler}}}, \ and\ \bibinfo {editor} {\bibfnamefont {Edward~C.}\
  \bibnamefont {{Tong}}}}\ (\bibinfo {year} {2016})\ p.\ \bibinfo {pages}
  {99040W}\BibitemShut {NoStop}%
\bibitem [{\citenamefont {Bianchini}\ and\ \citenamefont
  {Fabbian}(2022)}]{Bianchini:2022dqh}%
  \BibitemOpen
  \bibfield  {author} {\bibinfo {author} {\bibfnamefont {Federico}\
  \bibnamefont {Bianchini}}\ and\ \bibinfo {author} {\bibfnamefont {Giulio}\
  \bibnamefont {Fabbian}},\ }\bibfield  {title} {\enquote {\bibinfo {title}
  {{CMB spectral distortions revisited: A new take on \ensuremath{\mu}
  distortions and primordial non-Gaussianities from FIRAS data}},}\ }\href
  {\doibase 10.1103/PhysRevD.106.063527} {\bibfield  {journal} {\bibinfo
  {journal} {Phys. Rev. D}\ }\textbf {\bibinfo {volume} {106}},\ \bibinfo
  {pages} {063527} (\bibinfo {year} {2022})},\ \Eprint
  {http://arxiv.org/abs/2206.02762} {arXiv:2206.02762 [astro-ph.CO]}
  \BibitemShut {NoStop}%
\bibitem [{\citenamefont {Refregier}\ \emph {et~al.}(2000)\citenamefont
  {Refregier}, \citenamefont {Komatsu}, \citenamefont {Spergel},\ and\
  \citenamefont {Pen}}]{Refregier:2000xz}%
  \BibitemOpen
  \bibfield  {author} {\bibinfo {author} {\bibfnamefont {Alexandre}\
  \bibnamefont {Refregier}}, \bibinfo {author} {\bibfnamefont {Eiichiro}\
  \bibnamefont {Komatsu}}, \bibinfo {author} {\bibfnamefont {David~N.}\
  \bibnamefont {Spergel}}, \ and\ \bibinfo {author} {\bibfnamefont {Ue-Li}\
  \bibnamefont {Pen}},\ }\bibfield  {title} {\enquote {\bibinfo {title} {{Power
  spectrum of the Sunyaev-Zel'dovich effect}},}\ }\href {\doibase
  10.1103/PhysRevD.61.123001} {\bibfield  {journal} {\bibinfo  {journal} {Phys.
  Rev. D}\ }\textbf {\bibinfo {volume} {61}},\ \bibinfo {pages} {123001}
  (\bibinfo {year} {2000})},\ \Eprint {http://arxiv.org/abs/astro-ph/9912180}
  {arXiv:astro-ph/9912180} \BibitemShut {NoStop}%
\bibitem [{\citenamefont {Zhang}\ \emph {et~al.}(2004)\citenamefont {Zhang},
  \citenamefont {Pen},\ and\ \citenamefont {Trac}}]{Zhang:2004fh}%
  \BibitemOpen
  \bibfield  {author} {\bibinfo {author} {\bibfnamefont {Peng-Jie}\
  \bibnamefont {Zhang}}, \bibinfo {author} {\bibfnamefont {Ue-Li}\ \bibnamefont
  {Pen}}, \ and\ \bibinfo {author} {\bibfnamefont {Hy}~\bibnamefont {Trac}},\
  }\bibfield  {title} {\enquote {\bibinfo {title} {{The Intergalactic medium
  temperature and Compton y parameter}},}\ }\href {\doibase
  10.1111/j.1365-2966.2004.08328.x} {\bibfield  {journal} {\bibinfo  {journal}
  {Mon. Not. Roy. Astron. Soc.}\ }\textbf {\bibinfo {volume} {355}},\ \bibinfo
  {pages} {451} (\bibinfo {year} {2004})},\ \Eprint
  {http://arxiv.org/abs/astro-ph/0402115} {arXiv:astro-ph/0402115} \BibitemShut
  {NoStop}%
\bibitem [{\citenamefont {Dolag}\ \emph {et~al.}(2016)\citenamefont {Dolag},
  \citenamefont {Komatsu},\ and\ \citenamefont {Sunyaev}}]{Dolag:2015dta}%
  \BibitemOpen
  \bibfield  {author} {\bibinfo {author} {\bibfnamefont {Klaus}\ \bibnamefont
  {Dolag}}, \bibinfo {author} {\bibfnamefont {Eiichiro}\ \bibnamefont
  {Komatsu}}, \ and\ \bibinfo {author} {\bibfnamefont {Rashid}\ \bibnamefont
  {Sunyaev}},\ }\bibfield  {title} {\enquote {\bibinfo {title} {{SZ effects in
  the Magneticum Pathfinder Simulation: Comparison with the Planck, SPT, and
  ACT results}},}\ }\href {\doibase 10.1093/mnras/stw2035} {\bibfield
  {journal} {\bibinfo  {journal} {Mon. Not. Roy. Astron. Soc.}\ }\textbf
  {\bibinfo {volume} {463}},\ \bibinfo {pages} {1797--1811} (\bibinfo {year}
  {2016})},\ \Eprint {http://arxiv.org/abs/1509.05134} {arXiv:1509.05134
  [astro-ph.CO]} \BibitemShut {NoStop}%
\bibitem [{\citenamefont {Hill}\ \emph {et~al.}(2015)\citenamefont {Hill},
  \citenamefont {Battaglia}, \citenamefont {Chluba}, \citenamefont {Ferraro},
  \citenamefont {Schaan},\ and\ \citenamefont {Spergel}}]{Hill:2015tqa}%
  \BibitemOpen
  \bibfield  {author} {\bibinfo {author} {\bibfnamefont {J.~Colin}\
  \bibnamefont {Hill}}, \bibinfo {author} {\bibfnamefont {Nick}\ \bibnamefont
  {Battaglia}}, \bibinfo {author} {\bibfnamefont {Jens}\ \bibnamefont
  {Chluba}}, \bibinfo {author} {\bibfnamefont {Simone}\ \bibnamefont
  {Ferraro}}, \bibinfo {author} {\bibfnamefont {Emmanuel}\ \bibnamefont
  {Schaan}}, \ and\ \bibinfo {author} {\bibfnamefont {David~N.}\ \bibnamefont
  {Spergel}},\ }\bibfield  {title} {\enquote {\bibinfo {title} {{Taking the
  Universe\textquoteright{}s Temperature with Spectral Distortions of the
  Cosmic Microwave Background}},}\ }\href {\doibase
  10.1103/PhysRevLett.115.261301} {\bibfield  {journal} {\bibinfo  {journal}
  {Phys. Rev. Lett.}\ }\textbf {\bibinfo {volume} {115}},\ \bibinfo {pages}
  {261301} (\bibinfo {year} {2015})},\ \Eprint
  {http://arxiv.org/abs/1507.01583} {arXiv:1507.01583 [astro-ph.CO]}
  \BibitemShut {NoStop}%
\bibitem [{\citenamefont {De~Zotti}\ \emph {et~al.}(2016)\citenamefont
  {De~Zotti}, \citenamefont {Negrello}, \citenamefont {Castex}, \citenamefont
  {Lapi},\ and\ \citenamefont {Bonato}}]{DeZotti:2015awh}%
  \BibitemOpen
  \bibfield  {author} {\bibinfo {author} {\bibfnamefont {G.}~\bibnamefont
  {De~Zotti}}, \bibinfo {author} {\bibfnamefont {M.}~\bibnamefont {Negrello}},
  \bibinfo {author} {\bibfnamefont {G.}~\bibnamefont {Castex}}, \bibinfo
  {author} {\bibfnamefont {A.}~\bibnamefont {Lapi}}, \ and\ \bibinfo {author}
  {\bibfnamefont {M.}~\bibnamefont {Bonato}},\ }\bibfield  {title} {\enquote
  {\bibinfo {title} {{Another look at distortions of the Cosmic Microwave
  Background spectrum}},}\ }\href {\doibase 10.1088/1475-7516/2016/03/047}
  {\bibfield  {journal} {\bibinfo  {journal} {JCAP}\ }\textbf {\bibinfo
  {volume} {03}},\ \bibinfo {pages} {047} (\bibinfo {year} {2016})},\ \Eprint
  {http://arxiv.org/abs/1512.04816} {arXiv:1512.04816 [astro-ph.CO]}
  \BibitemShut {NoStop}%
\bibitem [{\citenamefont {Chluba}\ and\ \citenamefont
  {Sunyaev}(2012)}]{Chluba:2011hw}%
  \BibitemOpen
  \bibfield  {author} {\bibinfo {author} {\bibfnamefont {J.}~\bibnamefont
  {Chluba}}\ and\ \bibinfo {author} {\bibfnamefont {R.~A.}\ \bibnamefont
  {Sunyaev}},\ }\bibfield  {title} {\enquote {\bibinfo {title} {{The evolution
  of CMB spectral distortions in the early Universe}},}\ }\href {\doibase
  10.1111/j.1365-2966.2011.19786.x} {\bibfield  {journal} {\bibinfo  {journal}
  {Mon. Not. Roy. Astron. Soc.}\ }\textbf {\bibinfo {volume} {419}},\ \bibinfo
  {pages} {1294--1314} (\bibinfo {year} {2012})},\ \Eprint
  {http://arxiv.org/abs/1109.6552} {arXiv:1109.6552 [astro-ph.CO]} \BibitemShut
  {NoStop}%
\bibitem [{\citenamefont {Khatri}\ and\ \citenamefont
  {Sunyaev}(2012)}]{Khatri:2012tw}%
  \BibitemOpen
  \bibfield  {author} {\bibinfo {author} {\bibfnamefont {Rishi}\ \bibnamefont
  {Khatri}}\ and\ \bibinfo {author} {\bibfnamefont {Rashid~A.}\ \bibnamefont
  {Sunyaev}},\ }\bibfield  {title} {\enquote {\bibinfo {title} {{Beyond y and
  \textbackslash{}mu: the shape of the CMB spectral distortions in the
  intermediate epoch, 1.5x10\textasciicircum{}4 \ensuremath{<} z \ensuremath{<}
  2x10\textasciicircum{}5}},}\ }\href {\doibase 10.1088/1475-7516/2012/09/016}
  {\bibfield  {journal} {\bibinfo  {journal} {JCAP}\ }\textbf {\bibinfo
  {volume} {09}},\ \bibinfo {pages} {016} (\bibinfo {year} {2012})},\ \Eprint
  {http://arxiv.org/abs/1207.6654} {arXiv:1207.6654 [astro-ph.CO]} \BibitemShut
  {NoStop}%
\bibitem [{\citenamefont {Wong}\ \emph {et~al.}(2006)\citenamefont {Wong},
  \citenamefont {Seager},\ and\ \citenamefont {Scott}}]{Wong:2005yr}%
  \BibitemOpen
  \bibfield  {author} {\bibinfo {author} {\bibfnamefont {Wan~Yan}\ \bibnamefont
  {Wong}}, \bibinfo {author} {\bibfnamefont {Sara}\ \bibnamefont {Seager}}, \
  and\ \bibinfo {author} {\bibfnamefont {Douglas}\ \bibnamefont {Scott}},\
  }\bibfield  {title} {\enquote {\bibinfo {title} {{Spectral distortions to the
  cosmic microwave background from the recombination of hydrogen and
  helium}},}\ }\href {\doibase 10.1111/j.1365-2966.2006.10076.x} {\bibfield
  {journal} {\bibinfo  {journal} {Mon. Not. Roy. Astron. Soc.}\ }\textbf
  {\bibinfo {volume} {367}},\ \bibinfo {pages} {1666--1676} (\bibinfo {year}
  {2006})},\ \Eprint {http://arxiv.org/abs/astro-ph/0510634}
  {arXiv:astro-ph/0510634} \BibitemShut {NoStop}%
\bibitem [{\citenamefont {Rubino-Martin}\ \emph {et~al.}(2008)\citenamefont
  {Rubino-Martin}, \citenamefont {Chluba},\ and\ \citenamefont
  {Sunyaev}}]{Rubino-Martin:2007tua}%
  \BibitemOpen
  \bibfield  {author} {\bibinfo {author} {\bibfnamefont {J.~A.}\ \bibnamefont
  {Rubino-Martin}}, \bibinfo {author} {\bibfnamefont {J.}~\bibnamefont
  {Chluba}}, \ and\ \bibinfo {author} {\bibfnamefont {R.~A.}\ \bibnamefont
  {Sunyaev}},\ }\bibfield  {title} {\enquote {\bibinfo {title} {{Lines in the
  cosmic microwave background spectrum from the epoch of cosmological helium
  recombination}},}\ }\href {\doibase 10.1051/0004-6361:20078993} {\bibfield
  {journal} {\bibinfo  {journal} {Astron. Astrophys.}\ }\textbf {\bibinfo
  {volume} {485}},\ \bibinfo {pages} {377} (\bibinfo {year} {2008})},\ \Eprint
  {http://arxiv.org/abs/0711.0594} {arXiv:0711.0594 [astro-ph]} \BibitemShut
  {NoStop}%
\bibitem [{\citenamefont {Puchwein}\ \emph {et~al.}(2019)\citenamefont
  {Puchwein}, \citenamefont {Haardt}, \citenamefont {Haehnelt},\ and\
  \citenamefont {Madau}}]{Puchwein:2018arm}%
  \BibitemOpen
  \bibfield  {author} {\bibinfo {author} {\bibfnamefont {Ewald}\ \bibnamefont
  {Puchwein}}, \bibinfo {author} {\bibfnamefont {Francesco}\ \bibnamefont
  {Haardt}}, \bibinfo {author} {\bibfnamefont {Martin~G.}\ \bibnamefont
  {Haehnelt}}, \ and\ \bibinfo {author} {\bibfnamefont {Piero}\ \bibnamefont
  {Madau}},\ }\bibfield  {title} {\enquote {\bibinfo {title} {{Consistent
  modelling of the meta-galactic UV background and the thermal/ionization
  history of the intergalactic medium}},}\ }\href {\doibase
  10.1093/mnras/stz222} {\bibfield  {journal} {\bibinfo  {journal} {Mon. Not.
  Roy. Astron. Soc.}\ }\textbf {\bibinfo {volume} {485}},\ \bibinfo {pages}
  {47--68} (\bibinfo {year} {2019})},\ \Eprint
  {http://arxiv.org/abs/1801.04931} {arXiv:1801.04931 [astro-ph.GA]}
  \BibitemShut {NoStop}%
\bibitem [{\citenamefont {Cirelli}\ \emph {et~al.}(2011)\citenamefont
  {Cirelli}, \citenamefont {Corcella}, \citenamefont {Hektor}, \citenamefont
  {Hutsi}, \citenamefont {Kadastik}, \citenamefont {Panci}, \citenamefont
  {Raidal}, \citenamefont {Sala},\ and\ \citenamefont
  {Strumia}}]{Cirelli:2010xx}%
  \BibitemOpen
  \bibfield  {author} {\bibinfo {author} {\bibfnamefont {Marco}\ \bibnamefont
  {Cirelli}}, \bibinfo {author} {\bibfnamefont {Gennaro}\ \bibnamefont
  {Corcella}}, \bibinfo {author} {\bibfnamefont {Andi}\ \bibnamefont {Hektor}},
  \bibinfo {author} {\bibfnamefont {Gert}\ \bibnamefont {Hutsi}}, \bibinfo
  {author} {\bibfnamefont {Mario}\ \bibnamefont {Kadastik}}, \bibinfo {author}
  {\bibfnamefont {Paolo}\ \bibnamefont {Panci}}, \bibinfo {author}
  {\bibfnamefont {Martti}\ \bibnamefont {Raidal}}, \bibinfo {author}
  {\bibfnamefont {Filippo}\ \bibnamefont {Sala}}, \ and\ \bibinfo {author}
  {\bibfnamefont {Alessandro}\ \bibnamefont {Strumia}},\ }\bibfield  {title}
  {\enquote {\bibinfo {title} {{PPPC 4 DM ID: A Poor Particle Physicist
  Cookbook for Dark Matter Indirect Detection}},}\ }\href {\doibase
  10.1088/1475-7516/2012/10/E01} {\bibfield  {journal} {\bibinfo  {journal}
  {JCAP}\ }\textbf {\bibinfo {volume} {03}},\ \bibinfo {pages} {051} (\bibinfo
  {year} {2011})},\ \bibinfo {note} {[Erratum: JCAP 10, E01 (2012)]},\ \Eprint
  {http://arxiv.org/abs/1012.4515} {arXiv:1012.4515 [hep-ph]} \BibitemShut
  {NoStop}%
\bibitem [{\citenamefont {Aghanim}\ \emph {et~al.}(2020)\citenamefont {Aghanim}
  \emph {et~al.}}]{Planck:2018vyg}%
  \BibitemOpen
  \bibfield  {author} {\bibinfo {author} {\bibfnamefont {N.}~\bibnamefont
  {Aghanim}} \emph {et~al.} (\bibinfo {collaboration} {Planck}),\ }\bibfield
  {title} {\enquote {\bibinfo {title} {{Planck 2018 results. VI. Cosmological
  parameters}},}\ }\href {\doibase 10.1051/0004-6361/201833910} {\bibfield
  {journal} {\bibinfo  {journal} {Astron. Astrophys.}\ }\textbf {\bibinfo
  {volume} {641}},\ \bibinfo {pages} {A6} (\bibinfo {year} {2020})},\ \bibinfo
  {note} {[Erratum: Astron.Astrophys. 652, C4 (2021)]},\ \Eprint
  {http://arxiv.org/abs/1807.06209} {arXiv:1807.06209 [astro-ph.CO]}
  \BibitemShut {NoStop}%
\bibitem [{\citenamefont {Seager}\ \emph {et~al.}(2000)\citenamefont {Seager},
  \citenamefont {Sasselov},\ and\ \citenamefont {Scott}}]{Seager:1999km}%
  \BibitemOpen
  \bibfield  {author} {\bibinfo {author} {\bibfnamefont {Sara}\ \bibnamefont
  {Seager}}, \bibinfo {author} {\bibfnamefont {Dimitar~D.}\ \bibnamefont
  {Sasselov}}, \ and\ \bibinfo {author} {\bibfnamefont {Douglas}\ \bibnamefont
  {Scott}},\ }\bibfield  {title} {\enquote {\bibinfo {title} {{How exactly did
  the universe become neutral?}}}\ }\href {\doibase 10.1086/313388} {\bibfield
  {journal} {\bibinfo  {journal} {Astrophys. J. Suppl.}\ }\textbf {\bibinfo
  {volume} {128}},\ \bibinfo {pages} {407--430} (\bibinfo {year} {2000})},\
  \Eprint {http://arxiv.org/abs/astro-ph/9912182} {arXiv:astro-ph/9912182
  [astro-ph]} \BibitemShut {NoStop}%
\bibitem [{\citenamefont {Seager}\ \emph {et~al.}(1999)\citenamefont {Seager},
  \citenamefont {Sasselov},\ and\ \citenamefont {Scott}}]{Seager_1999}%
  \BibitemOpen
  \bibfield  {author} {\bibinfo {author} {\bibfnamefont {S.}~\bibnamefont
  {Seager}}, \bibinfo {author} {\bibfnamefont {D.~D.}\ \bibnamefont
  {Sasselov}}, \ and\ \bibinfo {author} {\bibfnamefont {D.}~\bibnamefont
  {Scott}},\ }\bibfield  {title} {\enquote {\bibinfo {title} {A new calculation
  of the recombination epoch},}\ }\href {\doibase 10.1086/312250} {\bibfield
  {journal} {\bibinfo  {journal} {The Astrophysical Journal}\ }\textbf
  {\bibinfo {volume} {523}},\ \bibinfo {pages} {L1--L5} (\bibinfo {year}
  {1999})}\BibitemShut {NoStop}%
\bibitem [{\citenamefont {Ferrara}\ \emph {et~al.}(2012)\citenamefont
  {Ferrara}, \citenamefont {Valdés}, \citenamefont {Yoshida},\ and\
  \citenamefont {Evoli}}]{MEDEAII}%
  \BibitemOpen
  \bibfield  {author} {\bibinfo {author} {\bibfnamefont {A.}~\bibnamefont
  {Ferrara}}, \bibinfo {author} {\bibfnamefont {M.}~\bibnamefont {Valdés}},
  \bibinfo {author} {\bibfnamefont {N.}~\bibnamefont {Yoshida}}, \ and\
  \bibinfo {author} {\bibfnamefont {C.}~\bibnamefont {Evoli}},\ }\bibfield
  {title} {\enquote {\bibinfo {title} {{Energy deposition by weakly interacting
  massive particles: a comprehensive study}},}\ }\href {\doibase
  10.1111/j.1365-2966.2012.20624.x} {\bibfield  {journal} {\bibinfo  {journal}
  {Monthly Notices of the Royal Astronomical Society}\ }\textbf {\bibinfo
  {volume} {422}},\ \bibinfo {pages} {420--433} (\bibinfo {year} {2012})},\
  \Eprint
  {http://arxiv.org/abs/http://oup.prod.sis.lan/mnras/article-pdf/422/1/420/18597009/mnras0422-0420.pdf}
  {http://oup.prod.sis.lan/mnras/article-pdf/422/1/420/18597009/mnras0422-0420.pdf}
  \BibitemShut {NoStop}%
\bibitem [{\citenamefont {Cadamuro}\ and\ \citenamefont
  {Redondo}(2012)}]{Cadamuro:2011fd}%
  \BibitemOpen
  \bibfield  {author} {\bibinfo {author} {\bibfnamefont {Davide}\ \bibnamefont
  {Cadamuro}}\ and\ \bibinfo {author} {\bibfnamefont {Javier}\ \bibnamefont
  {Redondo}},\ }\bibfield  {title} {\enquote {\bibinfo {title} {{Cosmological
  bounds on pseudo Nambu-Goldstone bosons}},}\ }\href {\doibase
  10.1088/1475-7516/2012/02/032} {\bibfield  {journal} {\bibinfo  {journal}
  {JCAP}\ }\textbf {\bibinfo {volume} {02}},\ \bibinfo {pages} {032} (\bibinfo
  {year} {2012})},\ \Eprint {http://arxiv.org/abs/1110.2895} {arXiv:1110.2895
  [hep-ph]} \BibitemShut {NoStop}%
\bibitem [{\citenamefont {Wadekar}\ and\ \citenamefont
  {Wang}(2021)}]{Wadekar:2021qae}%
  \BibitemOpen
  \bibfield  {author} {\bibinfo {author} {\bibfnamefont {Digvijay}\
  \bibnamefont {Wadekar}}\ and\ \bibinfo {author} {\bibfnamefont {Zihui}\
  \bibnamefont {Wang}},\ }\bibfield  {title} {\enquote {\bibinfo {title}
  {{Strong constraints on decay and annihilation of dark matter from heating of
  gas-rich dwarf galaxies}},}\ }\href@noop {} {\  (\bibinfo {year} {2021})},\
  \Eprint {http://arxiv.org/abs/2111.08025} {arXiv:2111.08025 [hep-ph]}
  \BibitemShut {NoStop}%
\bibitem [{\citenamefont {O'Hare}(2020)}]{AxionLimits}%
  \BibitemOpen
  \bibfield  {author} {\bibinfo {author} {\bibfnamefont {Ciaran}\ \bibnamefont
  {O'Hare}},\ }\href {\doibase 10.5281/zenodo.3932430} {\enquote {\bibinfo
  {title} {cajohare/axionlimits: Axionlimits},}\ }\bibinfo {howpublished}
  {\url{https://cajohare.github.io/AxionLimits/}} (\bibinfo {year}
  {2020})\BibitemShut {NoStop}%
\bibitem [{\citenamefont {Nakayama}\ and\ \citenamefont
  {Yin}(2022)}]{Nakayama:2022jza}%
  \BibitemOpen
  \bibfield  {author} {\bibinfo {author} {\bibfnamefont {Kazunori}\
  \bibnamefont {Nakayama}}\ and\ \bibinfo {author} {\bibfnamefont {Wen}\
  \bibnamefont {Yin}},\ }\bibfield  {title} {\enquote {\bibinfo {title}
  {{Anisotropic cosmic optical background bound for decaying dark matter in
  light of the LORRI anomaly}},}\ }\href {\doibase 10.1103/PhysRevD.106.103505}
  {\bibfield  {journal} {\bibinfo  {journal} {Phys. Rev. D}\ }\textbf {\bibinfo
  {volume} {106}},\ \bibinfo {pages} {103505} (\bibinfo {year} {2022})},\
  \Eprint {http://arxiv.org/abs/2205.01079} {arXiv:2205.01079 [hep-ph]}
  \BibitemShut {NoStop}%
\bibitem [{\citenamefont {Aghanim}\ \emph {et~al.}(2016)\citenamefont {Aghanim}
  \emph {et~al.}}]{Planck:2015bpv}%
  \BibitemOpen
  \bibfield  {author} {\bibinfo {author} {\bibfnamefont {N.}~\bibnamefont
  {Aghanim}} \emph {et~al.} (\bibinfo {collaboration} {Planck}),\ }\bibfield
  {title} {\enquote {\bibinfo {title} {{Planck 2015 results. XI. CMB power
  spectra, likelihoods, and robustness of parameters}},}\ }\href {\doibase
  10.1051/0004-6361/201526926} {\bibfield  {journal} {\bibinfo  {journal}
  {Astron. Astrophys.}\ }\textbf {\bibinfo {volume} {594}},\ \bibinfo {pages}
  {A11} (\bibinfo {year} {2016})},\ \Eprint {http://arxiv.org/abs/1507.02704}
  {arXiv:1507.02704 [astro-ph.CO]} \BibitemShut {NoStop}%
\bibitem [{\citenamefont {Bolliet}\ and\ \citenamefont
  {Chluba}(2023)}]{Bolliet:private_comm}%
  \BibitemOpen
  \bibfield  {author} {\bibinfo {author} {\bibfnamefont {Boris}\ \bibnamefont
  {Bolliet}}\ and\ \bibinfo {author} {\bibfnamefont {Jens}\ \bibnamefont
  {Chluba}},\ }\href@noop {} {}\bibinfo {howpublished} {private communication}
  (\bibinfo {year} {2023})\BibitemShut {NoStop}%
\bibitem [{\citenamefont {Langhoff}\ \emph {et~al.}(2022)\citenamefont
  {Langhoff}, \citenamefont {Outmezguine},\ and\ \citenamefont
  {Rodd}}]{Langhoff:2022bij}%
  \BibitemOpen
  \bibfield  {author} {\bibinfo {author} {\bibfnamefont {Kevin}\ \bibnamefont
  {Langhoff}}, \bibinfo {author} {\bibfnamefont {Nadav~Joseph}\ \bibnamefont
  {Outmezguine}}, \ and\ \bibinfo {author} {\bibfnamefont {Nicholas~L.}\
  \bibnamefont {Rodd}},\ }\bibfield  {title} {\enquote {\bibinfo {title} {{The
  Irreducible Axion Background}},}\ }\href@noop {} {\  (\bibinfo {year}
  {2022})},\ \Eprint {http://arxiv.org/abs/2209.06216} {arXiv:2209.06216
  [hep-ph]} \BibitemShut {NoStop}%
\bibitem [{\citenamefont {Capozzi}\ \emph {et~al.}(2023)\citenamefont
  {Capozzi}, \citenamefont {Ferreira}, \citenamefont {Lopez-Honorez},\ and\
  \citenamefont {Mena}}]{Capozzi:2023xie}%
  \BibitemOpen
  \bibfield  {author} {\bibinfo {author} {\bibfnamefont {Francesco}\
  \bibnamefont {Capozzi}}, \bibinfo {author} {\bibfnamefont {Ricardo~Z.}\
  \bibnamefont {Ferreira}}, \bibinfo {author} {\bibfnamefont {Laura}\
  \bibnamefont {Lopez-Honorez}}, \ and\ \bibinfo {author} {\bibfnamefont
  {Olga}\ \bibnamefont {Mena}},\ }\bibfield  {title} {\enquote {\bibinfo
  {title} {{CMB and Lyman-$\alpha$ constraints on dark matter decays to
  photons}},}\ }\href@noop {} {\  (\bibinfo {year} {2023})},\ \Eprint
  {http://arxiv.org/abs/2303.07426} {arXiv:2303.07426 [astro-ph.CO]}
  \BibitemShut {NoStop}%
\bibitem [{\citenamefont {Kluyver}\ \emph {et~al.}(2016)\citenamefont {Kluyver}
  \emph {et~al.}}]{Kluyver2016JupyterN}%
  \BibitemOpen
  \bibfield  {author} {\bibinfo {author} {\bibfnamefont {Thomas}\ \bibnamefont
  {Kluyver}} \emph {et~al.},\ }\bibfield  {title} {\enquote {\bibinfo {title}
  {Jupyter notebooks - a publishing format for reproducible computational
  workflows},}\ }in\ \href@noop {} {\emph {\bibinfo {booktitle} {ELPUB}}}\
  (\bibinfo {year} {2016})\BibitemShut {NoStop}%
\bibitem [{\citenamefont {Hunter}(2007)}]{Hunter:2007ouj}%
  \BibitemOpen
  \bibfield  {author} {\bibinfo {author} {\bibfnamefont {John~D.}\ \bibnamefont
  {Hunter}},\ }\bibfield  {title} {\enquote {\bibinfo {title} {{Matplotlib: A
  2D Graphics Environment}},}\ }\href {\doibase 10.1109/MCSE.2007.55}
  {\bibfield  {journal} {\bibinfo  {journal} {Comput. Sci. Eng.}\ }\textbf
  {\bibinfo {volume} {9}},\ \bibinfo {pages} {90--95} (\bibinfo {year}
  {2007})}\BibitemShut {NoStop}%
\bibitem [{\citenamefont {Harris}\ \emph {et~al.}(2020)\citenamefont {Harris}
  \emph {et~al.}}]{Harris:2020xlr}%
  \BibitemOpen
  \bibfield  {author} {\bibinfo {author} {\bibfnamefont {Charles~R.}\
  \bibnamefont {Harris}} \emph {et~al.},\ }\bibfield  {title} {\enquote
  {\bibinfo {title} {{Array programming with NumPy}},}\ }\href {\doibase
  10.1038/s41586-020-2649-2} {\bibfield  {journal} {\bibinfo  {journal}
  {Nature}\ }\textbf {\bibinfo {volume} {585}},\ \bibinfo {pages} {357--362}
  (\bibinfo {year} {2020})},\ \Eprint {http://arxiv.org/abs/2006.10256}
  {arXiv:2006.10256 [cs.MS]} \BibitemShut {NoStop}%
\bibitem [{\citenamefont {{Virtanen}}\ \emph {et~al.}(2020)\citenamefont
  {{Virtanen}}, \citenamefont {{Gommers}}, \citenamefont {{Oliphant}},
  \citenamefont {{Haberland}}, \citenamefont {{Reddy}}, \citenamefont
  {{Cournapeau}}, \citenamefont {{Burovski}}, \citenamefont {{Peterson}},
  \citenamefont {{Weckesser}}, \citenamefont {{Bright}}, \citenamefont {{van
  der Walt}}, \citenamefont {{Brett}}, \citenamefont {{Wilson}}, \citenamefont
  {{Millman}}, \citenamefont {{Mayorov}}, \citenamefont {{Nelson}},
  \citenamefont {{Jones}}, \citenamefont {{Kern}}, \citenamefont {{Larson}},
  \citenamefont {{Carey}}, \citenamefont {{Polat}}, \citenamefont {{Feng}},
  \citenamefont {{Moore}}, \citenamefont {{VanderPlas}}, \citenamefont
  {{Laxalde}}, \citenamefont {{Perktold}}, \citenamefont {{Cimrman}},
  \citenamefont {{Henriksen}}, \citenamefont {{Quintero}}, \citenamefont
  {{Harris}}, \citenamefont {{Archibald}}, \citenamefont {{Ribeiro}},
  \citenamefont {{Pedregosa}}, \citenamefont {{van Mulbregt}},\ and\
  \citenamefont {{SciPy 1. 0 Contributors}}}]{2020NatMe..17..261V}%
  \BibitemOpen
  \bibfield  {author} {\bibinfo {author} {\bibfnamefont {Pauli}\ \bibnamefont
  {{Virtanen}}}, \bibinfo {author} {\bibfnamefont {Ralf}\ \bibnamefont
  {{Gommers}}}, \bibinfo {author} {\bibfnamefont {Travis~E.}\ \bibnamefont
  {{Oliphant}}}, \bibinfo {author} {\bibfnamefont {Matt}\ \bibnamefont
  {{Haberland}}}, \bibinfo {author} {\bibfnamefont {Tyler}\ \bibnamefont
  {{Reddy}}}, \bibinfo {author} {\bibfnamefont {David}\ \bibnamefont
  {{Cournapeau}}}, \bibinfo {author} {\bibfnamefont {Evgeni}\ \bibnamefont
  {{Burovski}}}, \bibinfo {author} {\bibfnamefont {Pearu}\ \bibnamefont
  {{Peterson}}}, \bibinfo {author} {\bibfnamefont {Warren}\ \bibnamefont
  {{Weckesser}}}, \bibinfo {author} {\bibfnamefont {Jonathan}\ \bibnamefont
  {{Bright}}}, \bibinfo {author} {\bibfnamefont {St{\'e}fan~J.}\ \bibnamefont
  {{van der Walt}}}, \bibinfo {author} {\bibfnamefont {Matthew}\ \bibnamefont
  {{Brett}}}, \bibinfo {author} {\bibfnamefont {Joshua}\ \bibnamefont
  {{Wilson}}}, \bibinfo {author} {\bibfnamefont {K.~Jarrod}\ \bibnamefont
  {{Millman}}}, \bibinfo {author} {\bibfnamefont {Nikolay}\ \bibnamefont
  {{Mayorov}}}, \bibinfo {author} {\bibfnamefont {Andrew R.~J.}\ \bibnamefont
  {{Nelson}}}, \bibinfo {author} {\bibfnamefont {Eric}\ \bibnamefont
  {{Jones}}}, \bibinfo {author} {\bibfnamefont {Robert}\ \bibnamefont
  {{Kern}}}, \bibinfo {author} {\bibfnamefont {Eric}\ \bibnamefont {{Larson}}},
  \bibinfo {author} {\bibfnamefont {C.~J.}\ \bibnamefont {{Carey}}}, \bibinfo
  {author} {\bibfnamefont {{\.I}lhan}\ \bibnamefont {{Polat}}}, \bibinfo
  {author} {\bibfnamefont {Yu}~\bibnamefont {{Feng}}}, \bibinfo {author}
  {\bibfnamefont {Eric~W.}\ \bibnamefont {{Moore}}}, \bibinfo {author}
  {\bibfnamefont {Jake}\ \bibnamefont {{VanderPlas}}}, \bibinfo {author}
  {\bibfnamefont {Denis}\ \bibnamefont {{Laxalde}}}, \bibinfo {author}
  {\bibfnamefont {Josef}\ \bibnamefont {{Perktold}}}, \bibinfo {author}
  {\bibfnamefont {Robert}\ \bibnamefont {{Cimrman}}}, \bibinfo {author}
  {\bibfnamefont {Ian}\ \bibnamefont {{Henriksen}}}, \bibinfo {author}
  {\bibfnamefont {E.~A.}\ \bibnamefont {{Quintero}}}, \bibinfo {author}
  {\bibfnamefont {Charles~R.}\ \bibnamefont {{Harris}}}, \bibinfo {author}
  {\bibfnamefont {Anne~M.}\ \bibnamefont {{Archibald}}}, \bibinfo {author}
  {\bibfnamefont {Ant{\^o}nio~H.}\ \bibnamefont {{Ribeiro}}}, \bibinfo {author}
  {\bibfnamefont {Fabian}\ \bibnamefont {{Pedregosa}}}, \bibinfo {author}
  {\bibfnamefont {Paul}\ \bibnamefont {{van Mulbregt}}}, \ and\ \bibinfo
  {author} {\bibnamefont {{SciPy 1. 0 Contributors}}},\ }\bibfield  {title}
  {\enquote {\bibinfo {title} {{SciPy 1.0: fundamental algorithms for
  scientific computing in Python}},}\ }\href {\doibase
  10.1038/s41592-019-0686-2} {\bibfield  {journal} {\bibinfo  {journal} {Nature
  Methods}\ }\textbf {\bibinfo {volume} {17}},\ \bibinfo {pages} {261--272}
  (\bibinfo {year} {2020})},\ \Eprint {http://arxiv.org/abs/1907.10121}
  {arXiv:1907.10121 [cs.MS]} \BibitemShut {NoStop}%
\bibitem [{\citenamefont {da~Costa-Luis}(2019)}]{daCosta-Luis2019}%
  \BibitemOpen
  \bibfield  {author} {\bibinfo {author} {\bibfnamefont {Casper~O.}\
  \bibnamefont {da~Costa-Luis}},\ }\bibfield  {title} {\enquote {\bibinfo
  {title} {`tqdm`: A fast, extensible progress meter for python and cli},}\
  }\href {\doibase 10.21105/joss.01277} {\bibfield  {journal} {\bibinfo
  {journal} {Journal of Open Source Software}\ }\textbf {\bibinfo {volume}
  {4}},\ \bibinfo {pages} {1277} (\bibinfo {year} {2019})}\BibitemShut
  {NoStop}%
\bibitem [{\citenamefont {Rohatgi}(2022)}]{Rohatgi2022}%
  \BibitemOpen
  \bibfield  {author} {\bibinfo {author} {\bibfnamefont {Ankit}\ \bibnamefont
  {Rohatgi}},\ }\href {https://automeris.io/WebPlotDigitizer} {\enquote
  {\bibinfo {title} {Webplotdigitizer: Version 4.6},}\ } (\bibinfo {year}
  {2022})\BibitemShut {NoStop}%
\bibitem [{\citenamefont {{Svensson}}\ and\ \citenamefont
  {{Zdziarski}}(1990)}]{1990ApJ...349..415S}%
  \BibitemOpen
  \bibfield  {author} {\bibinfo {author} {\bibfnamefont {Roland}\ \bibnamefont
  {{Svensson}}}\ and\ \bibinfo {author} {\bibfnamefont {Andrzej}\ \bibnamefont
  {{Zdziarski}}},\ }\bibfield  {title} {\enquote {\bibinfo {title}
  {{Photon-Photon Scattering of Gamma Rays at Cosmological Distances}},}\
  }\href {\doibase 10.1086/168325} {\bibfield  {journal} {\bibinfo  {journal}
  {ApJ}\ }\textbf {\bibinfo {volume} {349}},\ \bibinfo {pages} {415} (\bibinfo
  {year} {1990})}\BibitemShut {NoStop}%
\bibitem [{\citenamefont {{Furlanetto}}\ and\ \citenamefont
  {{Stoever}}(2010)}]{2010MNRAS.404.1869F}%
  \BibitemOpen
  \bibfield  {author} {\bibinfo {author} {\bibfnamefont {Steven~R.}\
  \bibnamefont {{Furlanetto}}}\ and\ \bibinfo {author} {\bibfnamefont
  {Samuel~Johnson}\ \bibnamefont {{Stoever}}},\ }\bibfield  {title} {\enquote
  {\bibinfo {title} {{Secondary ionization and heating by fast electrons}},}\
  }\href {\doibase 10.1111/j.1365-2966.2010.16401.x} {\bibfield  {journal}
  {\bibinfo  {journal} {Monthly Notices of the Royal Astronomical Society}\
  }\textbf {\bibinfo {volume} {404}},\ \bibinfo {pages} {1869--1878} (\bibinfo
  {year} {2010})},\ \Eprint {http://arxiv.org/abs/0910.4410} {arXiv:0910.4410
  [astro-ph.CO]} \BibitemShut {NoStop}%
\end{thebibliography}%

\end{document}